\documentclass[letterpaper,11pt]{article}

\usepackage{fancyhdr}
\usepackage{amsmath}
\usepackage{amsbsy}
\usepackage{amssymb}
\usepackage{amsthm}
\usepackage{amscd}
\usepackage{amsfonts}
\usepackage{supertabular}
\usepackage{graphics}
\usepackage{graphicx}
\usepackage{verbatim}
\usepackage{subfigure}
\usepackage{epsfig}
\usepackage{xspace}
\usepackage{euscript}
\usepackage{alltt}
\usepackage{boxedminipage}
\usepackage{float}
\usepackage{times}
\usepackage[colorlinks]{hyperref}
\usepackage[square,numbers]{natbib}
\usepackage{setspace}
\usepackage{authblk}

\usepackage{wasysym}
\usepackage{algorithm}
\usepackage{algorithmic}

\def\Put(#1,#2)#3{\leavevmode\makebox(0,0){\put(#1,#2){#3}}}

\begin{document}

\title{Generalized loading-unloading contact laws for elasto-plastic spheres with bonding strength}
\author[1]{Marcial Gonzalez \thanks{marcial-gonzalez@purdue.edu}}
\affil[1]{School of Mechanical Engineering, Purdue University, \break West Lafayette, IN 47907, USA}
\maketitle

\begin{abstract}

We present generalized loading-unloading contact laws for elasto-plastic spheres with bonding strength. The proposed mechanistic contact laws are continuous at the onset of unloading by means of a regularization term, in the spirit of a cohesive zone model, that introduces a small and controllable error in the conditions for interparticle breakage. This continuity property is in sharp contrast with the behavior of standard mechanistic loading and unloading contact theories, which exhibit a discontinuity at the onset of unloading when particles form solid bridges during plastic deformation. The formulation depends on five material properties, namely two elastic properties (Young's modulus and Poisson's ratio), two plastic properties (a plastic stiffness and a power-law hardening exponent) and one fracture mechanics property (fracture toughness), and its predictions are in agreement with detailed finite-element simulations. The numerical robustness and efficiency of the proposed formulation are borne out by performing three-dimensional particle mechanics static calculations of microstructure evolution during the three most important steps of powder die-compaction, namely during compaction, unloading, and ejection. These simulations reveal the evolution, up to relative densities close to one, of microstructural features, process variables and compact mechanical attributes which are quantitatively similar to those experimentally observed and in remarkable agreement with the (semi-)empirical formulae reported in the literature.
\end{abstract}

\section{Introduction}

Many physical mechanisms are required to convert a powder bed confined inside a rigid die into a compressed solid compact by the sole application of a compaction force. Typically, the initial stage of this process is characterized by rearrangement of particles that leads to the formation of a closely packed granular system. In the subsequent stage, the porosity or the packing volume cannot be further reduced by particle rearrangement and therefore particles undergo brittle fracture or plastic deformation, or both \cite{Celik-2016, Alderborn-1996}. It is indeed these dissipative and irreversible processes, in which the volume of the powder bed is reduced, that ultimately give rise to compact formation inside the die. Specifically, fracture and permanent deformation generate particle-to-particle contact surface and thus the opportunity for bond formation.  The understanding of microstructure formation and evolution during this process is therefore of paramount importance to elucidate strength formation. Particle size, shape, and roughness affect the initial stage of compaction, but it is fragmentation and plastic deformation that dominate the synthesis of highly dense compacts \cite{Duberg-1985}. For polymeric solids, fracture and plastic behavior are dependent on the physical form of the material; that is amorphous polymers have a tendency to ductile, elasto-plastic deformation whereas crystalline polymers exhibit brittle failure at room temperature \cite{Kinloch-2013}. In addition, most materials exhibit a brittle-ductile transition temperature that is pressure dependent and strain rate dependent---high temperature and low strain rate promote ductile, plastic behavior, while low temperature and high strain rate promote brittle fracture \cite{Kinloch-2013}. It is worth noting that the formation of particle-to-particle contact surface then clearly depends on both material properties and process variables, such as compaction speed and temperature.

\begin{figure}[htbp]
    \centering
    \includegraphics[scale=0.49]{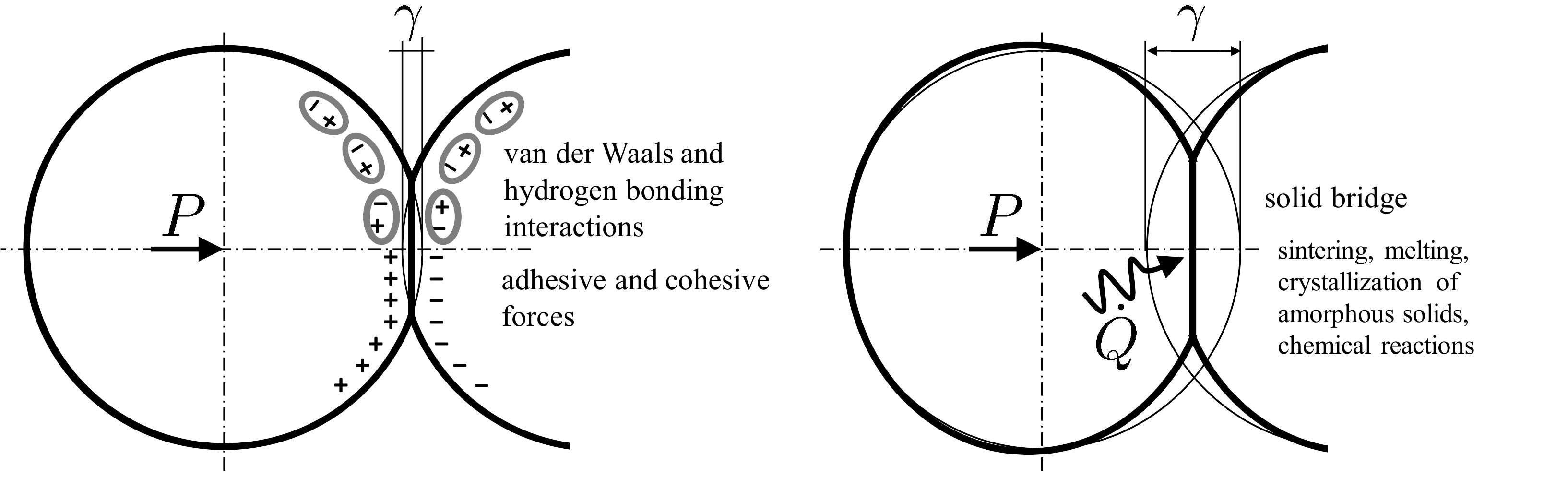}
    \caption{Different bonding mechanisms participating in the interaction between particles. Attractive interfacial interactions are dominant under small deformations whereas solid bridge formation occurs under large deformations.}
    \label{Fig-AdhesiveMechanisms}
\end{figure}

Bonding surface area can be regarded as the effective surface area that is involved in the interaction between particles. According to  Rumpf \cite{Rumpf-19xx}, the bonding mechanisms participating in this interaction can be classified into five different types: (i) formation of solid bridges (driven by processes such as sintering, melting, crystallization of amorphous solids, or chemical reactions); (ii) bonding between movable liquids (caused by capillary and surface tension forces in the presence of some moisture); (iii) non-freely movable binder bridges (resulting from adhesive and cohesive forces in binders such as those used in wet granulation); (iv) long-range attractive forces between solid particles (such as van der Waals and hydrogen bonding interactions); and (v) mechanical particle interlocking. For particles with low aspect ratio and low roughness, mechanical interlocking can be neglected. Among the remaining mechanisms, there is general agreement that solid bridge formation and attractive interfacial forces are the major contributions to strength formation. Figure~\ref{Fig-AdhesiveMechanisms} illustrates these different mechanisms and that attractive interfacial interactions are dominant under small deformations whereas solid bridge formation occurs under large deformations.

During powder compaction, regions of high plastic deformation are formed around particle contact interfaces. Plastic deformation dissipates energy as heat and thus locally increases the temperature. This change in temperature may, in turn, promote molecular movement and, consequently, the formation of a new solid region that bridges the particles in contact.  It is then the formation of an interconnected network of solid bridges that enables strength formation in the solid compact. The strength of these solid bridges will vary from material to material depending on the forces that hold the (poly)crystals and amorphous solids together \cite{Rumpf-19xxb,Down-1985,Mitchell-1984,Ahineck-1989}. It is worth noting that this process is irreversible, and thus it is not possible to divide the compacted system into its original particles. The system can only be separated by fracture of either solid bridges or particles, whichever is weaker. This irreversibility is one of the main differences between solid bridges and the attractive interfacial interactions. This observation also suggests to characterize the strength of the solid bridges with fracture mechanics properties and to characterize the strength of the solid compact by its tensile strength.

In this work, we will restrict attention to powder blends used by the pharmaceutical industry to fabricate solid tables, the most popular dosage form in use today. Therefore, it is worth noting that the contact surface created during compaction of these powders also allows for the formation of attractive forces such as van der Waals and hydrogen bonding interactions  \cite{Joesten-1974,Israelachvili-2011,Derjaguin-1960,Derjaguin-1956,Israelachvili-1973b}. These long-range forces are of lower energy than covalent bonding forces, and they are present in many excipients (such as sugars, celluloses, and starches) and active pharmaceutical ingredients. It is also important to note that some polymers can experience a change in physical form with relative humidity (e.g., lactose) and that some active ingredients can experience strain- and temperature-driven solid-state transitions and amorphization. These additional physical mechanisms clearly increase the complexity of the analysis and, even though necessary for specific powder blends \cite{Sebhatu-1994}, their study will be beyond the scope of this work. We will specifically restrict attention to the formation of solid bridges, as it is a physical mechanism that dominates the synthesis of many, but not all, pharmaceutical excipients. We will simplify the powder morphology to spherical particles, and we will consider that these particles are amenable to elasto-plastic deformation without brittle failure.

A quantitative elucidation of strength formation requires not only the identification of the deformation and bonding mechanisms of interest but also of the bonding surface involved in the process. Unfortunately, it is not possible to experimentally measure the actual interfacial area that is available during tableting \cite{Karehill-1993,Nystrom-1986}. However, one can assume that an upper bound for the bonding surface involved in the formation of solid bridges is the particle-to-particle contact area created during compaction. It bears emphasis that the presence of lubricants in the formulation can diminish the bonding surface, and thus the tablet strength \cite{Karehill-1993, deBoer-1978, Razavi-2018}. The most common lubricant used in pharmaceutical tablets is magnesium stearate, typically prepared as a small particle size ingredient. Before tableting, the lubricant is mixed with the particles in the formulation, partially coating their surface and thus altering the tribochemical properties of the particle-to-particle contact area. In the context of this work, we will assume that the lubricant alters the fracture mechanics properties of the solid bridges.

\begin{figure}[htbp]
    \centering
    \includegraphics[scale=0.30]{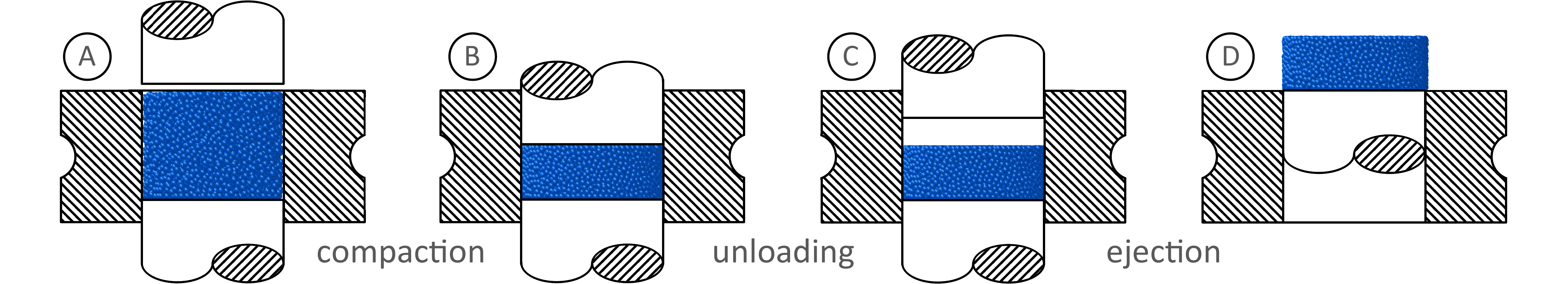}
    \caption{Transformation of the powder bed during the stages of die compaction: (A) die filling, (B) compaction, (C) unloading, and (D) tablet ejection.}
    \label{Fig-CompactionSchematics}
\end{figure}

In this paper, we report three-dimensional particle mechanics static calculations that enable us to predict microstructure evolution during compaction, unloading, and ejection---that is during the three most important steps of die-compaction of solid tablets (see Figure~\ref{Fig-CompactionSchematics})---of elasto-plastic spherical particles capable of forming solid bridges. To this end, we develop and employ generalized loading-unloading contact laws for elasto-plastic spheres with bonding strength. The proposed loading-unloading contact laws are continuous at the onset of unloading by means of a regularization term that introduces a small, controllable error in the solid bridge breakage force and the critical contact surface. The contact laws are explicit in terms of the relative position between the particles, and their strain path dependency is accounted for incrementally. The resulting formulation is then numerically robust and efficient, and mechanistically sound.

The paper is organized as follows. The generalized loading-unloading contact laws for elasto-plastic spheres with bonding strength are presented and validated in Section~\ref{Section-GeneralizedContactLaws}, after reviewing the state of the art in Section~\ref{Section-StateOfTheArt}. The particle mechanics approach used to generate a sequence of static equilibrium configurations of granular systems at high levels of confinement is presented in Section~\ref{Section-ParticleMechanicsAlgorithm}. The evolution of microstructural statistical features and of macroscopic effective properties during compaction, unloading and ejection is investigated in Section~\ref{Section-MicrostructureEvolution}. Specifically, we study the evolution of the mechanical coordination number (number of non-zero contact forces between a particle and its neighbors), punch and die-wall pressures, in-die elastic recovery, residual radial pressure and ejection pressure, the network of contact forces and granular fabric anisotropy, bonding surface area, Young's modulus and Poisson's ratio of the compacted solid, and a microstructure-mediated process-structure-property-performance interrelationship. Finally, concluding remarks are collected in Section~\ref{Section-ConcludingRemarks}.

\section{Loading and unloading contact laws for elasto-plastic spheres}
\label{Section-StateOfTheArt}

Loading contact laws for elasto-plastic spheres have been developed by Stor\r{a}kers and co-workers \cite{Storakers-1997a,Storakers-1997b} using a rigid plastic flow formulation \cite{Hill-1998} and assuming a power-law plastic hardening behavior, i.e.,  $\sigma = \kappa \epsilon^{1/m}$ where $\kappa$ is the plastic stiffness and $m$ is the plastic law exponent. These contact laws have been generalized to dissimilar particles \cite{Mesarovic-1999,Mesarovic-2000}. Specifically, for particles with the same hardening exponent or when one particle is assumed to be rigid, an analytical self-similar solution is derived by assuming a contact radius sufficiently small compared to the particle size and by neglecting elastic behavior. For particles with different hardening exponents, Skrinjar and Larsson \cite{Skrinjar-2004,Skrinjar-2007a,Skrinjar-2007b} derived and verified an approximate formulae based on the self-similar solution proposed by Stor\r{a}kers. These loading contact laws for elasto-plastic spheres are successful in simulating the deformation of soft metals, such as bronze and aluminum \cite{Olsson-2012}, and pharmaceutical excipients, such as microcrystalline cellulose and lactose monohydrate \cite{Yohannes-2016,Yohannes-2017}, and of harder materials, such as ceramics and cemented carbides, when both elastic and plastic deformations are properly accounted for during the loading phase \cite{Olsson-2013}.

Unloading contact laws for elasto-plastic spheres with bonding strength, or adhesion, have been developed by Mesarovic and Johnson \cite{Mesarovic-2000b} assuming elastic perfectly-plastic behavior and using a rigid punch decomposition \cite{Hill-1990}. Olsson and Larsson \cite{Olsson-2013} have extended these laws to elasto-plastic spheres that exhibit power-law plastic hardening behavior, and have verified their validity with detailed finite element simulations. This formulation assumes elastic behavior, approximated by Hooke's law, and Irwin's fracture mechanics to describe the elastic recovery of the deformed spheres and the breakage of the solid bridge.

We present next these loading and unloading contact laws for elasto-plastic spheres with bonding strength. Specifically, we consider two elasto-plastic spherical particles of radii $R_1$ and $R_2$, Young's moduli $E_1$ and $E_2$, Poisson's ratios $\nu_1$ and $\nu_2$, plastic stiffnesses $\kappa_1$ and $\kappa_2$, and plastic law exponent $m$, that deform plastically under loading and relax elastically under unloading. For particles located at ${\bf x}_1$ and ${\bf x}_2$, the relative position between them $\gamma$ (see Figure~\ref{Fig-AdhesiveMechanisms})
is give by
$$
     \gamma
    =
    R_1+R_2-\|{\bf x}_1-{\bf x}_2\|
$$
and the contact radius $a$ is given by
\begin{equation}
    a^2
    =
    \left\{
        \begin{array}{ll}
             2c^2 \bar{R} \gamma =: a_{\mbox{\tiny P}}^2
             &
             \mbox{plastic loading}
             \\
             \left[
             a_{\mbox{\tiny P}}^2
             -
             \left(\frac{4 \bar{E} (\gamma_{\mbox{\tiny P}}-\gamma)}
                        {3 n_{\mbox{\tiny P}} a_{\mbox{\tiny P}}^{1/m}}
             \right)^2
             \right]_+
             &
             \mbox{elastic (un)loading}
        \end{array}
     \right.
\end{equation}
where the effective radius $\bar{R}$, the effective elastic stiffness $\bar{E}$ and the plastic law coefficient $n_{\mbox{\tiny P}}$ are given by
$$
    \bar{R} = \left(\frac{1}{R_1}+\frac{1}{R_2}\right)^{-1}
$$
$$
    \bar{E} = \left(\frac{1-\nu_1^2}{E_1}+\frac{1-\nu_2^2}{E_2}\right)^{-1}
$$
$$
    n_{\mbox{\tiny P}}
    =
    \pi k \bar{R}^{-1/m}
    \left(\frac{1}{\kappa_{1}^{m}}+\frac{1}{\kappa_{2}^{m}}\right)^{-1/m}
$$
with $k=3 \times 6^{-1/m}$, $c^2=1.43~\mathrm{e}^{-0.97/m}$ \cite{Storakers-1994,Fleck-1997,Johnson-1985}, and $[ \cdot ]_+ = \max\{\cdot, 0\}$. The permanent plastic deformation is characterized by the plastic relative position $\gamma_{\mbox{\tiny P}}$ and the plastic contact radius $a_{\mbox{\tiny P}}$, which are related by $a_{\mbox{\tiny P}}^2=2c^2 \bar{R} \gamma_{\mbox{\tiny P}}$. In addition, the elasto-plastic spherical particles are capable of forming a solid bridge characterized by its fracture toughness $K_{Ic}$. Therefore, the plastic and elastic (un)loading force is defined by
\begin{equation}
    P
    =
    \left\{
        \begin{array}{l}
             n_{\mbox{\tiny P}}~a^{2+1/m}
             \hspace{3.42in}
             \mbox{plastic loading}
             \\
             \frac{2 n_{\mbox{\tiny P}}}{\pi} a_{\mbox{\tiny P}}^{2+1/m}
             \left[
                \arcsin\left(\frac{a}{a_{\mbox{\tiny P}}}\right)
                -
                \frac{a}{a_{\mbox{\tiny P}}} \sqrt{1-\left(\frac{a}{a_{\mbox{\tiny P}}}\right)^2}
             \right]
             -
             2 K_{Ic} \pi^{1/2} a^{3/2}
             \hspace{0.25in}
             \mbox{elastic (un)loading}
        \end{array}
     \right.
\end{equation}
This force acts in direction $({\bf x}_1 - {\bf x}_2)/\| {\bf x}_1 - {\bf x}_2 \|$ on particle 2 and in direction $({\bf x}_2 - {\bf x}_1)/\| {\bf x}_2 - {\bf x}_1 \|$ on particle 1.

\begin{figure}[htbp]
    \centering
    \begin{tabular}{ll}
    \includegraphics[scale=0.64, trim=0 0 22 0, clip]{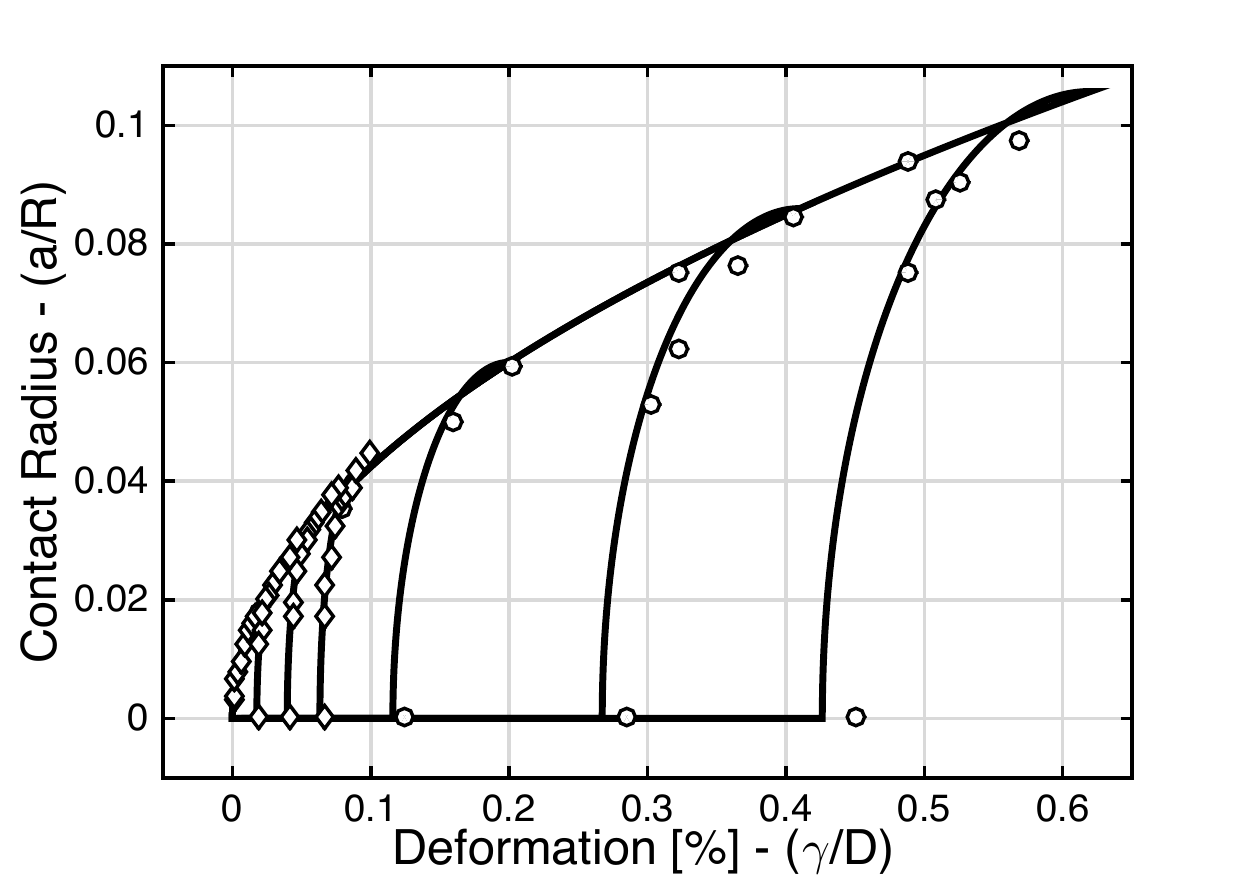}
    &
    \includegraphics[scale=0.64, trim=6 0 22 0, clip]{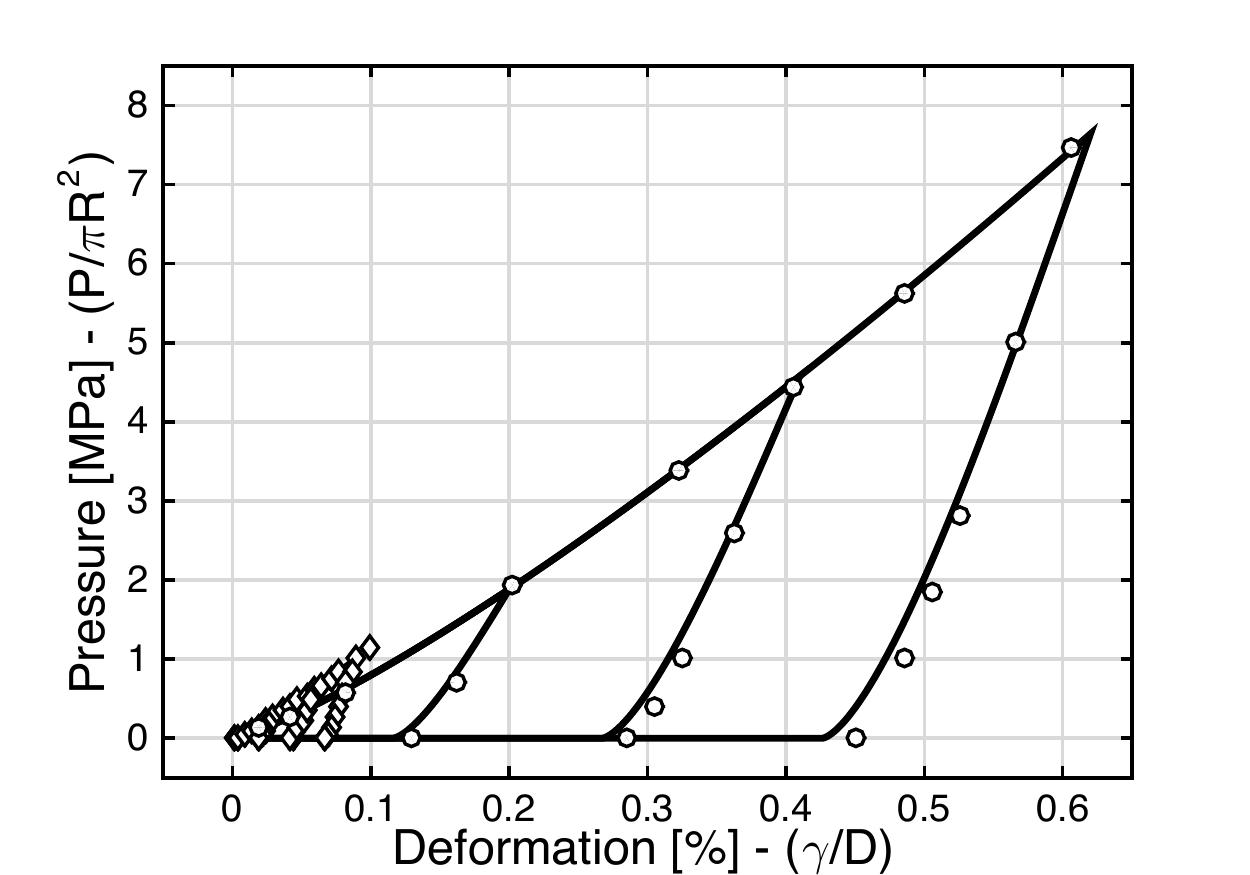}
    \\
    \small{(a)}
    &
    \small{(b)} 
    \\
    \includegraphics[scale=0.64, trim=0 0 22 0, clip]{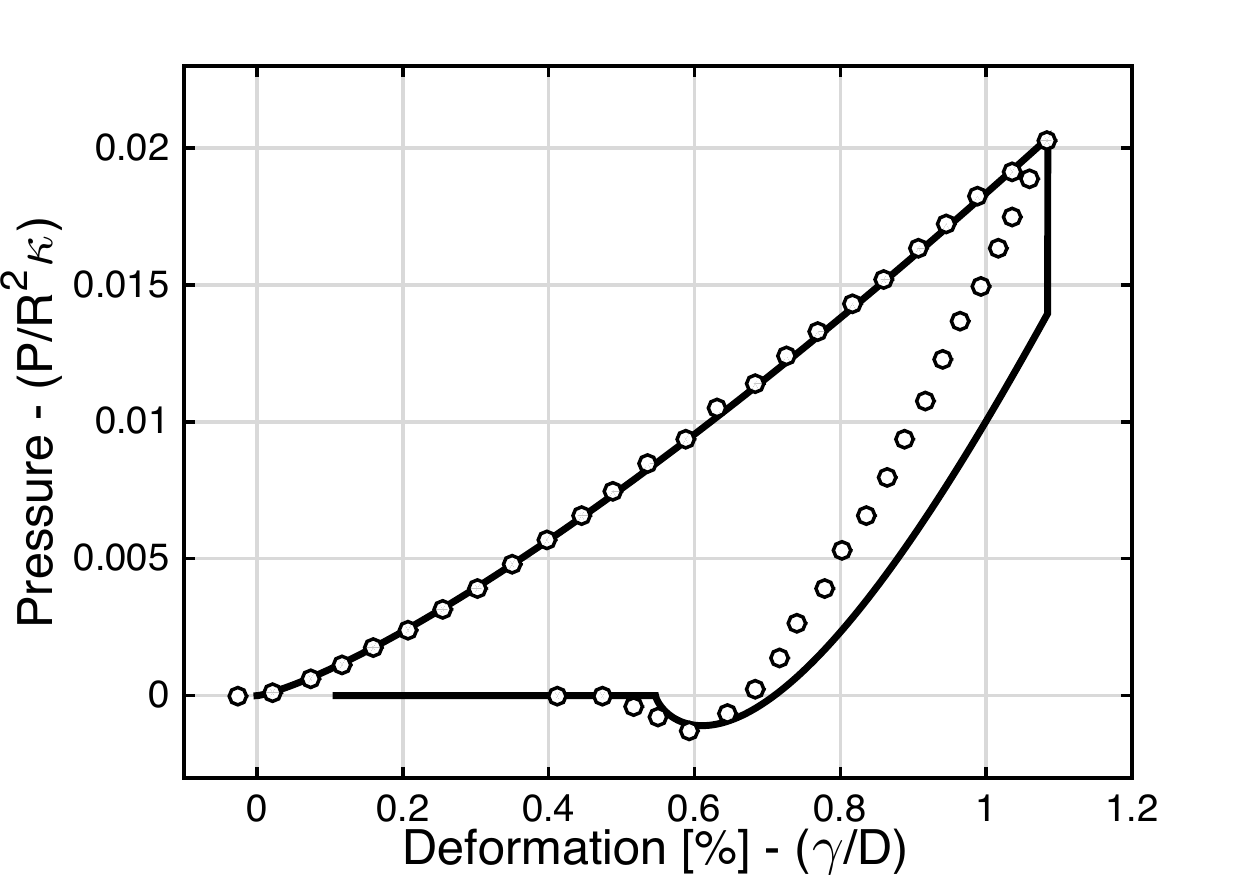}
    &
    \includegraphics[scale=0.64, trim=6 0 22 0, clip]{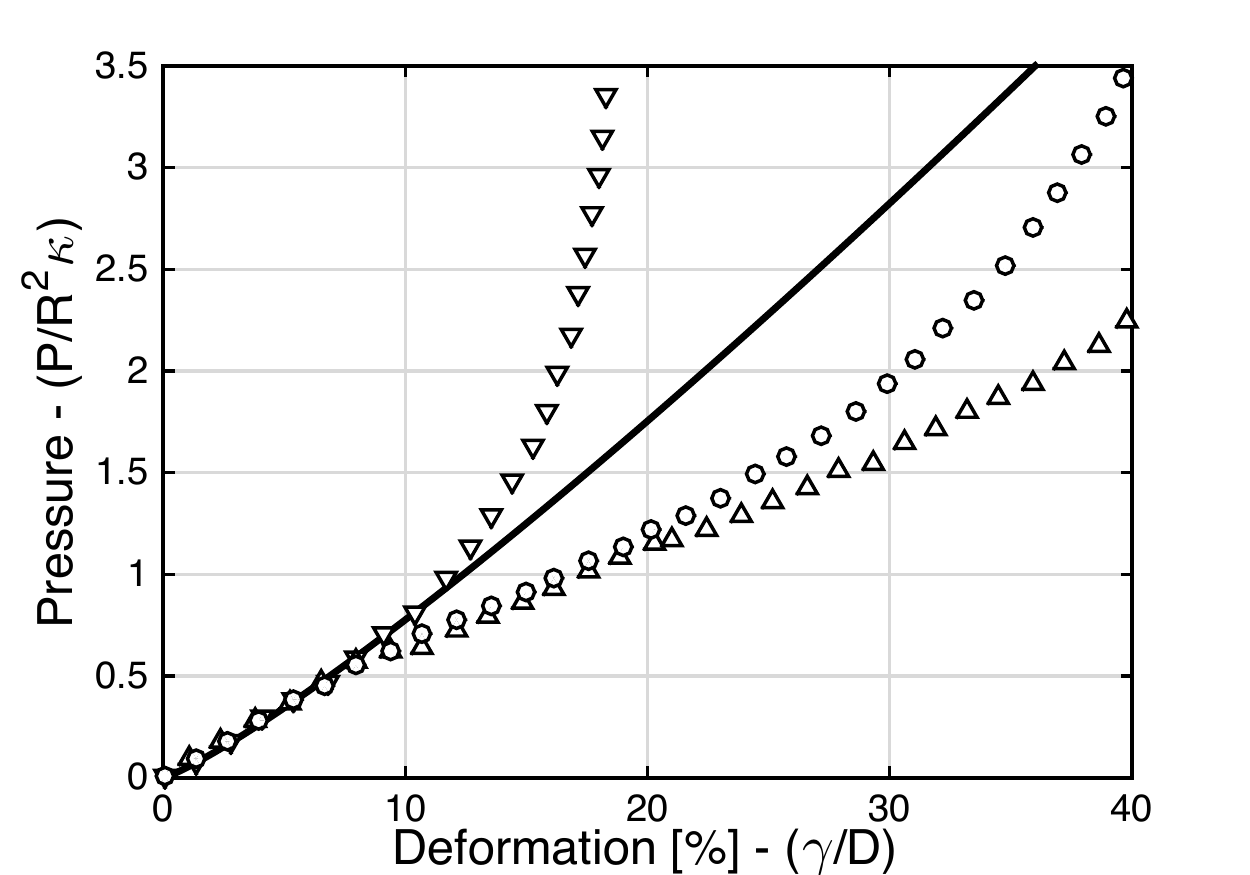}
    \\
    \small{(c)}
    &
    \small{(d)}    
    \end{tabular}
    \caption{Loading-unloading contact laws (solid lines) and finite element results (symbols) for elasto-plastic spheres. (a) Evolution of contact radius $a$, and (b) of contact force $P$, under diametrical compression, small deformations and no formation of solid bridges for ($\ocircle$) a spherical particle of $R=2$~mm and mechanical properties $E=80$~GPa, $\nu=0.3$, $\kappa=1.56$~GPa, $m=2.10$ and for ($\Diamond$) a spherical particle of $R=10$~mm and mechanical properties $E=200$~GPa, $\nu=0.32$, $\kappa=1.56$~GPa, $m=2.40$  \cite{Etsion-2005}. (c) Evolution of contact force $P$ under diametrical compression and small deformations for a spherical particle of $R=1.3~\mu$m, mechanical properties $E=233$~GPa, $\nu=0.3$, $\kappa=12.1$~GPa, $m=1.79$, that forms a solid bridge with fracture properties $K_{Ic}=0.48$~MPa~m$^{1/2}$ \cite{Du-2008}. (d) Evolution of contact force $P$ under large deformations and different loading configurations, namely diametrical compression ($\vartriangle $), diametrical compression and lateral confinement ($\ocircle$) and triaxial compression ($ \triangledown$), for a spherical particle of $R=5$~mm and mechanical properties $\kappa=15.5$~MPa, $m=2.86$  \cite{Harthong-2009}. }
    \label{Fig-SimilaritySlnValidation}
\end{figure}

This formulae is predictive at small deformations for particles that do not form solid bridges during plastic deformation, as it is depicted in Figures~\ref{Fig-SimilaritySlnValidation}(a) and \ref{Fig-SimilaritySlnValidation}(b) when compared with detailed finite element simulations of an elasto-plastic continuum solid \cite{Etsion-2005}. However, the formulation exhibits a discontinuity at the onset of unloading when particles form solid bridges during plastic deformation that it is not present in detailed finite element simulations as shown in Figure~\ref{Fig-SimilaritySlnValidation}(c). Specifically, there is a discontinuity at $a=a_{\mbox{\tiny P}}$, that is
$$
    P(a^+_{\mbox{\tiny P}})
    =
    \left\{
        \begin{array}{l}
	     n_{\mbox{\tiny P}}~a_{\mbox{\tiny P}}^{2+1/m} 
             \hspace{1.05in}
	     = 
	     P(a^-_{\mbox{\tiny P}})
             \hspace{0.7in}
             \mbox{if solid bridge is not formed}
             \\
             n_{\mbox{\tiny P}}~a_{\mbox{\tiny P}}^{2+1/m} - 2 K_{Ic} \pi^{1/2} a_{\mbox{\tiny P}}^{3/2}
             \ne 
             P(a^-_{\mbox{\tiny P}})
             \hspace{0.7in}
             \mbox{if solid bridge is formed}
        \end{array}
     \right.
$$
where $P(a^-_{\mbox{\tiny P}})$ and $P(a^+_{\mbox{\tiny P}})$ correspond to the contact force right before and after unloading, respectively. 

The loading-unloading contact laws proposed in this work, and described next in Section \ref{Section-GeneralizedContactLaws}, are continuous at the onset of unloading by means of a regularization term that introduces a small, controllable error in the solid bridge breakage force and the critical contact surface. Finally, at moderate to large deformations, detailed finite element simulations show a dependency of the response on the loading configuration and confinement of the particles (see Figure~\ref{Fig-SimilaritySlnValidation}(d) and \cite{Frenning-2013, Frenning-2015, Jonsson-2017, Tsigginos-2015}). This behavior is not captured by the above \emph{local} contact formulation and it calls for the development of plastic \emph{nonlocal} contact formulations (see \cite{Gonzalez-2012, Gonzalez-2016} for an elastic nonlocal contact formulation). The systematic development of nonlocal contact formulations for elasto-plastic spheres with bonding strength is a worthwhile direction of future research and, though beyond the scope of this work, it is currently being pursued by the author.

\section{Generalized loading-unloading contact laws for elasto-plastic spheres with bonding strength}
\label{Section-GeneralizedContactLaws}

\begin{figure}[htbp]
    \centering
    \begin{tabular}{ll}
    \includegraphics[scale=0.64]{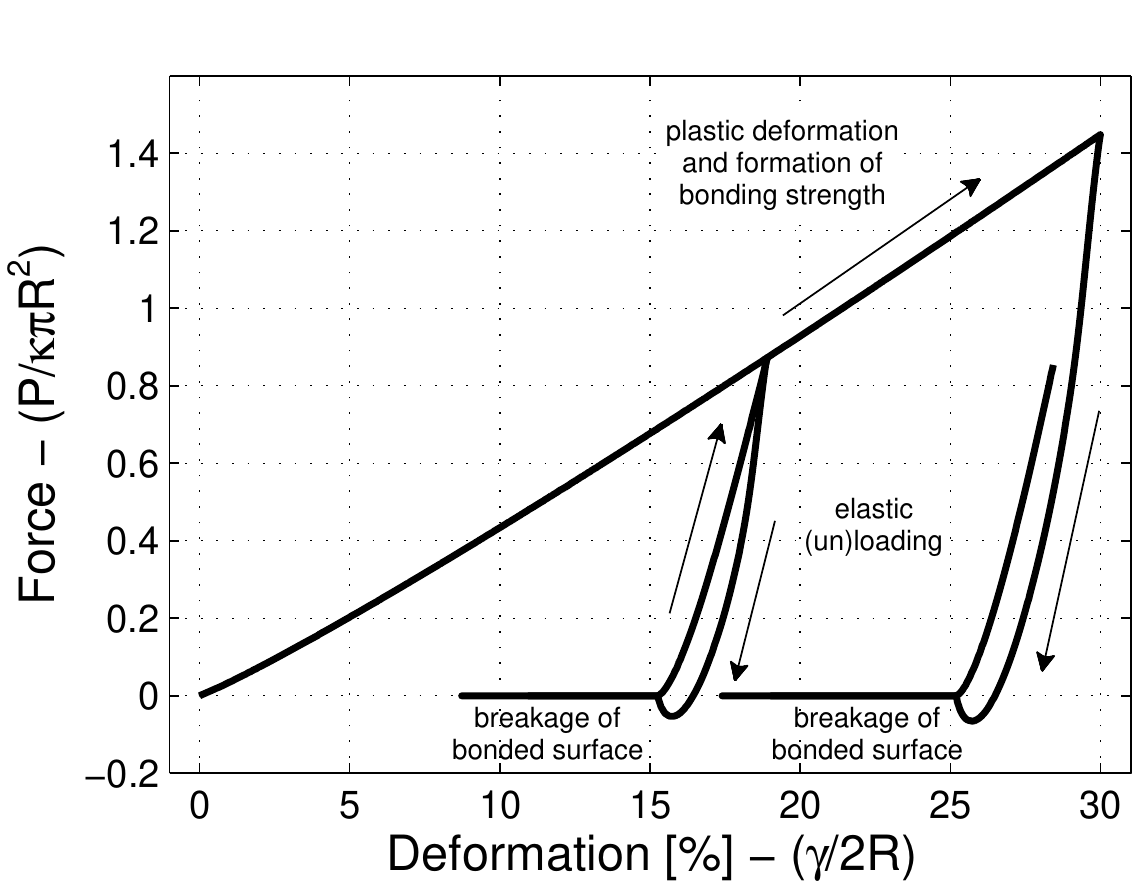}
    \Put(-181,229){\includegraphics[scale=0.250]{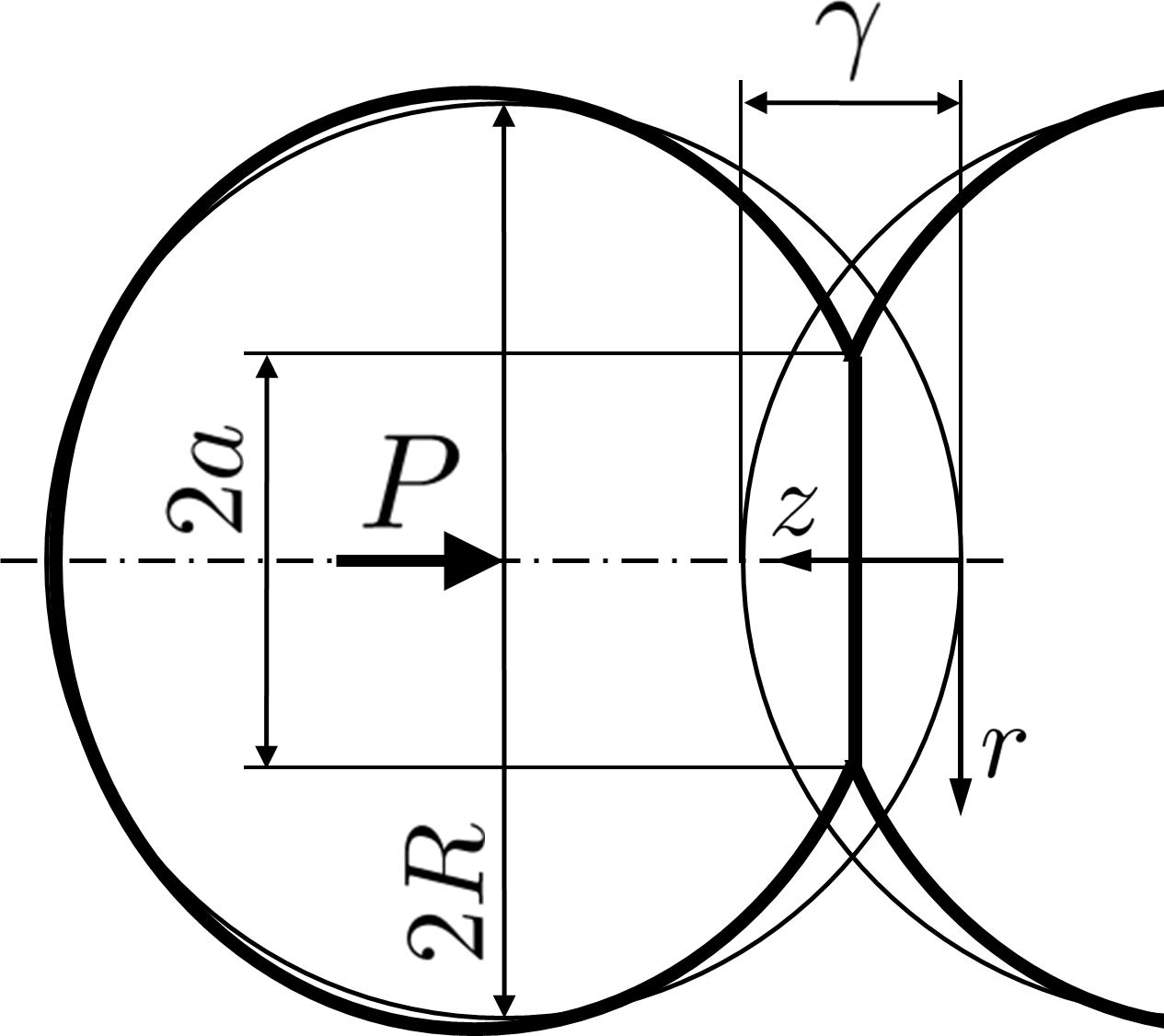}}
    &
    \includegraphics[scale=0.64]{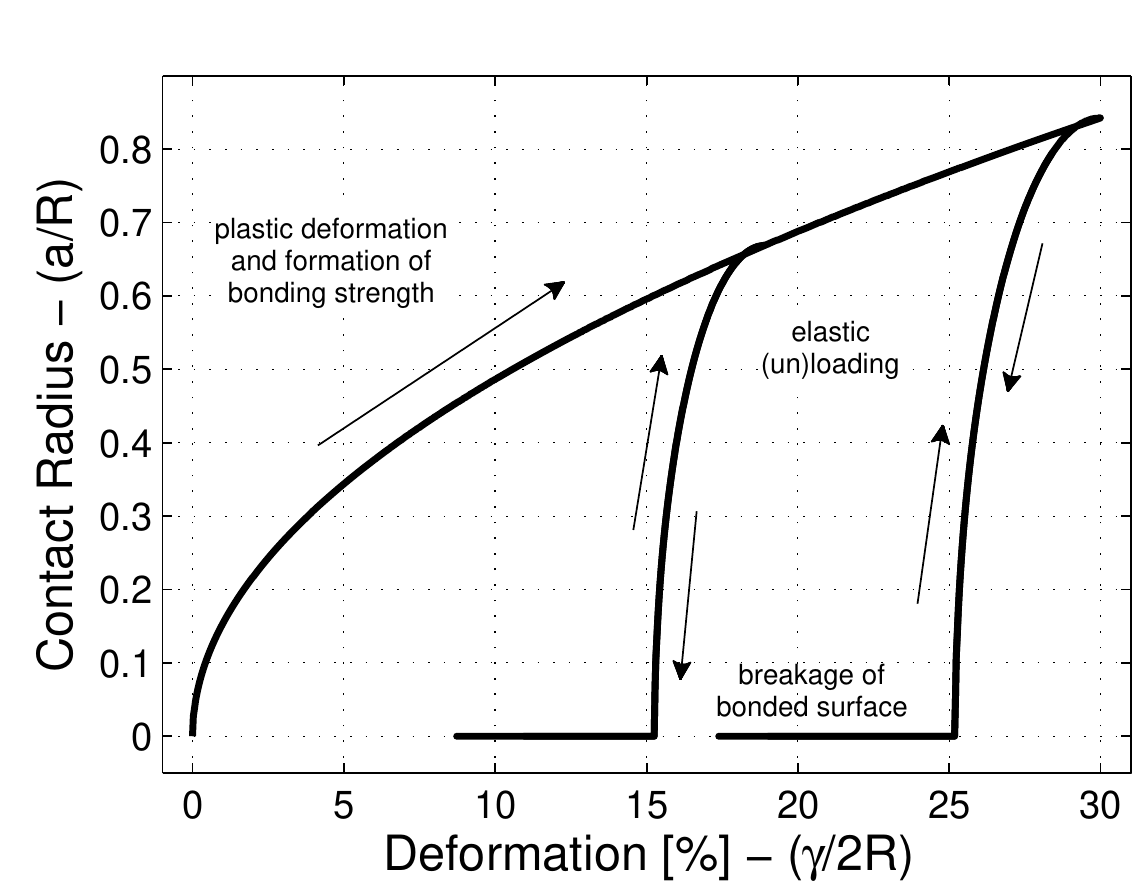}
    \\
    \small{(a)}
    &
    \small{(b)}
    \end{tabular}
    \caption{Generalized loading-unloading contact laws for elasto-plastic spheres with bonding strength. (a) Evolution of contact force $P$ under loading and two subsequent unloading-loading cycles that break the solid bridge. (b) Evolution of contact radius $a$. A regularization parameter $\xi_{\mbox{\tiny B}}$ equal to $0.01$ is used.}
    \label{Fig-GeneralizedContactLaw}
\end{figure}

We adopt the formulae developed by Mesarovi and co-workers for elasto-plastic spheres with power-law plastic hardening behavior \cite{Mesarovic-1999,Mesarovic-2000,Mesarovic-2000b}, presented in the previous section, and we propose a regularization of the contact force that does not modify the evolution of contact area $a$ which is give by
\begin{equation}
\label{Eqn-ContactRadius}
    a^2
    =
    \left\{
        \begin{array}{ll}
             2c^2 \bar{R} \gamma =: a_{\mbox{\tiny P}}^2
             &
             \mbox{plastic loading}
             \\
             \left[
             a_{\mbox{\tiny P}}^2
             -
             \left(\frac{4 \bar{E} (\gamma_{\mbox{\tiny P}}-\gamma)}
                        {3 n_{\mbox{\tiny P}} a_{\mbox{\tiny P}}^{1/m}}
             \right)^2
             \right]_+
             &
             \mbox{elastic (un)loading}
        \end{array}
     \right.
\end{equation}
with $\gamma_{\mbox{\tiny P}}=a_{\mbox{\tiny P}}^2/2c^2 \bar{R}$. It is worth noting that a solid bridge breaks during unloading at $\gamma = a_{\mbox{\tiny P}}^2/2c^2\bar{R}-3 n_{\mbox{\tiny P}} a_{\mbox{\tiny P}}^{1+1/m}/4 \bar{E}$ (see Figure \ref{Fig-GeneralizedContactLaw}b). The regularized contact force $P$ is defined by
\begin{equation}
\label{Eqn-ContactForce}
    P
    =
    \left\{
        \begin{array}{l}
             n_{\mbox{\tiny P}}~a^{2+1/m}
             \hspace{2.91in}
             \mbox{plastic loading}
             \\
             \frac{2 n_{\mbox{\tiny P}}}{\pi} a_{\mbox{\tiny P}}^{2+1/m}
             \left[
                \arcsin\left(\frac{a}{a_{\mbox{\tiny P}}}\right)
                -
                \frac{a}{a_{\mbox{\tiny P}}} \sqrt{1-\left(\frac{a}{a_{\mbox{\tiny P}}}\right)^2}
             \right]
             \\
             \hspace{1.37in}
             -
             2 K_{Ic} \pi^{1/2} a^{3/2}
             \frac{(1+\xi_{\mbox{\tiny B}})^2 [a_{\mbox{\tiny B}}-a]_{\mbox{\tiny +}}}
                   {(1+\xi_{\mbox{\tiny B}})a_{\mbox{\tiny B}}-a}
             \hspace{0.35in}
             \mbox{elastic (un)loading}
        \end{array}
     \right.
\end{equation}
where $\xi_{\mbox{\tiny B}}>0$ is the regularization parameter and $a_{\mbox{\tiny B}}$ is the radius of the bonded area, or solid bridge, which evolves as follows
\begin{equation}
\label{Eqn-EvolutionOfInternalVariable}
    \left\{
        \begin{array}{ll}
             a_{\mbox{\tiny B}} := a_{\mbox{\tiny P}}
             &
             \mbox{if mechano-chemical conditions are favorable, i.e., when $\dot{a}_{\mbox{\tiny P}}>0$}
             \\
	     a_{\mbox{\tiny B}} := 0
	     &
             \mbox{if solid bridge is broken, i.e., when $a = 0$}
             \\
             \dot{a}_{\mbox{\tiny B}} = 0
             &
             \mbox{otherwise, i.e., the size of the bonded area does not change}
        \end{array}
     \right.
\end{equation}
Furthermore, the fracture toughness of the solid bridge is given by $K_{Ic}=\sqrt{2 G\bar{E}}$, where the dissipated energy $G$ includes interfacial fracture energy $\omega$ (i.e., surface and field forces at direct contact) and plastic or other type of dissipation $G_p$, that is
$$
    K_{Ic}=\sqrt{\frac{(\omega_1+\omega_2+G_p) 2 E_1 E_2}{(1-\nu_1^2)E_2+(1-\nu_2^2)E_1}}
$$
A solid bridge between two particles of the same material then reduces to $K_{Ic}=\sqrt{G E/(1-\nu^2)}$. 

It bears emphasis that the contact force is  continuous at $a = a_{\mbox{\tiny P}}$, and it is equal to zero at $a=0$, for any $\xi_{\mbox{\tiny B}} >0$ and any value of $a_{\mbox{\tiny B}}$. Therefore, this generalized loading-unloading contact laws for elasto-plastic spheres with bonding strength are continuous at the onset of unloading after formation of a solid bridge, and at the onset of plastic loading after breakage of a solid bridge or bonded surface (see Figure \ref{Fig-GeneralizedContactLaw}a). This is achieved by means of a regularization term that introduces a small, controllable error in the solid bridge breakage force and the critical contact surface---that we will study in Section~\ref{Section-ErrorAnalysis}, after introducing a set of non-dimensional parameters in Section~\ref{Section-NonDimensional}. The proposed contact laws are validated in Section~\ref{Section-Validation}, followed by a reinterpretation of the regularization term as a cohesive zone model and by a sensitivity analysis in Sections \ref{Section-CohesizeZone} and \ref{Section-Sensitivity}, respectively.

\subsection{Non-dimensional analysis}
\label{Section-NonDimensional}

We recast the generalized loading-unloading contact laws for elasto-plastic spheres with bonding strength presented above using the following non-dimensional parameters: (i) elastic recovery ${\gamma}/{\gamma_{\mbox{\tiny P}}}$ and ${a}/{a_{\mbox{\tiny P}}}$, (ii) plastic deformation ${a_{\mbox{\tiny P}}}/{\bar{R}}$, (iii) bonded surface ${a_{\mbox{\tiny B}}}/{a_{\mbox{\tiny P}}}$, (iv) contact force ${P}{/\pi k \bar{\kappa} \bar{R}^2}$, (v) ratio of elastic to plastic stiffness $\psi={2 \bar{E}}/{3 \pi k \bar{\kappa}c^2}$, (vi) ratio of bonding to stored elastic energy $\chi = {4\pi K_{Ic}^2}/{n_{\mbox{\tiny P}}^2 a_{\mbox{\tiny P}}^{1+2/m}}$. Therefore, the non-dimensional contact radius then simplifies to
\begin{equation}
    \dfrac{a}{a_{\mbox{\tiny P}}}
    =
    \left\{
        \begin{array}{ll}
            1
             &
             \mbox{plastic loading, i.e.,}~\gamma/\gamma_{\mbox{\tiny P}}=1, a_{\mbox{\tiny P}} := \sqrt{2c^2 \bar{R} \gamma_{\mbox{\tiny P}}}
             \\
             \left[
             1
             -
             \left(\psi
             	\left(1-\frac{\gamma}{\gamma_{\mbox{\tiny P}}}\right)
		\left(\frac{a_{\mbox{\tiny P}}}{\bar{R}}\right)^{1-1/m}
             \right)^2
             \right]_+^{1/2}
             &
             \mbox{elastic (un)loading, i.e.,}~\gamma/\gamma_{\mbox{\tiny P}}<1
        \end{array}
     \right.
\end{equation}
and the non-dimensional contact force is
\begin{equation}
    \frac{P}{\pi k \bar{\kappa} \bar{R}^2}
    =
    \left\{
        \begin{array}{l}
             \left(\frac{a_{\mbox{\tiny P}}}{\bar{R}}\right)^{2+1/m}
             \hspace{2.91in}
             \mbox{plastic loading}
             \\
             \frac{2}{\pi}  \left(\frac{a_{\mbox{\tiny P}}}{\bar{R}}\right)^{2+1/m}
             \left[
                \arcsin\left(\frac{a}{a_{\mbox{\tiny P}}}\right)
                -
                \frac{a}{a_{\mbox{\tiny P}}} \sqrt{1-\left(\frac{a}{a_{\mbox{\tiny P}}}\right)^2}
             \right]
             \\
             \hspace{0.67in}
             -
             \chi
                \left(\frac{a_{\mbox{\tiny P}}}{\bar{R}}\right)^{2+1/m}    \left(\frac{a}{a_{\mbox{\tiny P}}}\right)^{3/2}
                \frac{(1+\xi_{\mbox{\tiny B}})^2 \left[a_{\mbox{\tiny B}}/a_{\mbox{\tiny P}}-a/a_{\mbox{\tiny P}}\right]_+}
                   {(1+\xi_{\mbox{\tiny B}})a_{\mbox{\tiny B}}/a_{\mbox{\tiny P}}-a/a_{\mbox{\tiny P}}}
             \hspace{0.22in}
             \mbox{elastic (un)loading}
        \end{array}
     \right.
\end{equation}
It is worth noting that bonding to stored elastic energy ratio $\chi$ is $16(\pi-2)/\pi^2 \approx 1.85$ times larger than the bonding energy to elastic energy ratio in \cite{Mesarovic-2000b}.

\subsection{Error analysis and optimal selection of the regularization parameter}
\label{Section-ErrorAnalysis}

\begin{figure}[htbp]
    \centering
    \begin{tabular}{ll}
    \includegraphics[scale=0.64, trim=0 0 27 0, clip]{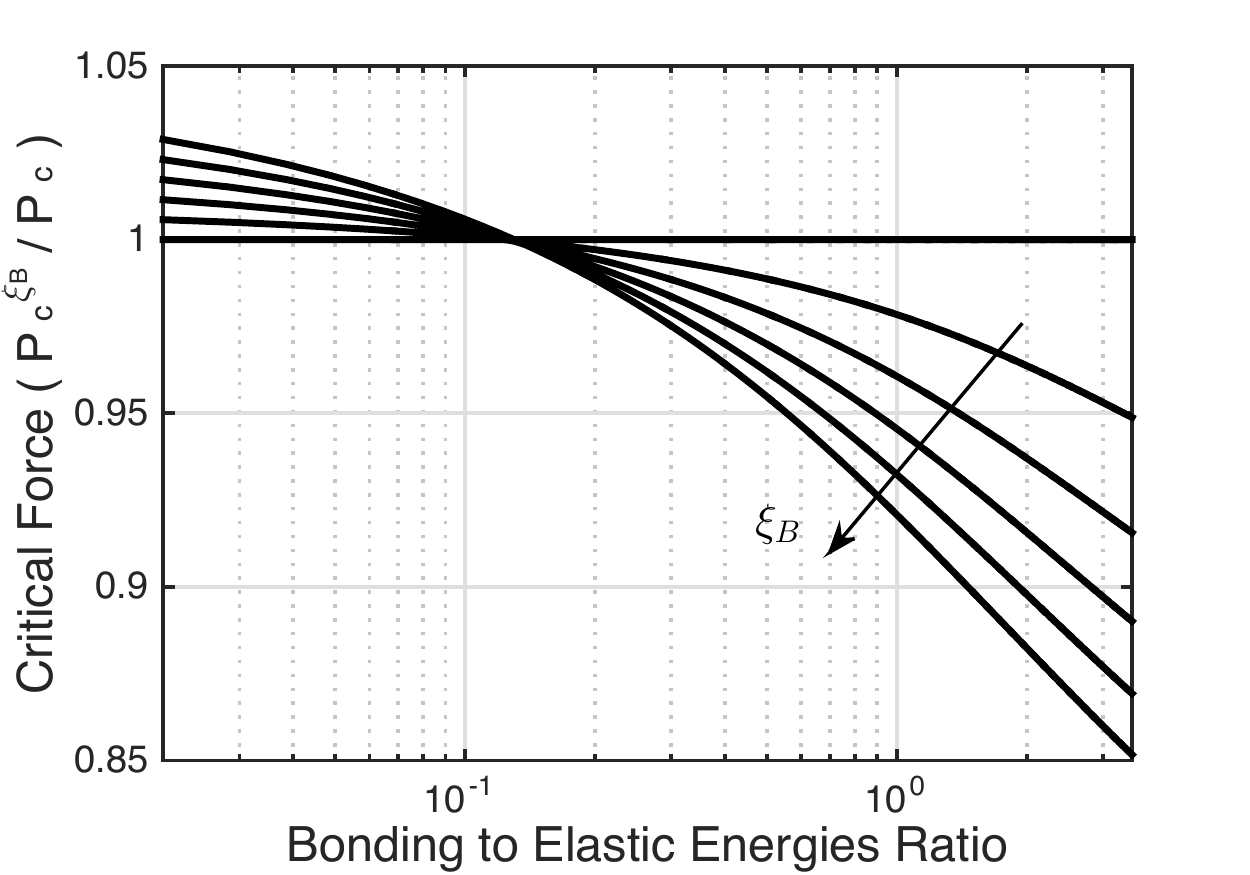}
    &
    \includegraphics[scale=0.64, trim=0 0 27 0, clip]{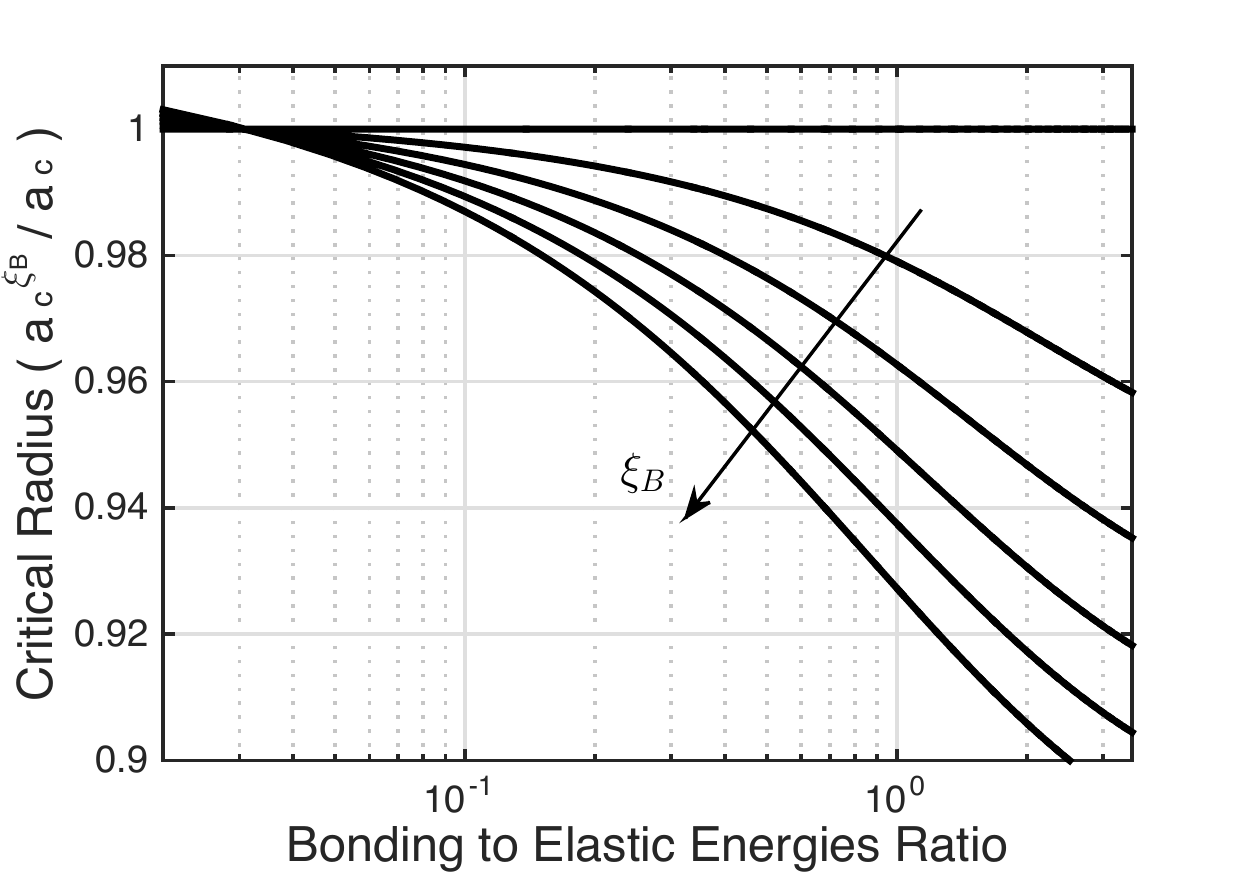}
    \\
    \small{(a)}
    &
    \small{(b)}    
    \end{tabular}
    \caption{Error incurred in the critical force ${P_{\mbox{\tiny C}}^{\xi_{\mbox{\tiny B}}}}$ and the critical contact radius ${a_{\mbox{\tiny C}}^{\xi_{\mbox{\tiny B}}}}$ for different regularization parameters $\xi_{\mbox{\tiny B}}=\{0, 0.005, 0.01, 0.015, 0.02, 0.025\}$ and bonding to elastic energies ratio $\chi$.}
    \label{Fig-ErrorAnalysis001}
\end{figure}

The proposed generalized loading-unloading contact laws introduce a controllable error in the solid bridge breakage force and the critical contact surface. In order to study these errors, we first derive the critical force $P_{\mbox{\tiny C}}^{\xi_{\mbox{\tiny B}}}$ and contact radius $a_{\mbox{\tiny C}}^{\xi_{\mbox{\tiny B}}}$ for a give regularization parameter $\xi_{\mbox{\tiny B}}$ assuming, for simplicity, ${a_{\mbox{\tiny B}}}/{a_{\mbox{\tiny P}}}=1$. The critical point occurs at the maximum tensile force, that is at 
\begin{equation}
    \frac{\partial P}{\partial \gamma}
    =
    0
    =
    \frac{16 a_{\mbox{\tiny P}}\bar{E}}{3\pi(\frac{a}{a_{\mbox{\tiny P}}})^{1/2}}
    \left[
        \left(\frac{a}{a_{\mbox{\tiny P}}}\right)^{3/2}
        -
        \frac{3\pi^{3/2} K_{Ic}}{4 n_{\mbox{\tiny P}} a_{\mbox{\tiny P}}^{1/2+1/m}}
        \times
        \frac{(1-\frac{a}{a_{\mbox{\tiny P}}})^2+\xi_{\mbox{\tiny B}}(1-\frac{5}{3}\frac{a}{a_{\mbox{\tiny P}}})}
             {(1-\frac{a}{a_{\mbox{\tiny P}}}+\xi_{\mbox{\tiny B}})^2(1+\xi_{\mbox{\tiny B}})^{-2}}
        \sqrt{
            1-\left(\frac{a}{a_{\mbox{\tiny P}}}\right)^2}
    \right]
\end{equation}
and thus, after performing a power series expansion in $\xi_{\mbox{\tiny B}}$, the critical contact radius or critical \emph{pull-off} force is given by the solution of
\begin{equation}
    \left(\frac{a_{\mbox{\tiny C}}}{a_{\mbox{\tiny P}}}\right)^3
    =
    \frac{9\pi^3 K_{Ic}^2}{16 n_{\mbox{\tiny P}}^2 a_{\mbox{\tiny P}}^{1+2/m}}
    \left[
        1
        -
        \left(\frac{a_{\mbox{\tiny C}}}{a_{\mbox{\tiny P}}}\right)^2
        -
        \frac{2(1+a_{\mbox{\tiny C}}/a_{\mbox{\tiny P}})(3-2a_{\mbox{\tiny C}}/a_{\mbox{\tiny P}})(3a_{\mbox{\tiny C}}/a_{\mbox{\tiny P}}-1)}
             {3(1-a_{\mbox{\tiny C}}/a_{\mbox{\tiny P}})}
        \xi_{\mbox{\tiny B}}
        +
        \mathcal{O}(\xi_{\mbox{\tiny B}}^2)
    \right]
\end{equation}
Specifically, the critical contact radius for the regularized contact law, $a_{\mbox{\tiny C}}^{\xi_{\mbox{\tiny B}}}$, is give by the solution of the following equation
$$
        \left(\frac{a_{\mbox{\tiny C}}^{\xi_{\mbox{\tiny B}}}}{a_{\mbox{\tiny P}}}\right)^{3/2}
        -
        \frac{3\pi^{3/2} K_{Ic}}{4 n_{\mbox{\tiny P}} a_{\mbox{\tiny P}}^{1/2+1/m}}
        \times
        \frac{(1-{a_{\mbox{\tiny C}}^{\xi_{\mbox{\tiny B}}}}/{a_{\mbox{\tiny P}}})^2+\xi_{\mbox{\tiny B}}(1-\frac{5}{3}{a_{\mbox{\tiny C}}^{\xi_{\mbox{\tiny B}}}}/{a_{\mbox{\tiny P}}})}
             {(1-{a_{\mbox{\tiny C}}^{\xi_{\mbox{\tiny B}}}}/{a_{\mbox{\tiny P}}}+\xi_{\mbox{\tiny B}})^2(1+\xi_{\mbox{\tiny B}})^{-2}}
        \sqrt{
            1-\left(\frac{a_{\mbox{\tiny C}}^{\xi_{\mbox{\tiny B}}}}{a_{\mbox{\tiny P}}}\right)^2}
        =
        0
$$
which reduces to $({a_{\mbox{\tiny C}}^{\xi_{\mbox{\tiny B}}}}/{a_{\mbox{\tiny P}}})^{3} = \chi (3\pi/8)^2  (1-({a_{\mbox{\tiny C}}^{\xi_{\mbox{\tiny B}}}}/{a_{\mbox{\tiny P}}})^2)$ for $\xi_{\mbox{\tiny B}}=0$ (cf. \cite{Mesarovic-2000}). Finally, the corresponding critical force is given by equation~\eqref{Eqn-ContactForce}, i.e., by $P_{\mbox{\tiny C}}^{\xi_{\mbox{\tiny B}}} = P(a_{\mbox{\tiny C}}^{\xi_{\mbox{\tiny B}}})$, and, naturally, the correct limiting behavior is retained, i.e., $a_{\mbox{\tiny C}} = a_{\mbox{\tiny C}}^{\xi_{\mbox{\tiny B}} \rightarrow 0}$ and $P_{\mbox{\tiny C}} = P_{\mbox{\tiny C}}^{\xi_{\mbox{\tiny B}} \rightarrow 0}$. Figure~\ref{Fig-ErrorAnalysis001} shows the error incurred in the critical force and the critical contact radius for different regularization parameters and bonding to elastic energies ratios.

\begin{figure}[htbp]
    \centering
    \includegraphics[scale=0.64]{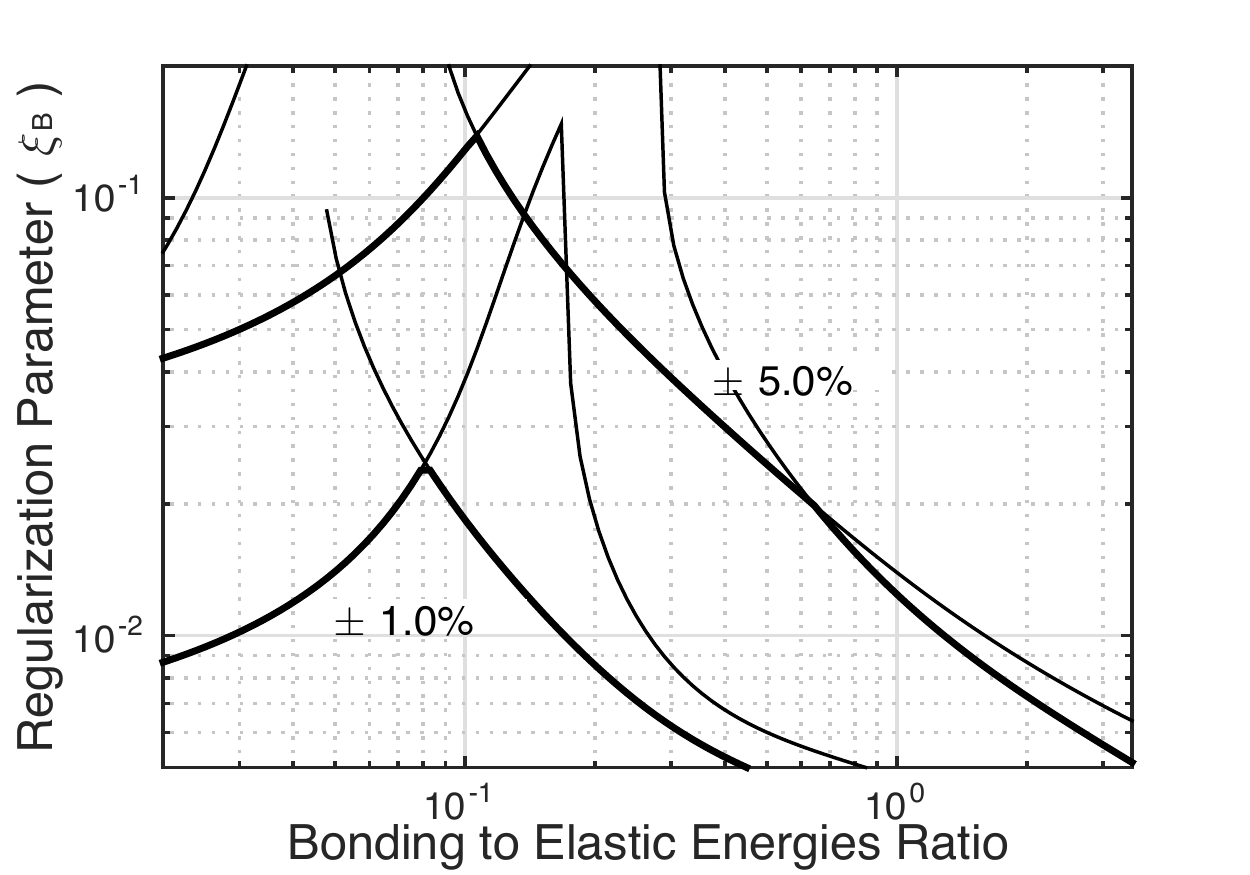}
    \caption{ Optimal regularization parameter $\bar{\xi}_{\mbox{\tiny B}}$ for introducing less than or equal to $1\%$ and $5\%$ error in ${a_{\mbox{\tiny C}}^{\xi_{\mbox{\tiny B}}}}$ and in ${P_{\mbox{\tiny C}}^{\xi_{\mbox{\tiny B}}}}$ simultaneously. Thin solid lines correspond to the optimal values of $\xi_{\mbox{\tiny B}}$ that keep ${a_{\mbox{\tiny C}}^{\xi_{\mbox{\tiny B}}}}$ within given bounds. Thin dashed lines correspond to the optimal values of $\xi_{\mbox{\tiny B}}$ that keep ${P_{\mbox{\tiny C}}^{\xi_{\mbox{\tiny B}}}}$ within given bounds.}
    \label{Fig-ErrorAnalysis002}
\end{figure}

We next define the optimal regulation parameter $ \bar{\xi}_{\mbox{\tiny B}}$ as follows
$$
	 \bar{\xi}_{\mbox{\tiny B}} := \min \left\{  \xi_{\mbox{\tiny B}}>0 \mbox{~~s.t.~~} \epsilon = | 1 - a_{\mbox{\tiny C}}^{\xi_{\mbox{\tiny B}}} / a_{\mbox{\tiny C}} | 
	 							\mbox{~~,~~}
	 							\xi_{\mbox{\tiny B}}>0 \mbox{~~s.t.~~} \epsilon = | 1 - P_{\mbox{\tiny C}}^{\xi_{\mbox{\tiny B}}} / P_{\mbox{\tiny C}} |
							   \right\}
$$
where $\epsilon$  is the maximum relative error incurred in the critical contact radius and the critical force. Figure~\ref{Fig-ErrorAnalysis002} shows that a regularization parameter of ${\xi}_{\mbox{\tiny B}}=0.01$ ensures a moderate error in the prediction of the critical point over a wide range of bonding to elastic energy ratio conditions. It is worth noting that this controllable error is the cost we pay to achieve a numerically robust and efficient as well as a mechanistically sound formulation. 

\subsection{Validation}
\label{Section-Validation}

\begin{figure}[htbp]
    \centering
    \begin{tabular}{ll}
    \includegraphics[scale=0.64, trim=5 0 22 0, clip]{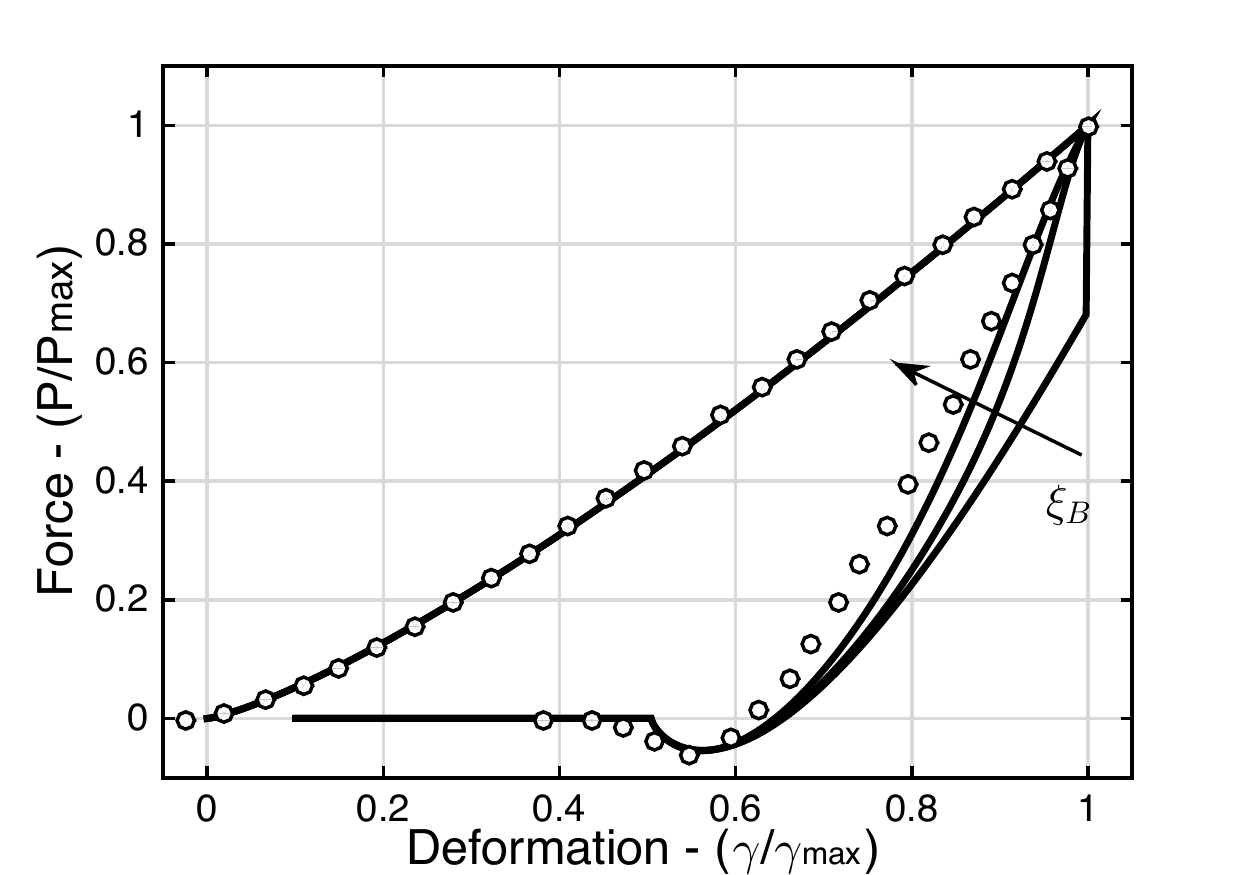}
    &
    \includegraphics[scale=0.64, trim=5 0 22 0, clip]{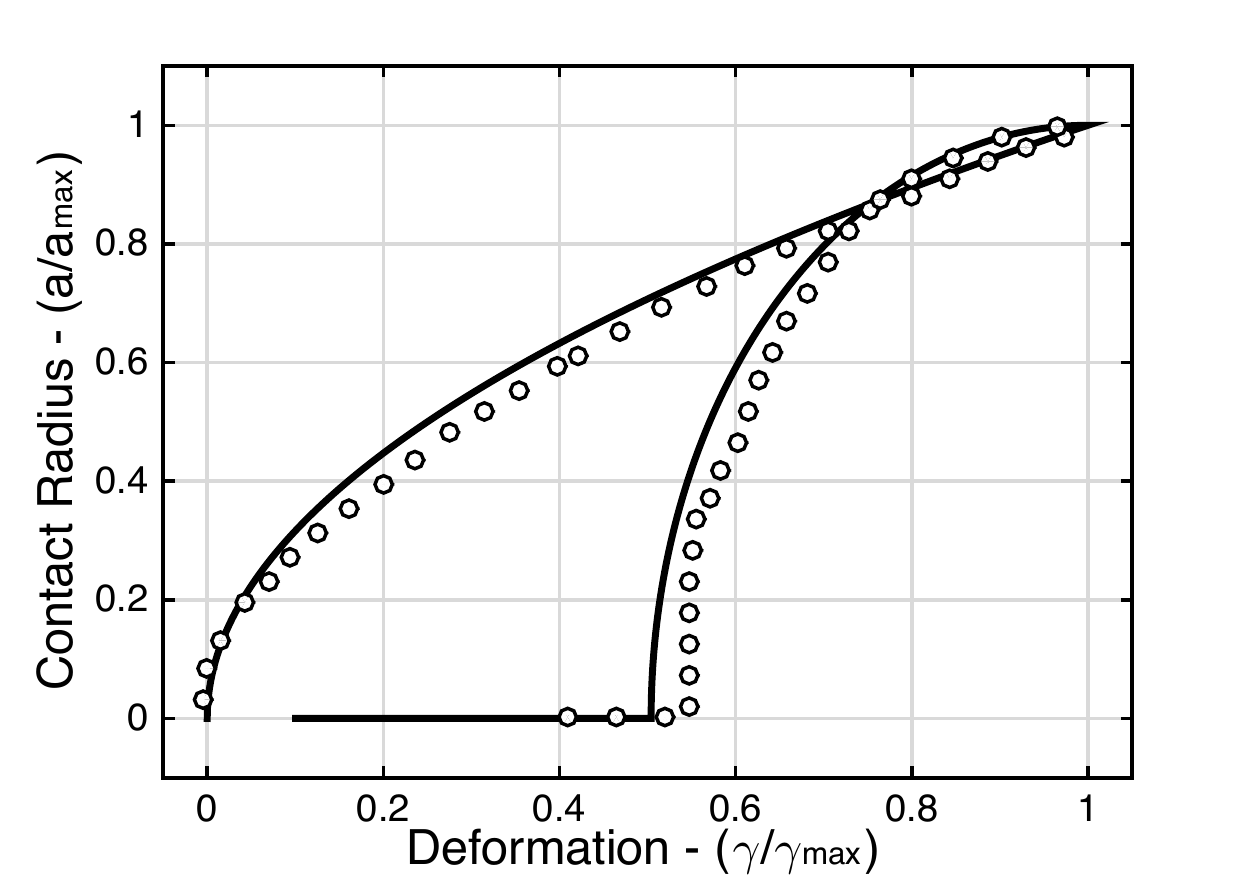}
    \\
    \small{(a)}
    &
    \small{(b)}    
    \end{tabular}
    \caption{Detailed finite element simulations (symbols) performed by Du et al. \cite{Du-2008} and generalized loading-unloading contact law predictions (solid line) for $R=1.3~\mu$m, $E=233$~GPa, $\nu=0.3$, $\kappa=12.1$~GPa, $m=1.79$, and $K_{Ic}=0.48$~MPa~m$^{1/2}$---the bonding energy to elastic energy ratio is $\chi=0.0493$. Three different values of $\xi_{\mbox{\tiny B}}$ are use $0$, $0.01$, and $0.05$. Dimensionless force vs deformation (a) and dimensionless contact radius vs deformation (b) correspond to a single loading-unloading cycle with $\gamma_{\textrm{max}}=28.2$~nm.}
    \label{Fig-Validation-Du2007}
\end{figure}

We compare next the proposed formulation with detailed finite element simulations with Lennard-Jones stresses at the interface \cite{Du-2007} performed by Du et al. \cite{Du-2008}. The finite element simulations correspond to an elasto-plastic spherical particle with radius $R=1.3~\mu$m, Young's modulus $E=233$~GPa, Poisson's ratio $\nu=0.3$, yield stress equal to 1.94~GPa and linear hardening equal to 2$\%$ of Young's modulus. We approximate the bi-linear elasto-plastic law by an exponential plastic law with plastic stiffness $\kappa=12.1$~GPa and plastic law exponent $m=1.79$. The bond interaction is represented by the Lennard-Jones potential between two parallel surfaces, which we approximate by an equivalent fracture toughness $K_{Ic}=0.48$~MPa~m$^{1/2}$. The maximum separation is $\gamma_{\textrm{max}}=28.2$~nm. Figure~\ref{Fig-Validation-Du2007} shows Du's finite element calculations and the predictions of the generalized loading-unloading contact law for elasto-plastic spheres with bonding strength using three different values of $\xi_{\mbox{\tiny B}}$ and adopting ${a_{\mbox{\tiny B}}}/{a_{\mbox{\tiny P}}}=1$. It is evident in the figure that the proposed loading-unloading contact law is continuous at the onset of unloading by means of the regularization term (cf. Figure~\ref{Fig-SimilaritySlnValidation}c).

\subsection{Regularization as a cohesive zone model}
\label{Section-CohesizeZone}

We show next that the proposed regularization, which provides continuity in the contact force at the onset of unloading, is in the spirit of a cohesive zone model. Specifically, we show that the relationship between interfacial separation traction $\hat{T}$ and separation displacement $\hat{\gamma}$ follows a typical cohesive law curve under strain control, for $\xi_{\mbox{\tiny B}}>0$. For simplicity, we restrict attention to ${a_{\mbox{\tiny B}}}/{a_{\mbox{\tiny P}}}=1$ but, in general, the non-dimensional bonded surface is $0 \le a_{\mbox{\tiny B}}/a_{\mbox{\tiny P}} \le 1$. 

\begin{figure}[htbp]
    \centering
    \begin{tabular}{ll}
    \includegraphics[scale=0.64, trim=11 0 15 0, clip]{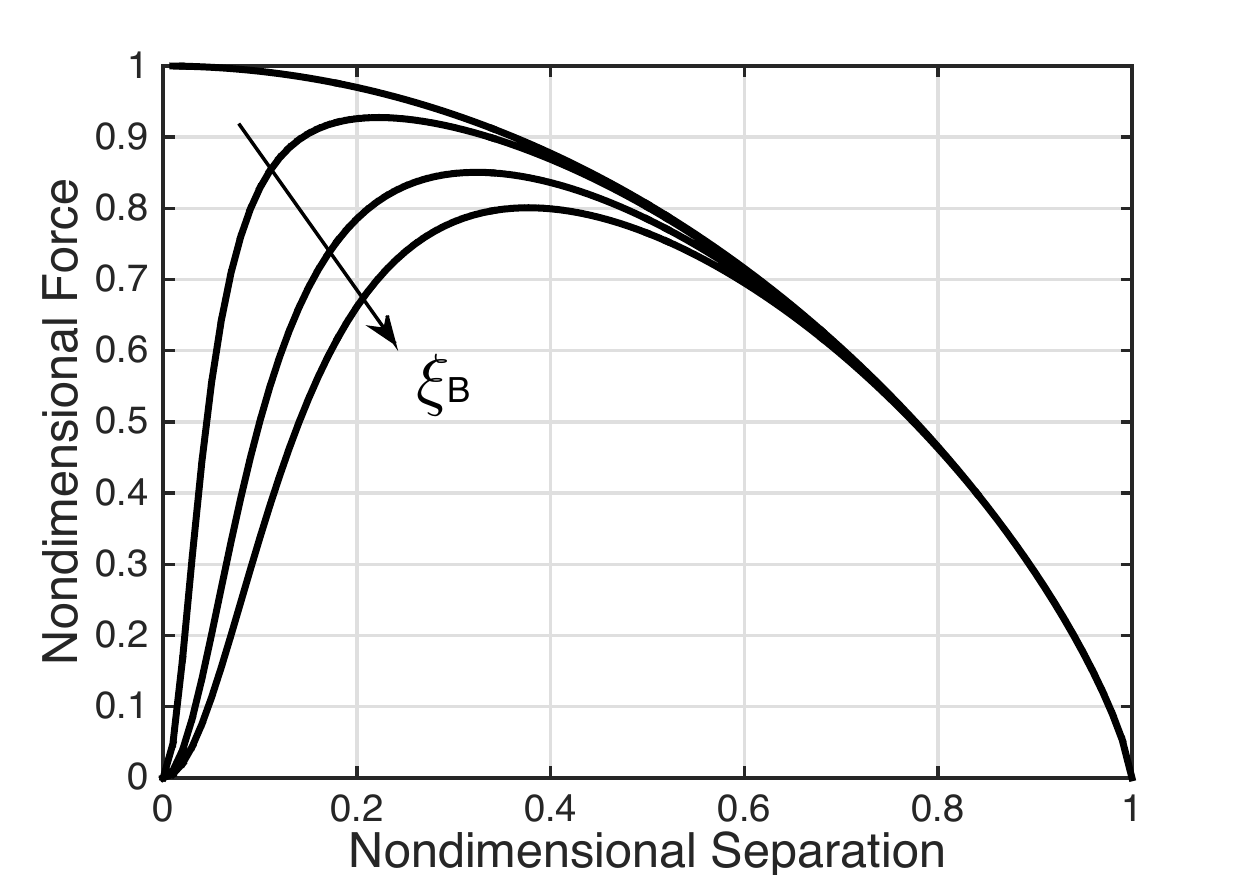}
    &
    \includegraphics[scale=0.64, trim=11 0 15 0, clip]{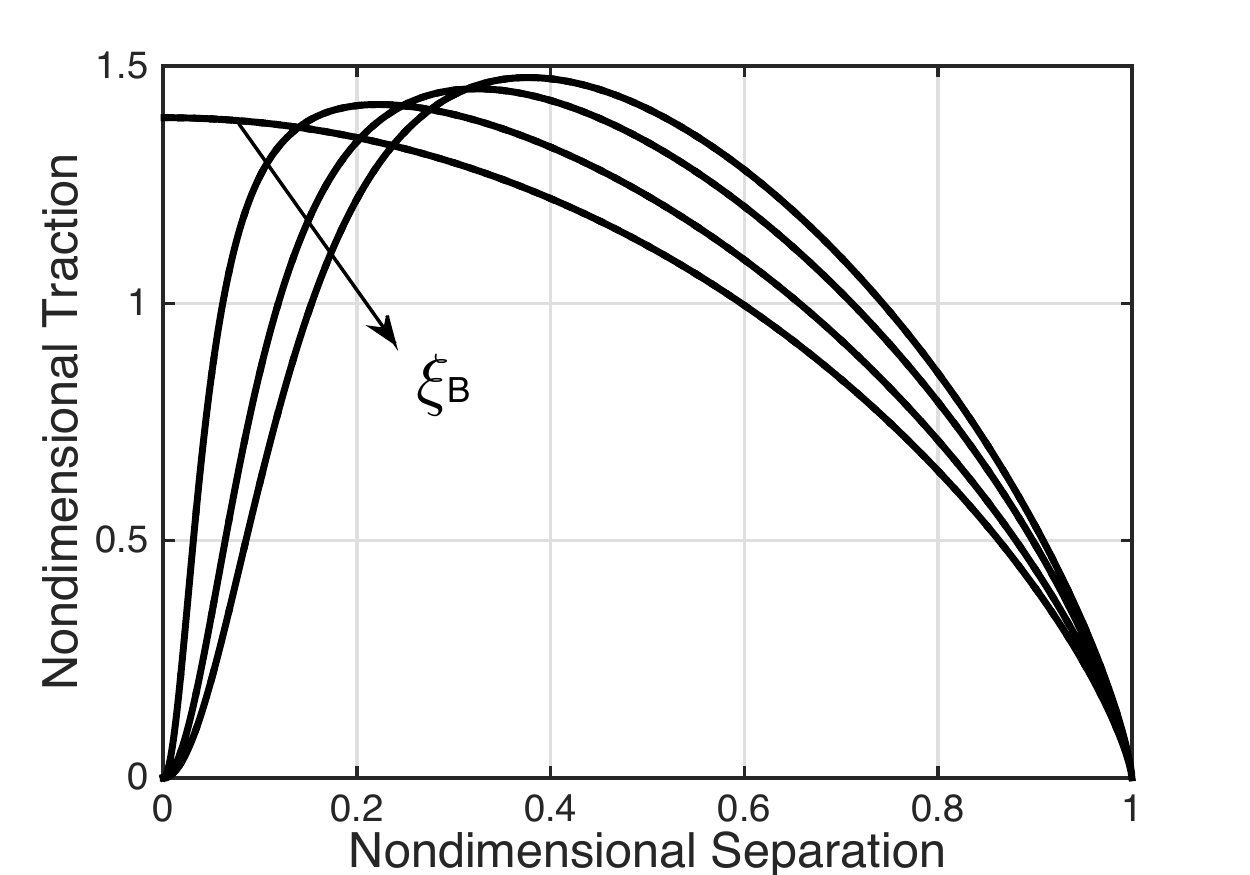}
    \\
    \small{(a)}
    &
    \small{(b)}
    \end{tabular}
   \caption{Non-dimensional separation force $\hat{P}$ (a) and non-dimensional separation traction $\hat{T}$ (b) as a function of the non-dimensional separation displacement $\hat{\gamma}$ for different regularization parameters $\xi_{\mbox{\tiny B}}=\{0, 0.001, 0.05, 0.01\}$.}
    \label{Fig-CohesiveZoneModel}
\end{figure}

It is worth noting that the separation force $ P_{\mbox{\tiny{B}}}$ is the term in the unloading contact force that corresponds to separation and breakage of the solid bridge. Similarly, the separation displacement is zero at the onset of unloading and equal to the critical separation $\Delta\gamma_c$ at solid bridge breakage. Therefore, a non-dimensional separation force $\hat{P}$ and a non-dimensional separation displacement $\hat{\gamma}$ are defined as 
$$ 
\hat{P} 
=
\frac{  P_{\mbox{\tiny{B}}} }{  2 K_{Ic} \pi^{1/2}  a_{\mbox{\tiny P}}^{3/2}  }
=
\frac{({a}/{a_{\mbox{\tiny P}}})^{3/2}(1+\xi_{\mbox{\tiny B}})^2 [1-{a}/{a_{\mbox{\tiny P}}}]_{\mbox{\tiny +}}}
       {(1+\xi_{\mbox{\tiny B}})-{a}/{a_{\mbox{\tiny P}}}}
$$
$$
\hat{\gamma} 
= 
\frac{(\gamma_{\mbox{\tiny P}}-\gamma)}
       {\Delta\gamma_c}
\mbox{\hspace{.25in} with \hspace{.25in}}
\Delta\gamma_c = \frac{3 n_{\mbox{\tiny P}} a_{\mbox{\tiny P}}^{1+1/m}}{4 \bar{E} }
$$
and thus they are related by
\begin{equation}
    \hat{P}
    =
	\frac{\left(1+\xi_{\mbox{\tiny B}}\right)^2 \left(1-\sqrt{1-\hat{\gamma}^2}\right) \left(1-\hat{\gamma}^2\right)^{3/4}}
            {1+\xi_{\mbox{\tiny B}}-\sqrt{1-\hat{\gamma}^2}}
\end{equation}
We next define a separation traction $T_{\mbox{\tiny B}}$ as
$$ 
T_{\mbox{\tiny{B}}}
= 
\frac{  P_{\mbox{\tiny{B}}}  K_{Ic} }{  3   n_{\mbox{\tiny P}}    \pi^{1/2}  a_{\mbox{\tiny P}}^{5/2+1/m}      \mathbb{C}(\xi_{\mbox{\tiny B}}) }
$$
where the correction factor $\mathbb{C}(\xi_{\mbox{\tiny B}})$ enforces $G=\int_0^{\Delta\gamma_c}   T_{\mbox{\tiny{B}}}  \mathrm{d}(\gamma_{\mbox{\tiny{P}}}-\gamma)$ and is equal to
$$
\mathbb{C}(\xi_{\mbox{\tiny B}}) 
= 
\int_0^1 	
\frac{\left(1+\xi_{\mbox{\tiny B}}\right)^2 \left(1-\sqrt{1-\hat{\gamma}^2}\right) \left(1-\hat{\gamma}^2\right)^{3/4}}
            {1+\xi_{\mbox{\tiny B}}-\sqrt{1-\hat{\gamma}^2}}
 \mathrm{d}\hat{\gamma}
$$
The nondimensional separation traction thus simplifies to
\begin{equation} 
\hat{T} 
=
\frac{  T_{\mbox{\tiny{B}}}  3   a_{\mbox{\tiny P}}^{1+1/m}   n_{\mbox{\tiny P}} }{  2 K_{Ic}^2 }
=
\frac{1}{\mathbb{C}(\xi_{\mbox{\tiny B}}) }
\frac{\left(1+\xi_{\mbox{\tiny B}}\right)^2 \left(1-\sqrt{1-\hat{\gamma}^2}\right) \left(1-\hat{\gamma}^2\right)^{3/4}}
       {1+\xi_{\mbox{\tiny B}}-\sqrt{1-\hat{\gamma}^2}}
\end{equation}
Figure~\ref{Fig-CohesiveZoneModel} shows the non-dimensional separation force $\hat{P}$ and non-dimensional separation traction $\hat{T}$ as a function of the non-dimensional separation displacement $\hat{\gamma}$ for different regularization parameters $\xi_{\mbox{\tiny B}}$. The similarity between a typical cohesive traction-separation curve under stain control and the traction-separation curves depicted in the figure is evident, for $\xi_{\mbox{\tiny B}}>0$.

\subsection{Sensitivity analysis}
\label{Section-Sensitivity}

\begin{figure}[htbp]
    \centering
    \begin{tabular}{ccc}
    \includegraphics[scale=0.415]{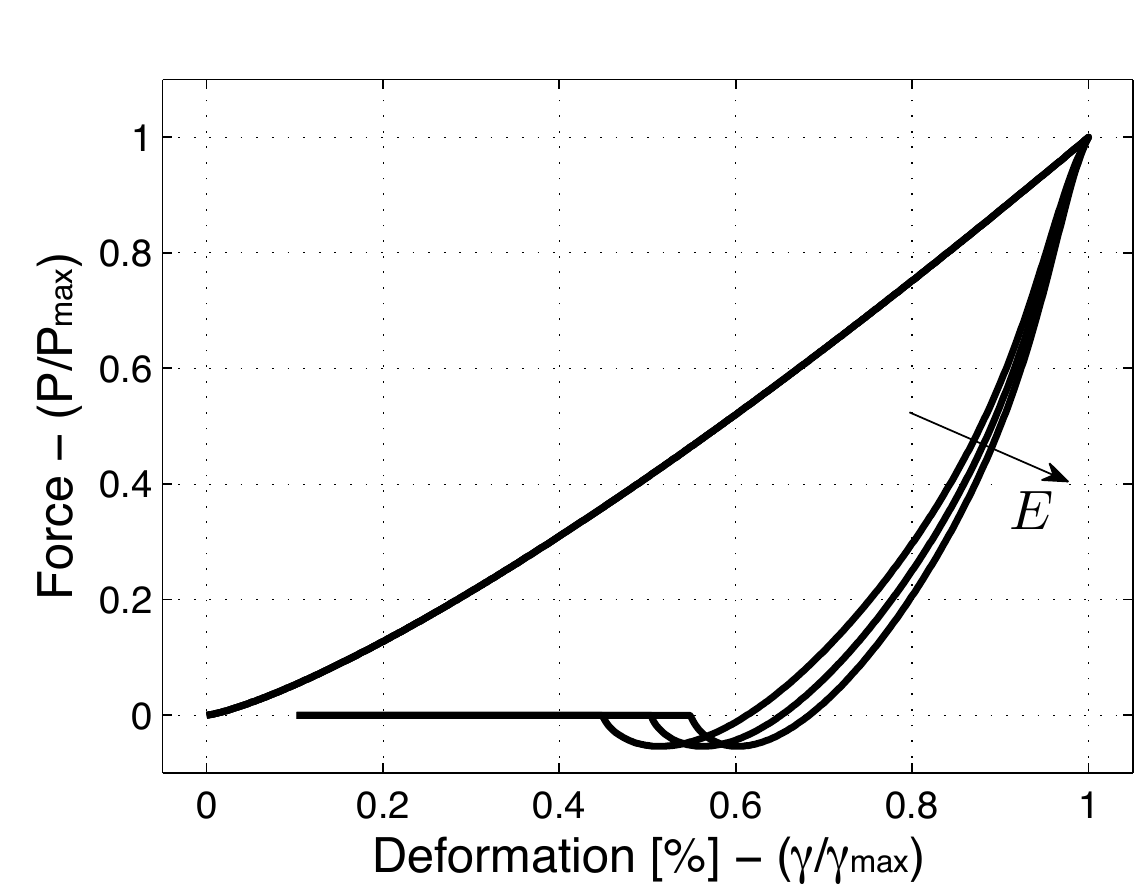}
    &
    \includegraphics[scale=0.415]{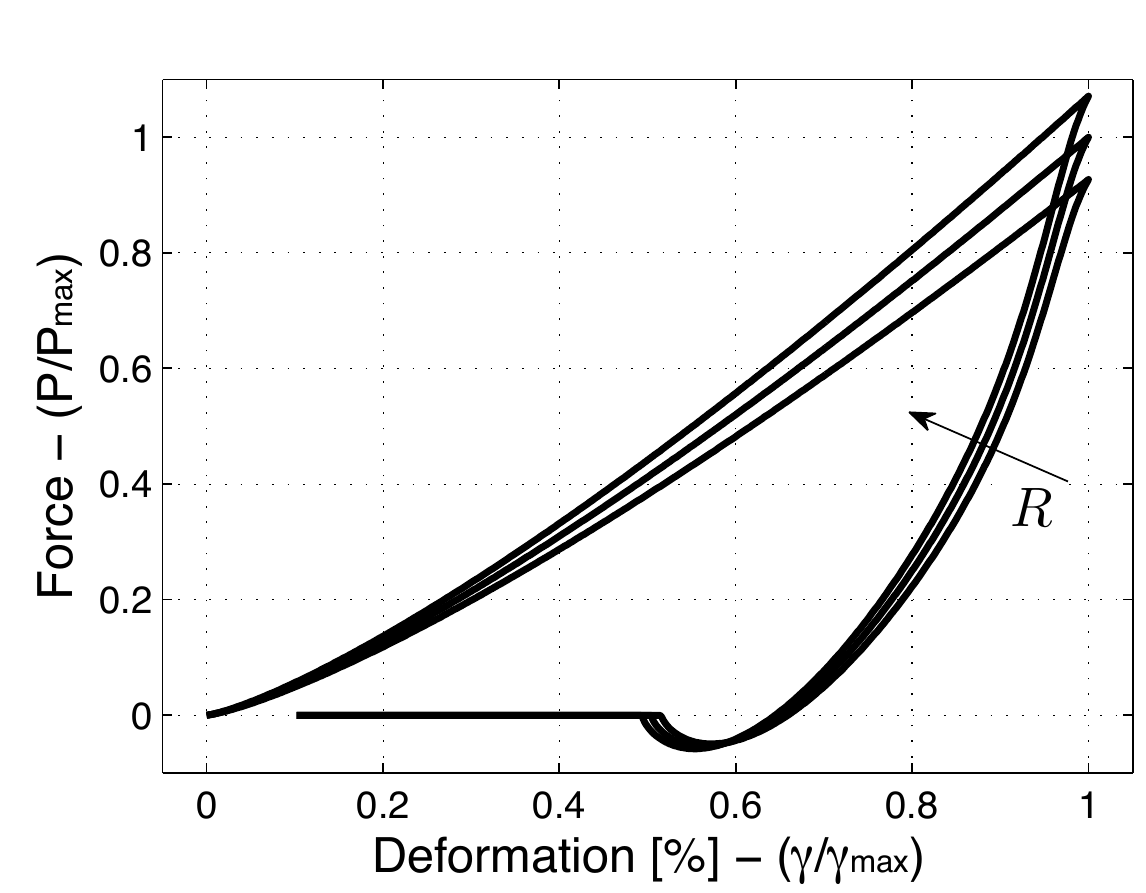}
    &
    \includegraphics[scale=0.415]{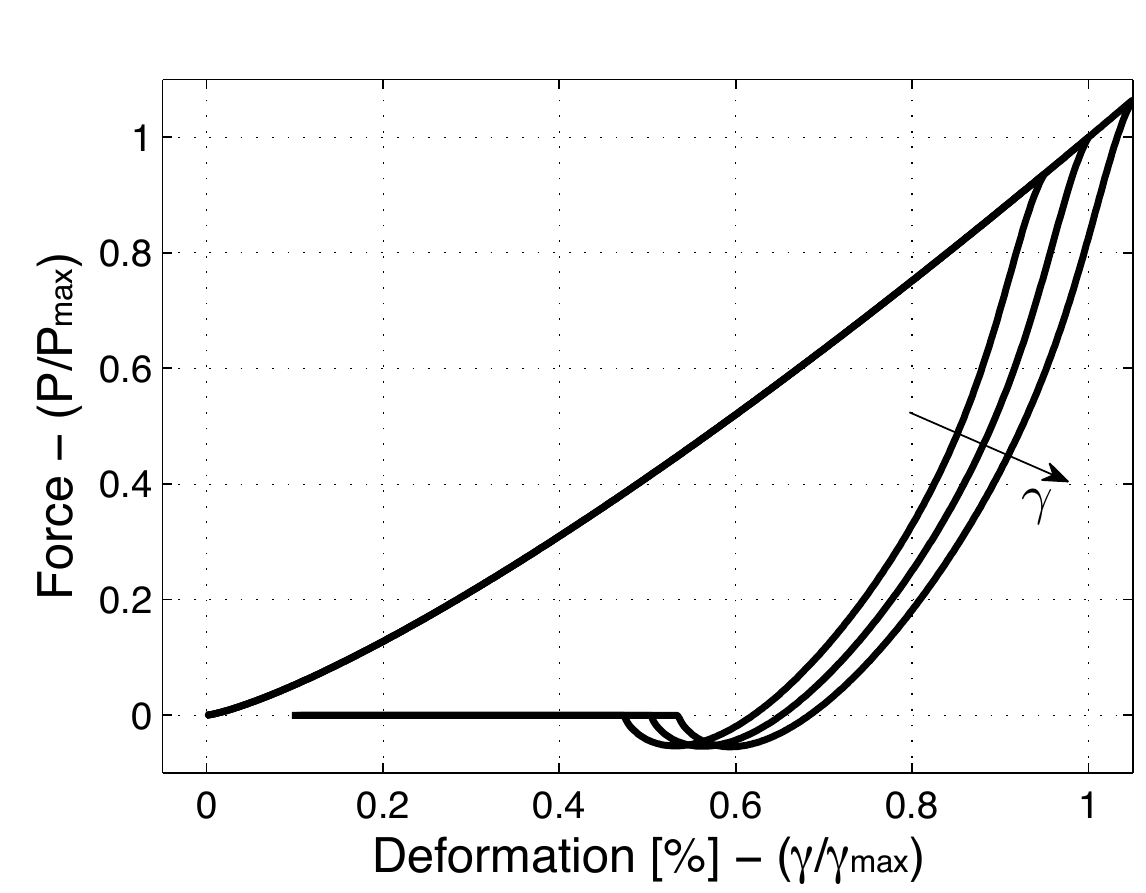}
    \\
    \includegraphics[scale=0.415]{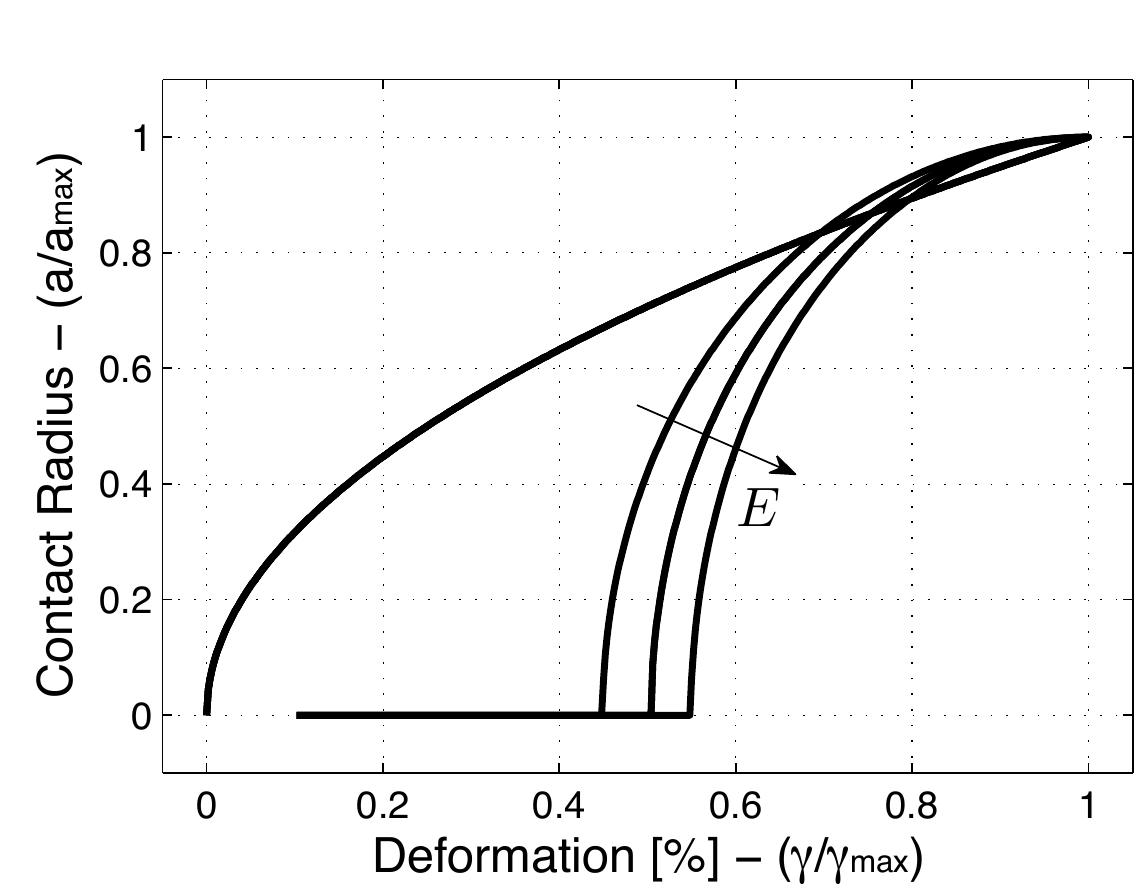}
    &
    \includegraphics[scale=0.415]{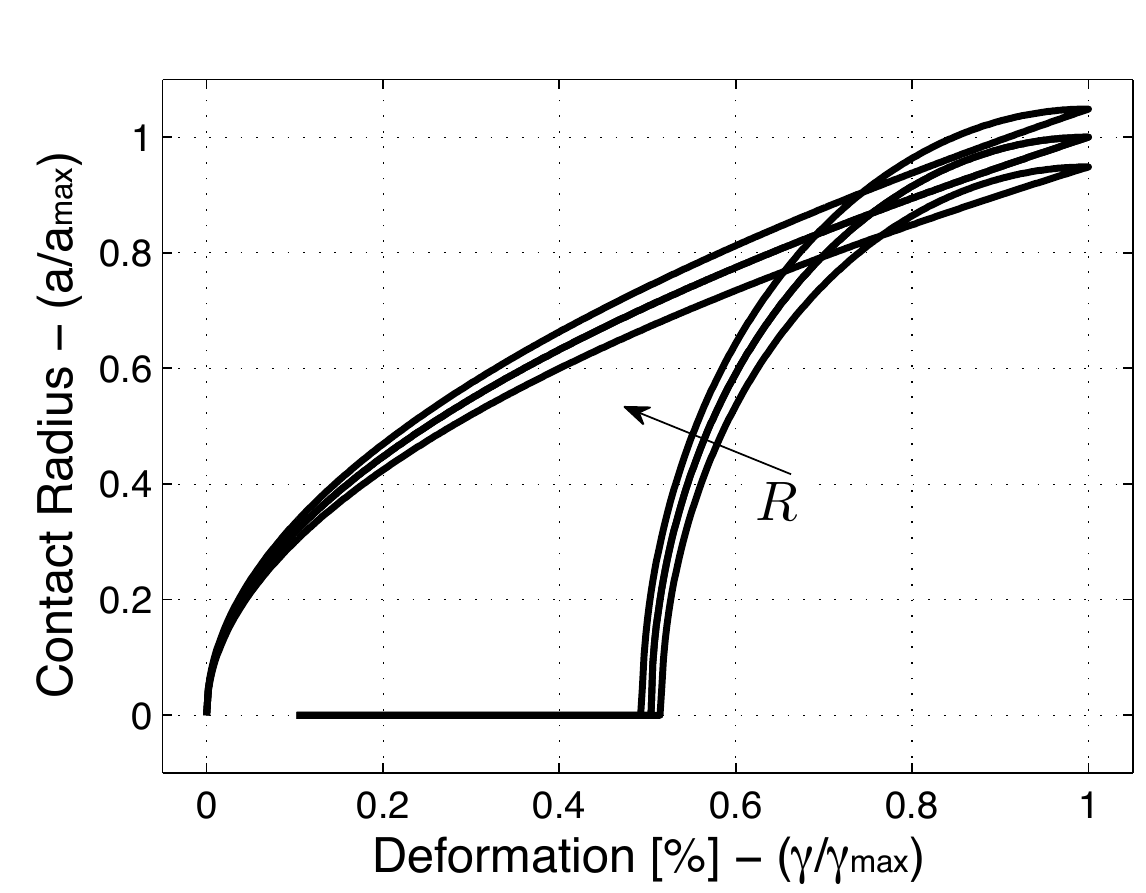}
    &
    \includegraphics[scale=0.415]{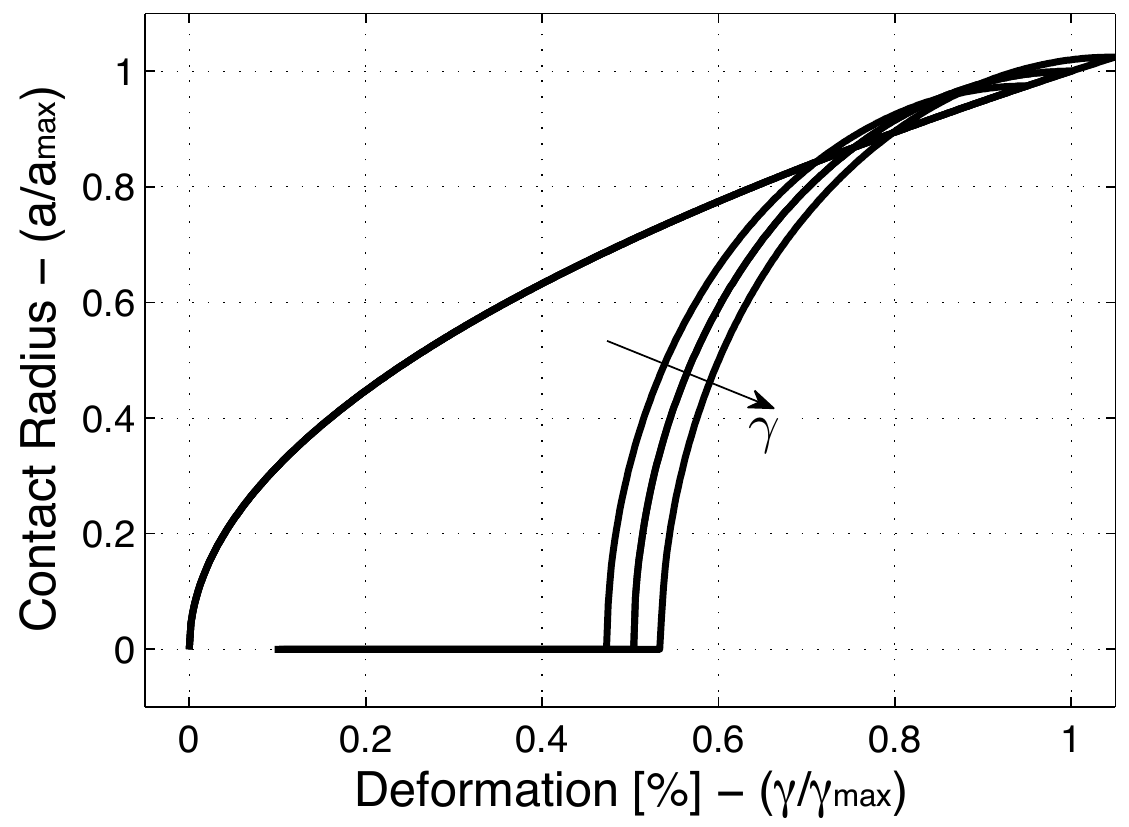}
    \\
    \footnotesize{(a) $E$~$\pm10\%$}
    &
    \footnotesize{(b) $R$~$\pm10\%$}
    &
    \footnotesize{(c) $\gamma$~$\pm5\%$}
	\\ 
    \hline
    \includegraphics[scale=0.415]{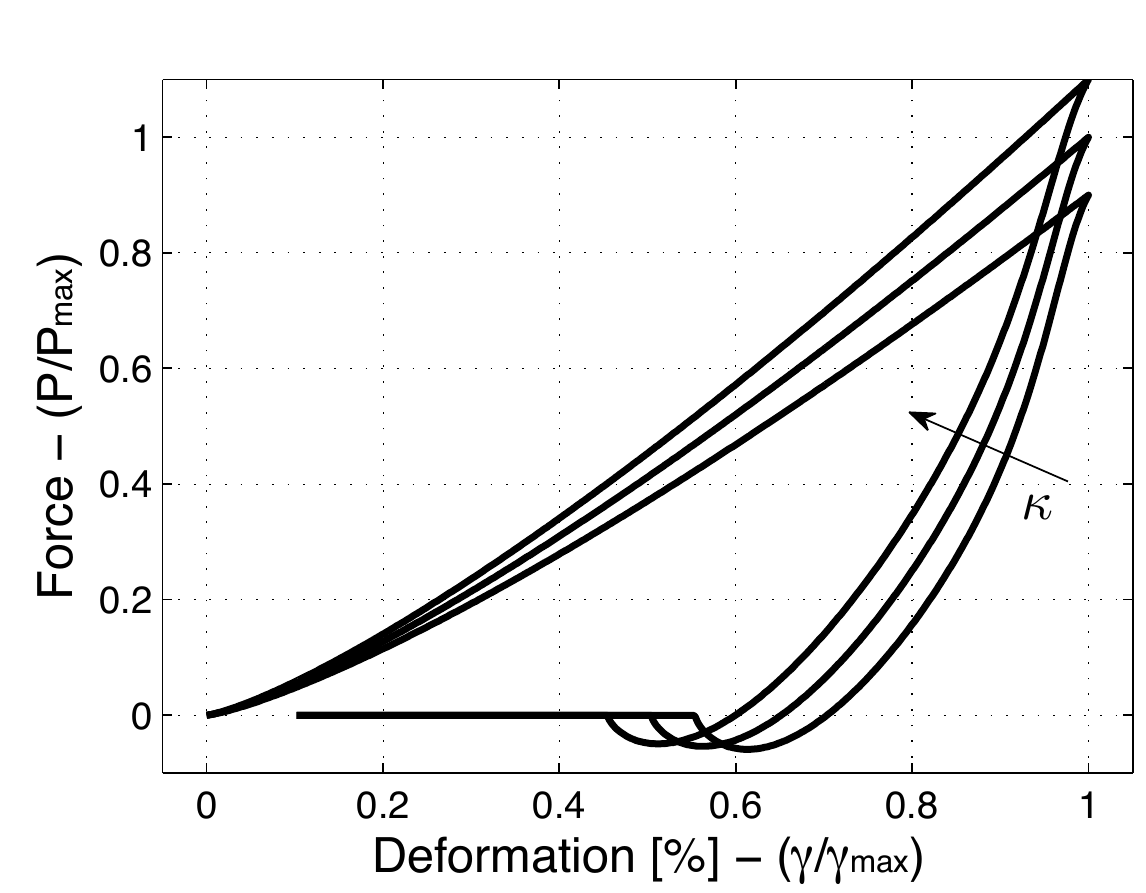}
    &
    \includegraphics[scale=0.415]{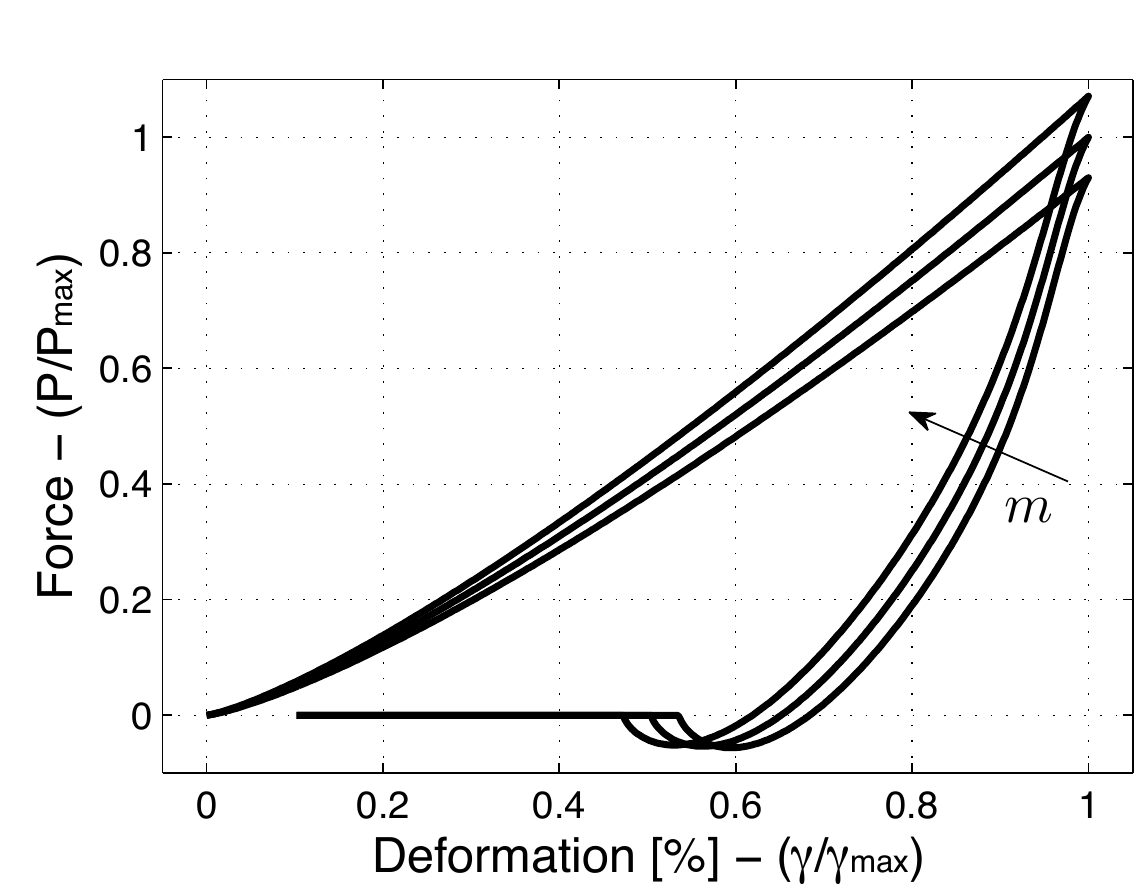}
    &
    \includegraphics[scale=0.415]{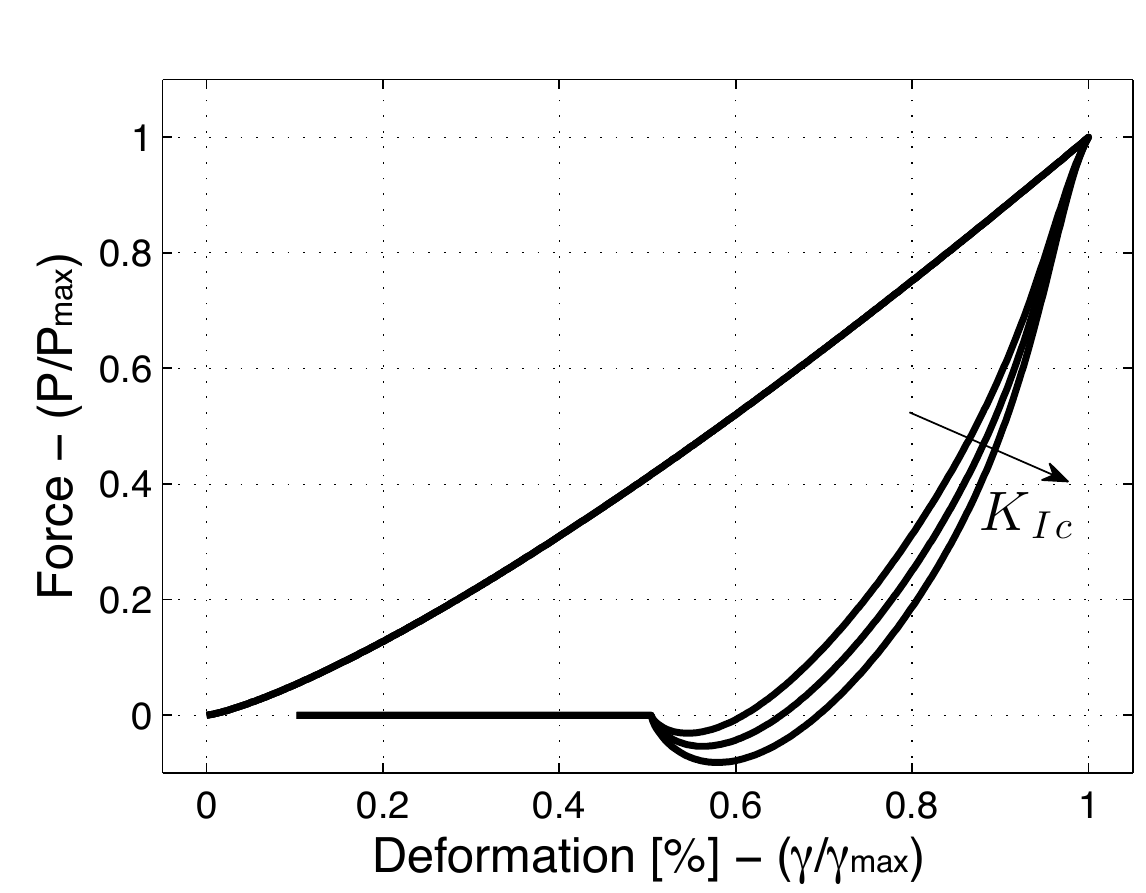}
    \\
    \includegraphics[scale=0.415]{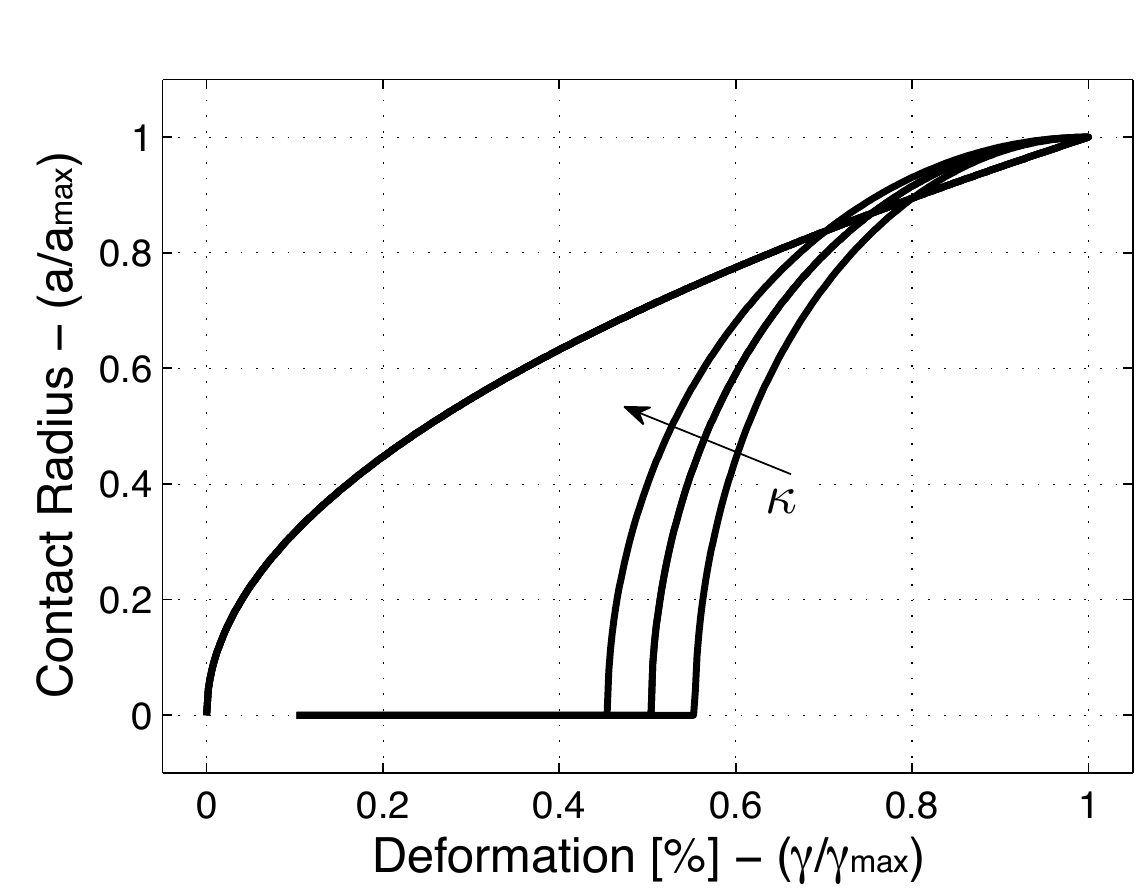}
    &
    \includegraphics[scale=0.415]{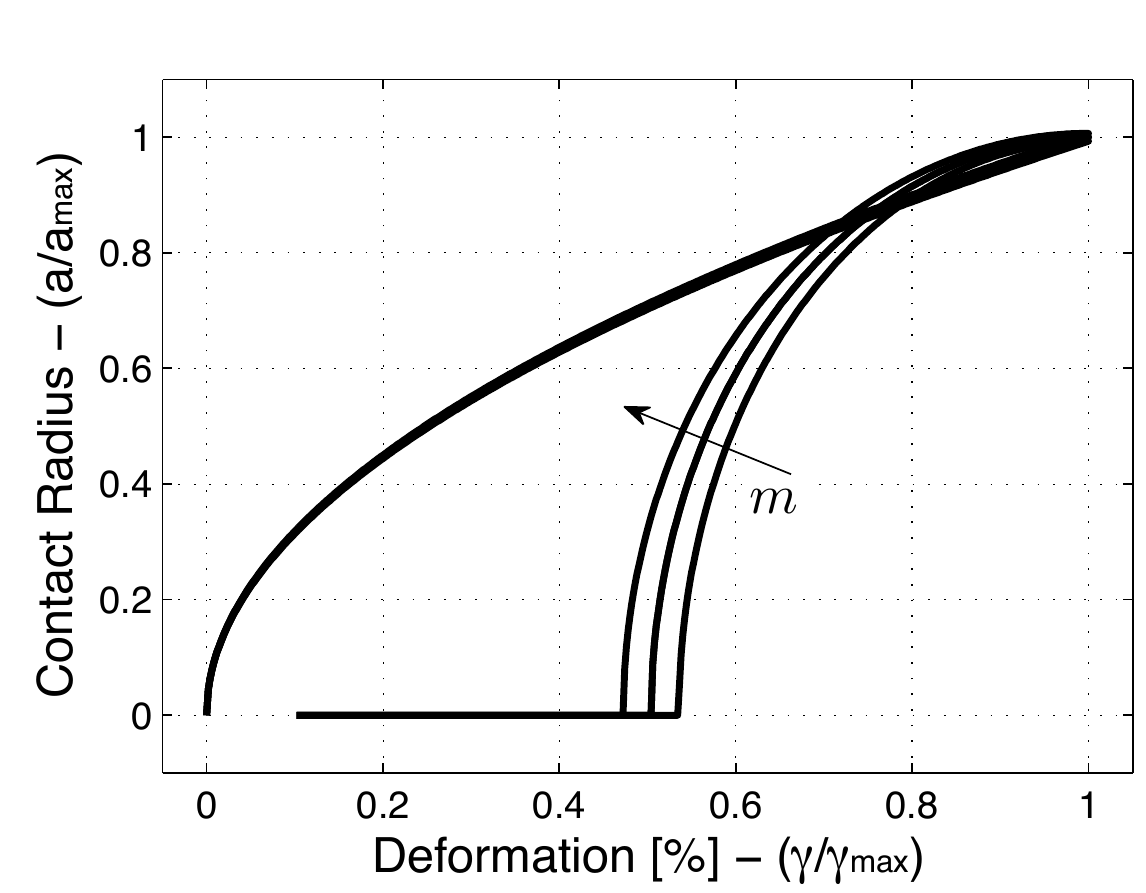}
    &
    \includegraphics[scale=0.415]{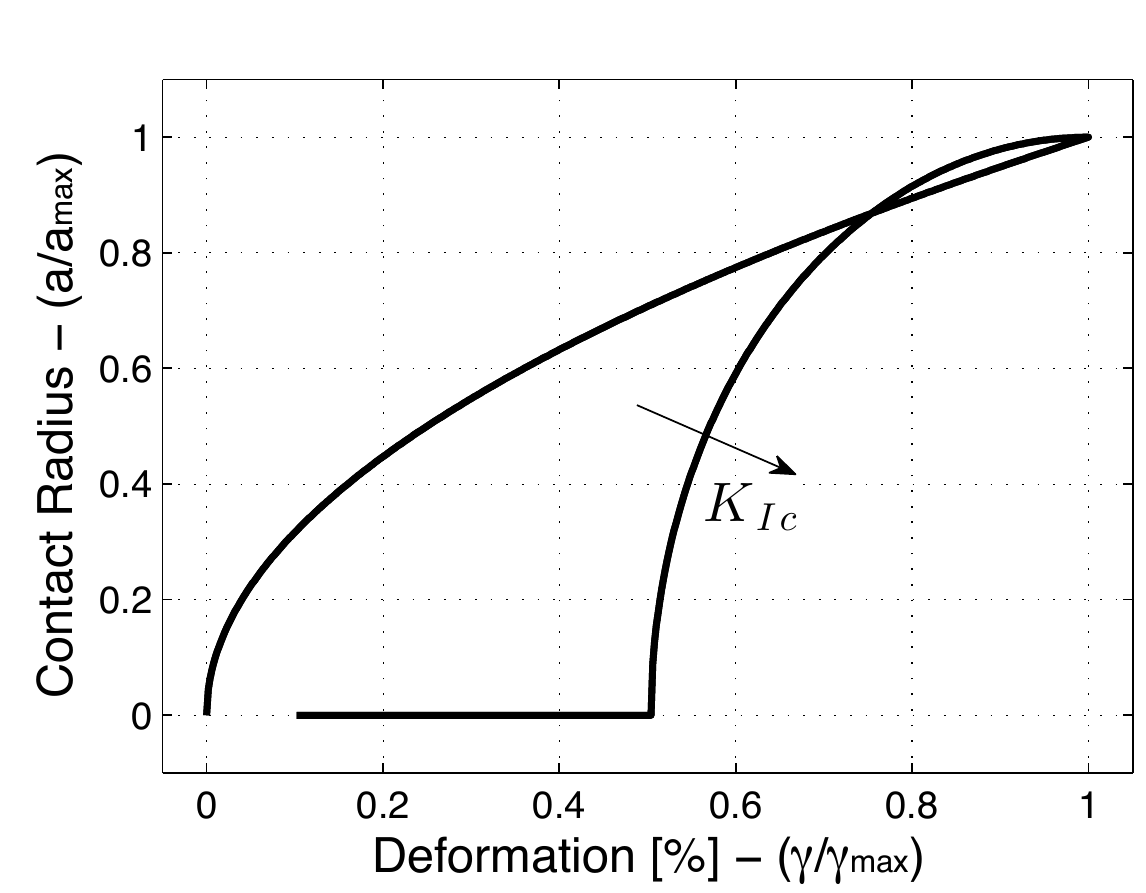}
    \\
    \footnotesize{(d) $\kappa$~$\pm10\%$}
    &
    \footnotesize{(e) $m$~$\pm2.5\%$}
    &
    \footnotesize{(f) $K_{Ic}$~$\pm25\%$}
    \end{tabular}
    \caption{Sensitivity analysis. Reference values $P_\mathrm{max}$ and $a_\mathrm{max}$ correspond to Fig.~\ref{Fig-Validation-Du2007} and \cite{Du-2008}, that is $R=1.3~\mu$m, $E=233$~GPa, $\nu=0.3$, $\kappa=12.1$~GPa, $m=1.79$, $K_{Ic}=0.48$~MPa~m$^{1/2}$, $\gamma_{\textrm{max}}=28.2$~nm. The bonding energy to elastic energy ratio is $\chi=0.0493$.}
    \label{Fig-SensitivityAnalysis}
\end{figure}

The proposed loading-unloading contact law depends on five material properties, namely two elastic properties ($E$ and $\nu$), two plastic properties ($\kappa$, $m$) and one fracture mechanics property ($K_{Ic}$). In order to gain insight into the role of these parameters and the coupled mechanisms involved, we performed a sensitivity analysis whose results are presented in Figure~\ref{Fig-SensitivityAnalysis}. It is interesting to note that the bonding surface can be controlled by changing the Young's modulus $E$ without changing the peak force, that is without changing the compaction force (see Fig.~\ref{Fig-SensitivityAnalysis}a). Bonding surface can also be manipulated by changing the plastic stiffness $\kappa$ (Fig.~\ref{Fig-SensitivityAnalysis}d), plastic law exponent $m$ (Fig.~\ref{Fig-SensitivityAnalysis}e) and the porosity or relative density through $\gamma$  (Fig.~\ref{Fig-SensitivityAnalysis}c) but, in contrast, this inevitably results in a change of the compaction force. This is valuable insight for product and process design integration, since, as mentioned above, compact strength is directly correlated to the bonding surface created during the compaction.

\section{Particle mechanics approach to powder compaction}
\label{Section-ParticleMechanicsAlgorithm}

The particle mechanics approach for granular systems under high confinement, developed by Gonzalez and Cuiti\~{n}o \cite{Gonzalez-2016}, describes each individual particle in the powder bed, and the collective rearrangement and deformation of the particles that result in a compacted specimen. This approach has been used to predict the microstructure evolution during die-compaction of elastic spherical particles up to relative densities close to one. By employing a nonlocal contact formulation that remains predictive at high levels of confinement \cite{Gonzalez-2012}, this study demonstrated that the coordination number depends on the level of compressibility of the particles and thus its scaling behavior is not independent of material properties as previously thought. The study also revealed that distributions of contact forces between particles and between particles and walls, although similar at jamming onset, are very different at full compaction---being particle-wall forces are in remarkable agreement with experimental measurements reported in the literature. 

In this work, we extend the particle mechanics approach to the treatment of internal variables (i.e., $\mathbf{a}_{\mbox{\tiny P}}$ and $\mathbf{a}_{\mbox{\tiny B}}$) and their equations of evolution (i.e., Equations \ref{Eqn-ContactRadius} and \ref{Eqn-EvolutionOfInternalVariable}) under quasi-static evolution. Therefore, an equilibrium configuration is defined by the solution of a system of nonlinear equations that corresponds to static equilibrium of the granular system, that is sum of all elasto-plastic contact forces acting on each particle equals zero, that is
\begin{equation}
\label{Eqn-SOE}
    \sum\nolimits_{j \in \mathcal{N}_i} P\left( a(R_i+R_j-\|\mathbf{x}_i - \mathbf{x}_j\|, a_{\mbox{\tiny P},ij}), a_{\mbox{\tiny P},ij}, a_{\mbox{\tiny B},ij} \right)
                              \tfrac{\mathbf{x}_i - \mathbf{x}_j}{\|\mathbf{x}_i - \mathbf{x}_j\|}
    = \mathbf{0}
\end{equation}
where $\mathbf{x}_i$ and $\mathcal{N}_i$ are the position and all the neighbors of particle $i$, respectively, $a_{\mbox{\tiny P},ij}=a_{\mbox{\tiny P},ji}$ and $a_{\mbox{\tiny B},ij}=a_{\mbox{\tiny B},ji}$ by definition, $P(a, a_{\mbox{\tiny P}}, a_{\mbox{\tiny B}})$ is given by equation \eqref{Eqn-ContactForce}, and $a(\gamma, a_{\mbox{\tiny P}})$ is given by equation \eqref{Eqn-ContactRadius}. A sequential strategy is proposed to treat the nonlinear problem (see Algorithm \ref{Alg-SequentialStrategy}). The equations of static equilibrium are solved for $\mathbf{x}=(\mathbf{x}_1^T, ..., \mathbf{x}_N^T)^T$, for given internal variables $ \mathbf{a}_{\mbox{\tiny P}}=(a_{\mbox{\tiny P},12}, a_{\mbox{\tiny P},13}, ...,a_{\mbox{\tiny P},N-1,N})^T$ and  $ \mathbf{a}_{\mbox{\tiny B}}=(a_{\mbox{\tiny B},12}, a_{\mbox{\tiny B},13}, ...,a_{\mbox{\tiny B},N-1,N})^T$, by employing a trust-region method \cite{Coleman-1996,Conn-2000} that successfully overcomes the characteristic ill-posedness of the problem (e.g., due to metastability \cite{Mehta-2007}). The basic trust-region algorithm requires the solution of a minimization problem to determine the step between iterations, namely the trust-region step. This minimization problem is of the form $\min\{\psi(\mathbf{s}):\|\mathbf{s}\| \leq \Delta\}$, where $\mathbf{s}={^{n+1}\mathbf{x}}-{^{n}\mathbf{x}}$ is the trust-region step, $\Delta$ is a trust-region radius and $\psi$ is a quadratic function that represents a local model of the objective function about ${^{n}\mathbf{x}}$, that is 
$$
\psi(\mathbf{s})= \tfrac{1}{2} \| {^n}\mathbf{F} + {^n}\mathbf{K} \mathbf{s} \|^2 
= 
\tfrac{1}{2} \left< {^n}\mathbf{F},{^n}\mathbf{F} \right> + \left< {^n}\mathbf{K} \mathbf{s}, {^n}\mathbf{F} \right> + \tfrac{1}{2} \left< {^n}\mathbf{K} \mathbf{s}, {^n}\mathbf{K} \mathbf{s}\right>
$$ 
with ${^n}\mathbf{F}$ and ${^n}\mathbf{K}$ the global force vector and stiffness matrix at ${^{n}\mathbf{x}}$---the first term in the above equation is not required in the minimization problem. It is worth noting that the trust-region step is not necessarily in the direction of a quasi-Newton step and that the trust-region radius acts as a regularization term that controls the growth in the size of the least squares solution observed in most ill-posed \cite{Vicente-1996}. Trading accuracy for performance, Byrd \cite{Byrd-1988}, among others, proposed to approximate the minimization problem by restricting the problem to a two-dimensional subspace. Furthermore, the two-dimensional subspace can be determined by a preconditioned conjugate gradient process and the trust-region radius can be adjusted over the iterative process (see, e.g., \cite{More-1983}). Here we adopt the implementation available in MATLAB R2016a Optimization Toolbox.

It is worth noting that an equilibrium configuration is not obtained by artificially damped or cooled-down dynamic processes but rather by iterative solvers that follow the energy landscape around the solution of static equilibrium. The strain path dependency of the contact law is accounted for incrementally by updating internal variables $\mathbf{a}_{\mbox{\tiny P}}$ and $\mathbf{a}_{\mbox{\tiny B}}$ at the new equilibrium configuration. Specifically, new particle-to-particle contact radii $a_{ij}^{\mbox{\tiny new}}$ are computed at the converged equilibrium configuration and internal variables are updated as follows:  
\begin{equation}
    \left\{
        \begin{array}{ll}
             a_{\mbox{\tiny P},ij}^{\mbox{\tiny new}} \leftarrow a_{ij}^{\mbox{\tiny new}}~,~~~a_{\mbox{\tiny B},ij}^{\mbox{\tiny new}} \leftarrow a_{ij}^{\mbox{\tiny new}}
             &
             \mbox{if $a_{ij}^{\mbox{\tiny new}} > a_{\mbox{\tiny P},ij}$~~~~plastic loading and bond formation}
             \\
	     a_{\mbox{\tiny B},ij}^{\mbox{\tiny new}} \leftarrow 0
	     &
             \mbox{if $a_{ij}^{\mbox{\tiny new}} = 0$~~~~~~~~~solid bridge is broken}
             \\
             a_{\mbox{\tiny P},ij}^{\mbox{\tiny new}} \leftarrow a_{\mbox{\tiny P},ij}~,~~~a_{\mbox{\tiny B},ij}^{\mbox{\tiny new}} \leftarrow a_{\mbox{\tiny B},ij}
             &
             \mbox{otherwise~~~~~~~~~~elastic (un)loading}
        \end{array}
     \right.
\end{equation}

\begin{algorithm}[htbp]
\caption{ParticleMechanicsApproach: sequential strategy for solving the equilibrium problem and for updating history-dependent internal variables.}
\label{Alg-SequentialStrategy}
\begin{algorithmic}[1]
\REQUIRE Initial guess for particles' coordinates ${^1}\mathbf{x}$, current state of internal variables $\{ \mathbf{a}_{\mbox{\tiny P}}, \mathbf{a}_{\mbox{\tiny B}} \}$, TOL1, trust-region radius $\Delta$ and the simplicial complex $X$
\STATE $\text{Error} \leftarrow \text{TOL1}$
\STATE $n \leftarrow 1$
\WHILE{$\text{Error} \geq \text{TOL1}$}
\STATE \textit{/$\ast$~~Compute global force and global stiffness.~~$\ast$/}
\STATE $\{{^n}\mathbf{F},{^n}\mathbf{K}\} \leftarrow \mathbf{ContactFormulation}({^n}\mathbf{x},\mathbf{a}_{\mbox{\tiny P}},\mathbf{a}_{\mbox{\tiny B}},X)$
\STATE \textit{/$\ast$~~Update coordinates with the trust-region step obtained by restricting the problem to a two-dimensional subspace \cite{Byrd-1988}.~~$\ast$/}
\STATE ${^n\mathbf{s}} \leftarrow \text{argmin}_\mathbf{s} ~  \left< {^n}\mathbf{K} \mathbf{s}, {^n}\mathbf{F} \right> + \tfrac{1}{2} \left< {^n}\mathbf{K} \mathbf{s}, {^n}\mathbf{K} \mathbf{s}\right>$ ~~subject to~~ $\|\mathbf{s}\| \le \Delta$
\STATE ${^{n+1}\mathbf{x}} \leftarrow {^{n}\mathbf{x}} + {^n\mathbf{s}}$
\STATE \textit{/$\ast$~~Compute a measure of convergence.~~$\ast$/}
\STATE $\text{Error} \leftarrow \| {^n\mathbf{s}} \|$
\STATE $n \leftarrow n+1$
\ENDWHILE
\STATE \textit{/$\ast$~~Update internal variables based on converged solution.~~$\ast$/}
\STATE $\{  \mathbf{a}_{\mbox{\tiny P}}^{\mbox{\tiny new}}, \mathbf{a}_{\mbox{\tiny B}}^{\mbox{\tiny new}} \} \leftarrow \mathbf{UpdateInternalVariables}({^n}\mathbf{x},\mathbf{a}_{\mbox{\tiny P}},\mathbf{a}_{\mbox{\tiny B}})$
\RETURN $\{ {^n\mathbf{x}}, \mathbf{a}_{\mbox{\tiny P}}^{\mbox{\tiny new}}, \mathbf{a}_{\mbox{\tiny B}}^{\mbox{\tiny new}}\}$ 
\end{algorithmic}
\end{algorithm}

\section{Microstructure formation and evolution during compaction, unloading and ejection}
\label{Section-MicrostructureEvolution}

We next report three-dimensional particle mechanics static calculations that enable us to study microstructure evolution during die-compaction up to relative densities close to one, unloading and ejection of elasto-plastic spherical particles with bonding strength. We employ the generalized loading-unloading contact laws presented in Section~\ref{Section-GeneralizedContactLaws} which result in a numerically robust and efficient formulation. The contact laws are continuous at the onset of unloading by means of a regularization term, they are explicit in terms of the relative position between the particles, and their strain path dependency is accounted for incrementally. Here we adopt a regularization parameter equal to $\xi_{\mbox{\tiny B}}=0.01$.
  
Three different relative densities are defined and used in the study, namely (i) the maximum relative density of the compact inside the die $\rho^{\mbox{\tiny in-die}}_{\mbox{\tiny max}}$ which occurs at the shortest gap between the two punches (i.e., (B) in Fig.~\ref{Fig-CompactionSchematics}), (ii) the minimum relative density of the compact inside the die  $\rho^{\mbox{\tiny in-die}}_{\mbox{\tiny min}}$ which occurs right after separation of the upper punch from the compact (i.e., (C) in Fig.~\ref{Fig-CompactionSchematics}), and (iii) the relative density of the tablet out of the die $\rho^{\mbox{\tiny tablet}}$ which is approximately equal to $\rho^{\mbox{\tiny in-die}}_{\mbox{\tiny min}}$ (i.e., (D) in Fig.~\ref{Fig-CompactionSchematics}).  
 
\begin{table}[htbp]
    \centering \footnotesize
{\onehalfspacing
    \begin{tabular}{c|cc|cc|c}
        \hline
        &
        \multicolumn{2}{p{0.20\linewidth}|}{\centering Elastic Deformation}
        &
        \multicolumn{2}{p{0.20\linewidth}|}{\centering Plastic Deformation}
        &
        \multicolumn{1}{p{0.34\linewidth}}{\centering Bonding \& Fracture}
        \\
        \hline
        &
        $E$ & $\nu$ & $\kappa$ & $m$ & $K_{Ic}$      \\
        \hline
        Material 1
        &
        5~GPa & 0.25 & 150~MPa & 2.00 & 1.26~MPa~m$^{1/2}$  ($\omega=150$~J/m$^2$,$G_p=0$)\\
        \hline
        Material 2
        &
        30~GPa & 0.25 & 900~MPa & 2.00 & 6.19~MPa~m$^{1/2}$ ($\omega=600$~J/m$^2$,$G_p=0$)  \\
        \hline
    \end{tabular}
    }
    \caption{Material properties.}
    \label{Table-MaterialProperties}
\end{table}
  
\begin{figure}[htbp]
    \centering
    \begin{tabular}{ll}
        \includegraphics[scale=2.5]{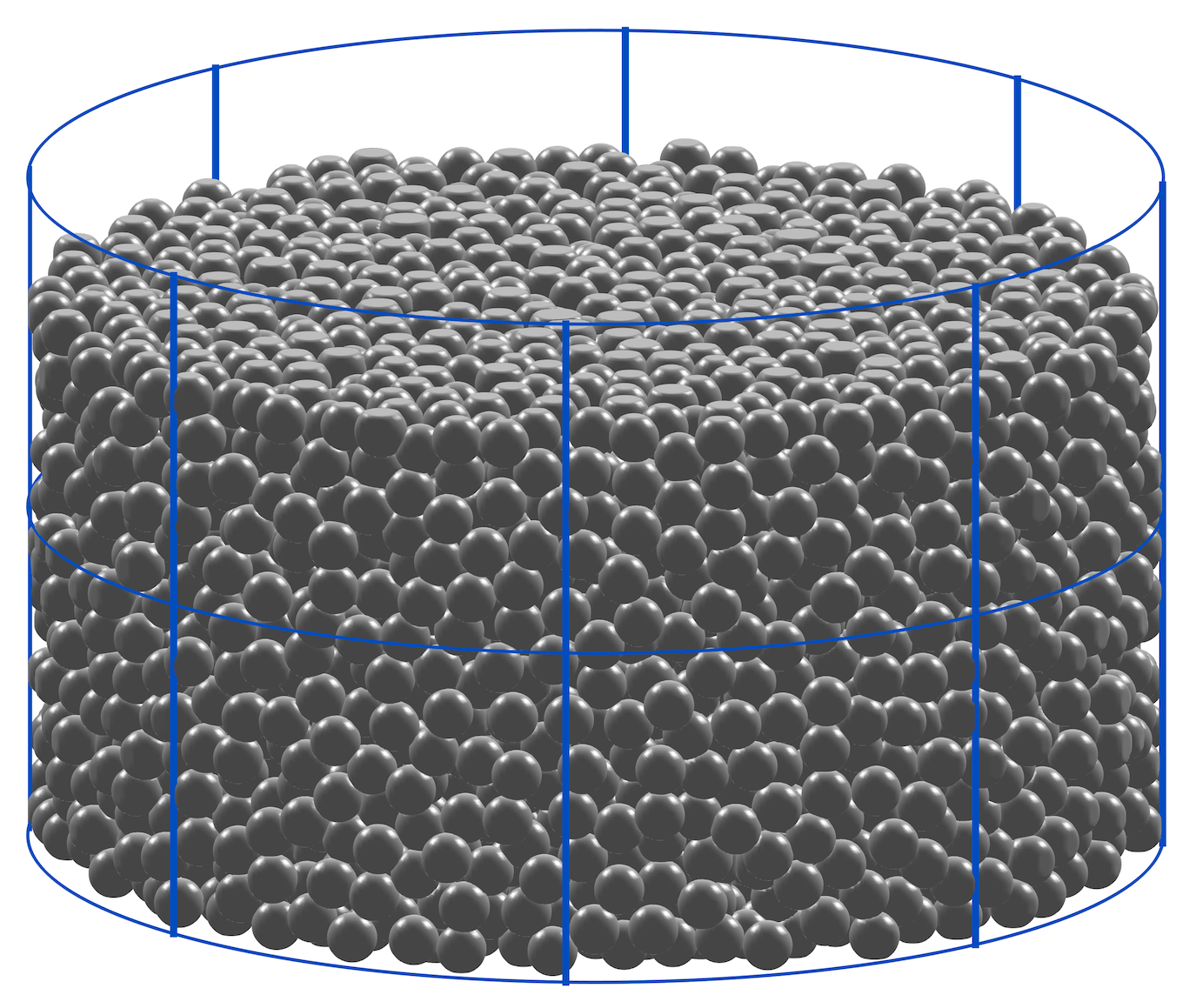}
        &
        \includegraphics[scale=2.5]{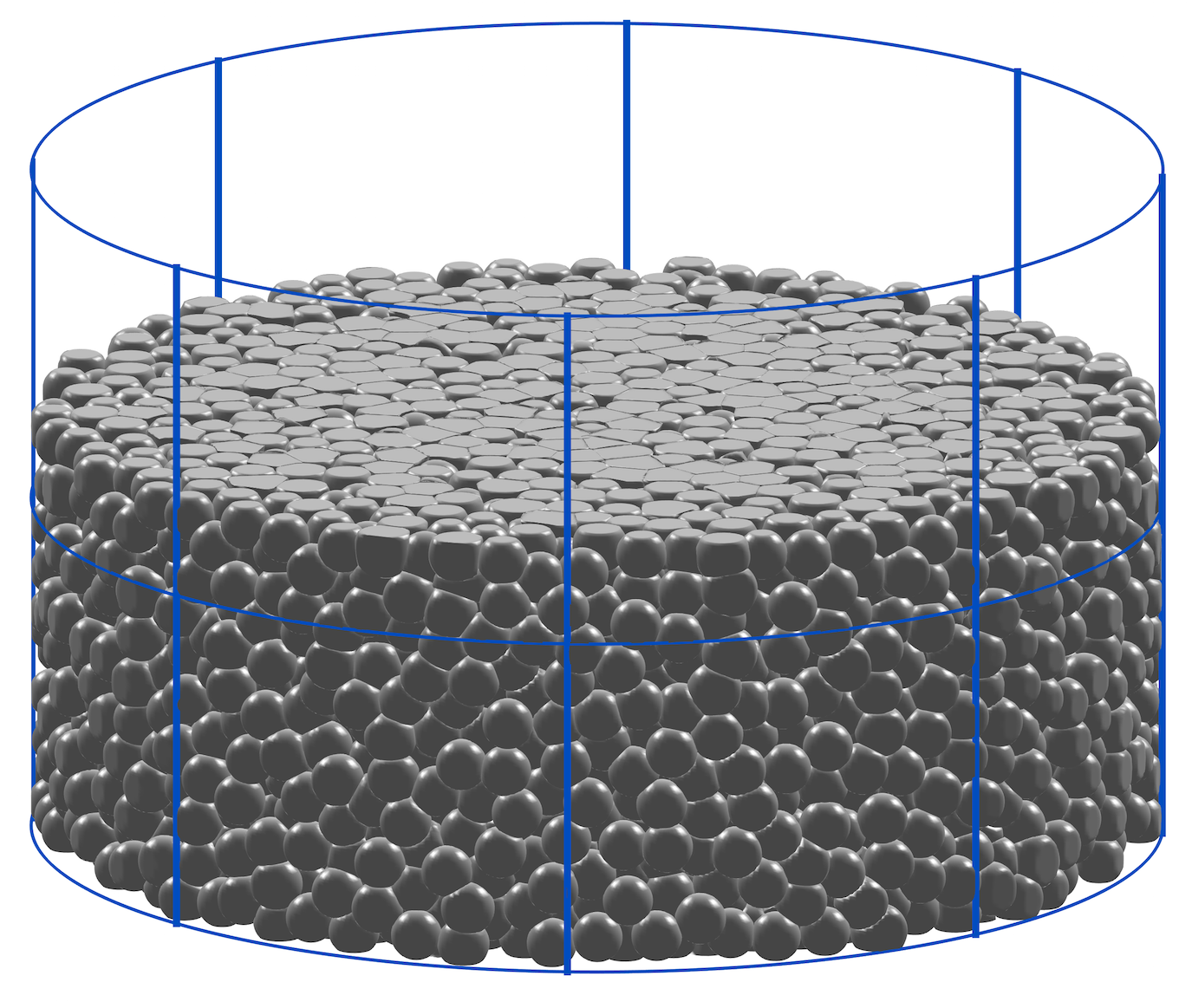}
    \\
    \small{(a)} 
    &
    \small{(b)}        
    \\
    \end{tabular}
    \caption{Compacted granular bed of Material 1 at  $\rho^{\mbox{\tiny in-die}}_{\mbox{\tiny max}} = 0.7323$ (a) and at $\rho^{\mbox{\tiny in-die}}_{\mbox{\tiny max}} = 0.9523$ (b).}
    \label{Fig-FullSystem}
\end{figure}

\begin{figure}[htbp]
    \centering
    \begin{tabular}{lllllll}
        \includegraphics[scale=0.57, trim=100 40 85 40, clip]{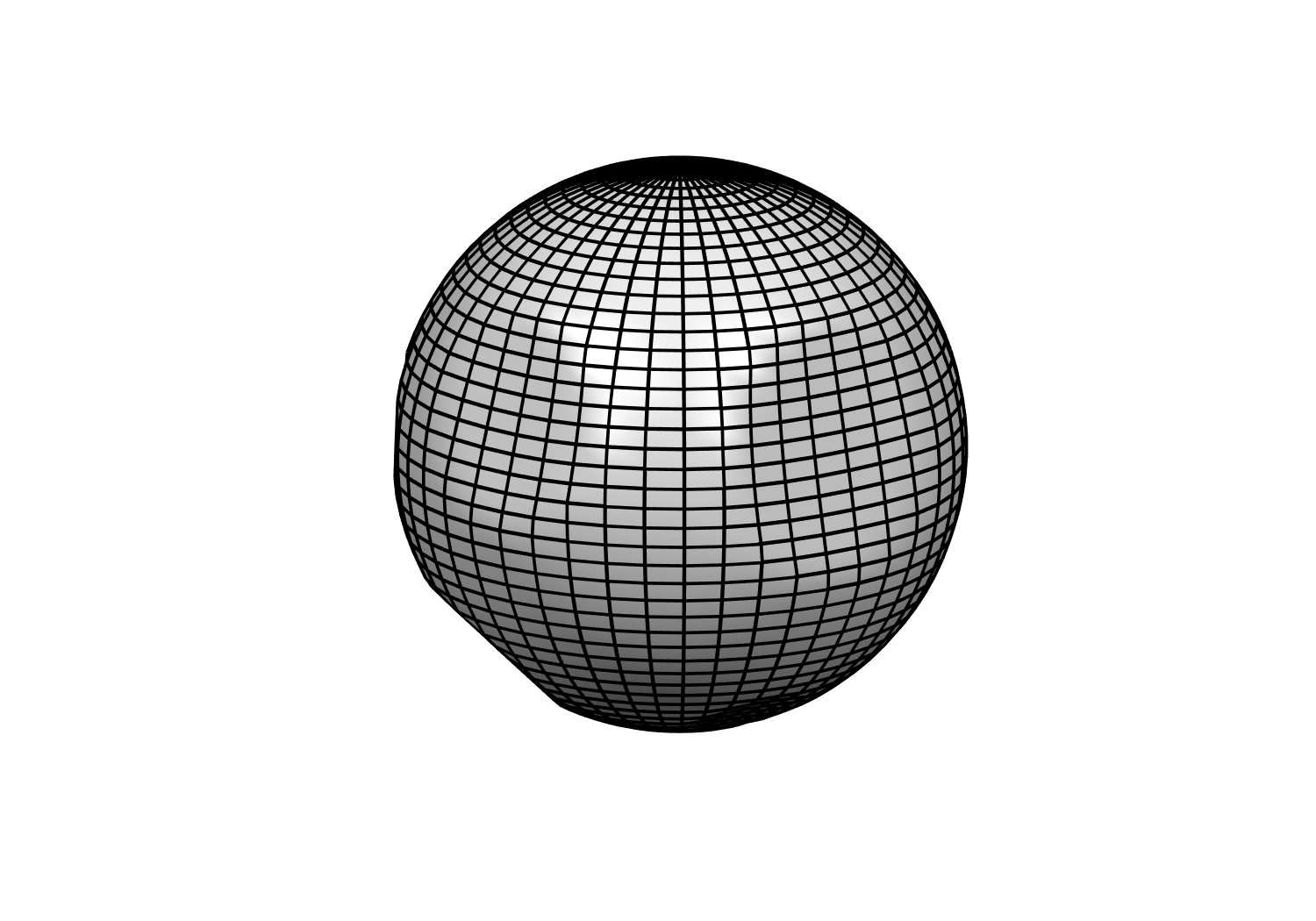}
        &
        \includegraphics[scale=0.57, trim=100 40 85 40, clip]{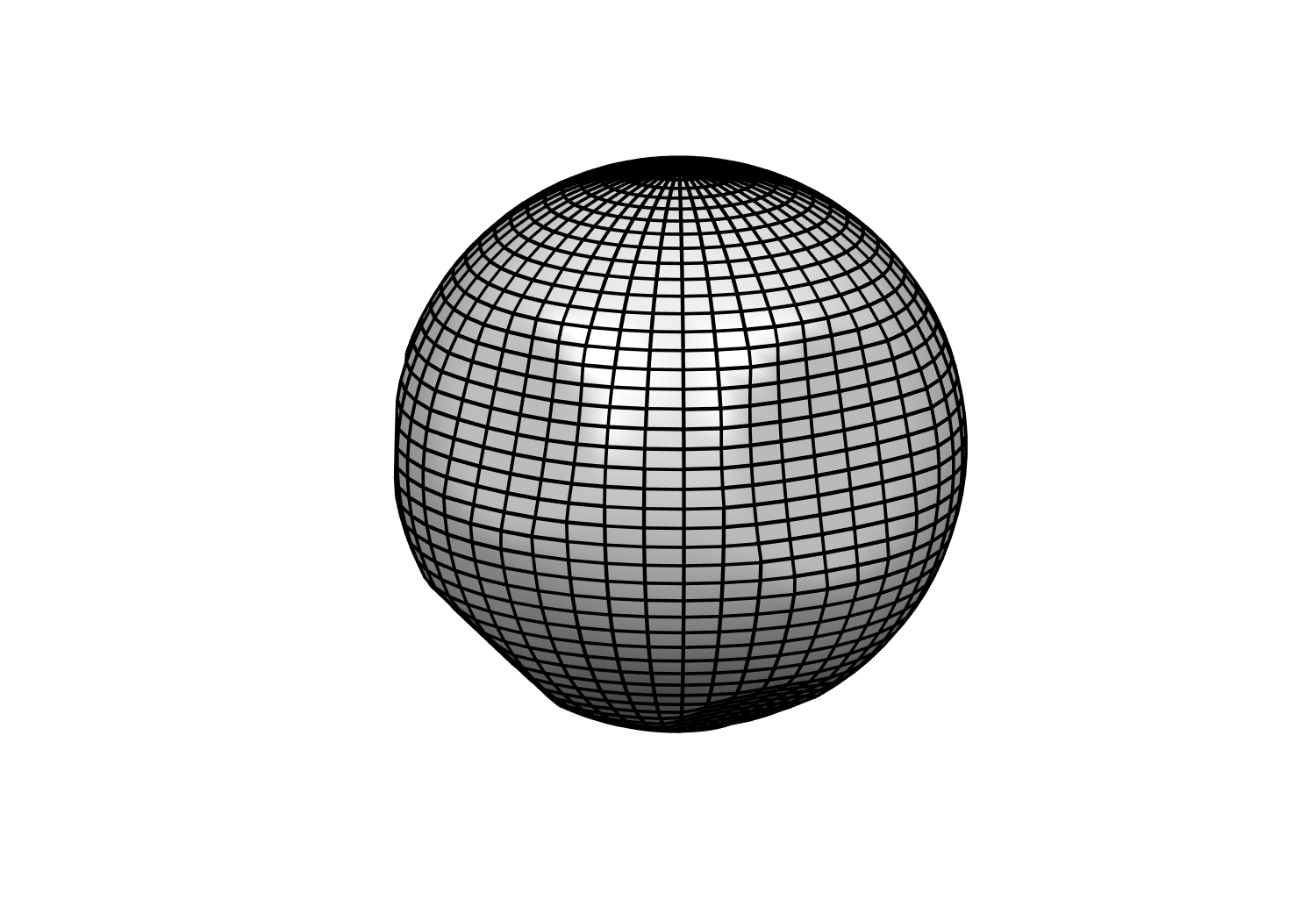}
        &
        \includegraphics[scale=0.57, trim=100 40 85 40, clip]{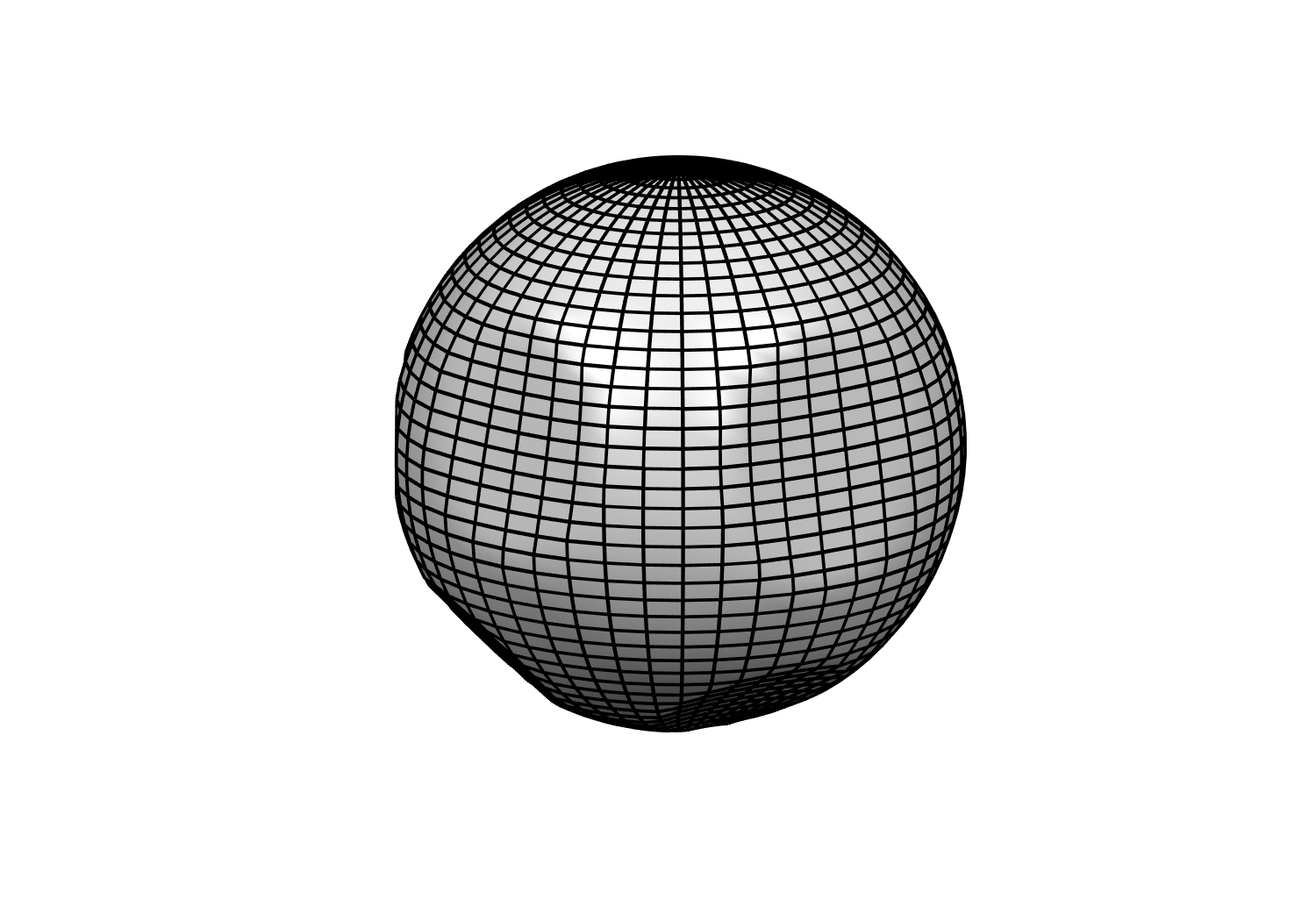}
        &
        \includegraphics[scale=0.57, trim=100 40 85 40, clip]{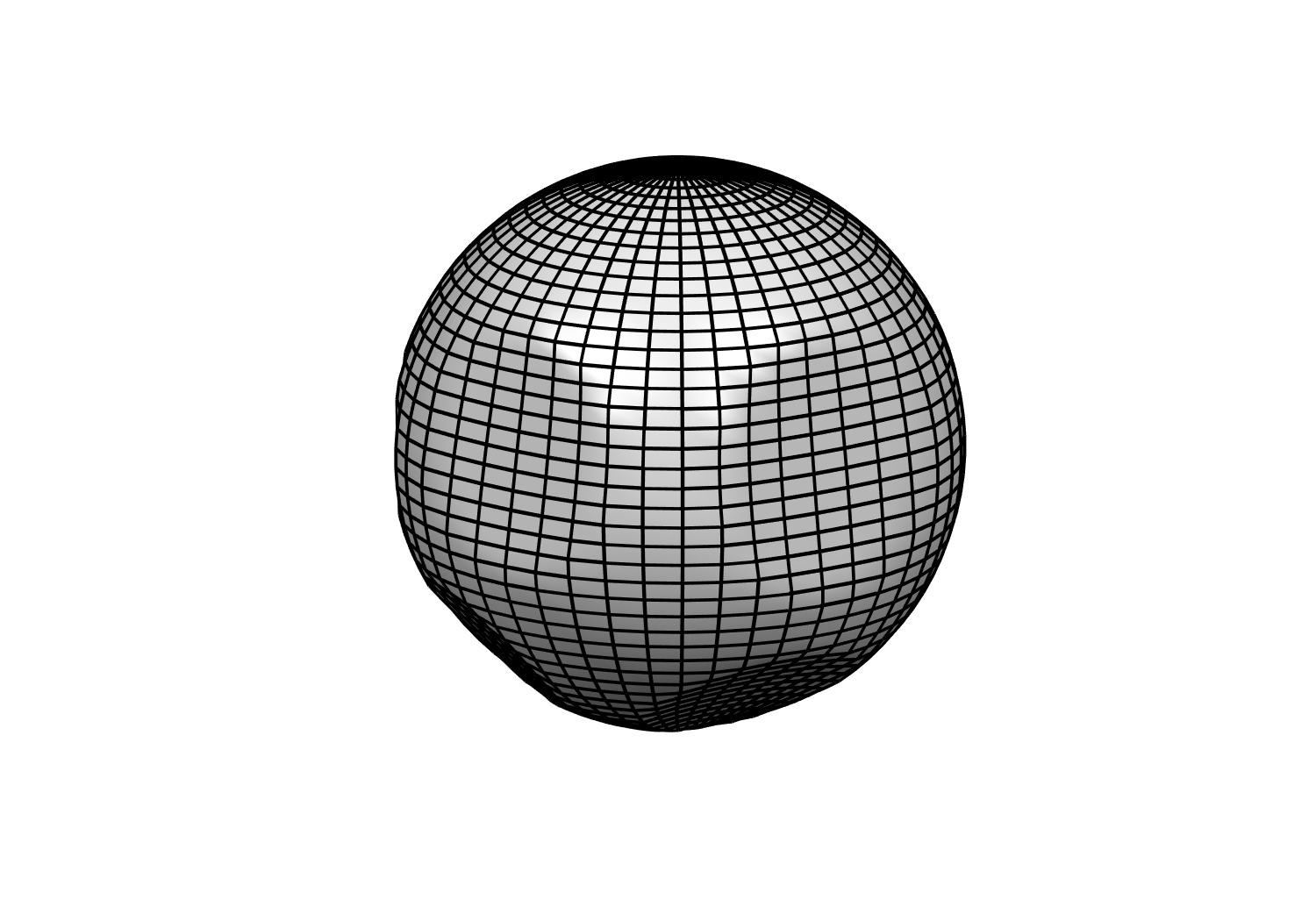}
        \\
        \small{(a) $\rho^{\mbox{\tiny in-die}}_{\mbox{\tiny max}} = 0.7323$}
        &
        \small{(b) $\rho^{\mbox{\tiny in-die}}_{\mbox{\tiny max}} = 0.7573$}
        &
        \small{(c) $\rho^{\mbox{\tiny in-die}}_{\mbox{\tiny max}} = 0.7841$}
        &
        \small{(d) $\rho^{\mbox{\tiny in-die}}_{\mbox{\tiny max}} = 0.8128$}
        \\
        \includegraphics[scale=0.57, trim=100 40 85 40, clip]{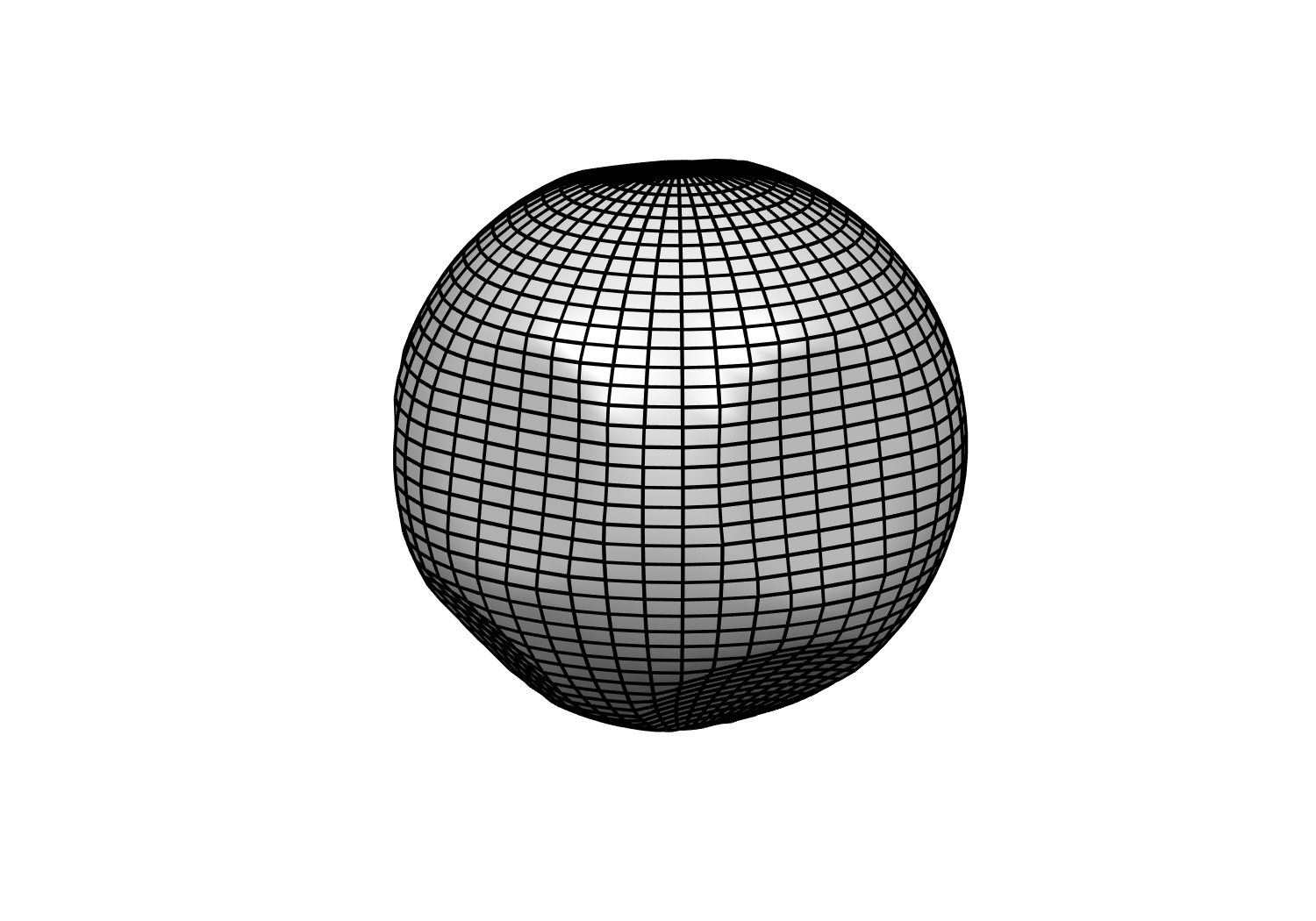}
        &
        \includegraphics[scale=0.57, trim=100 40 85 40, clip]{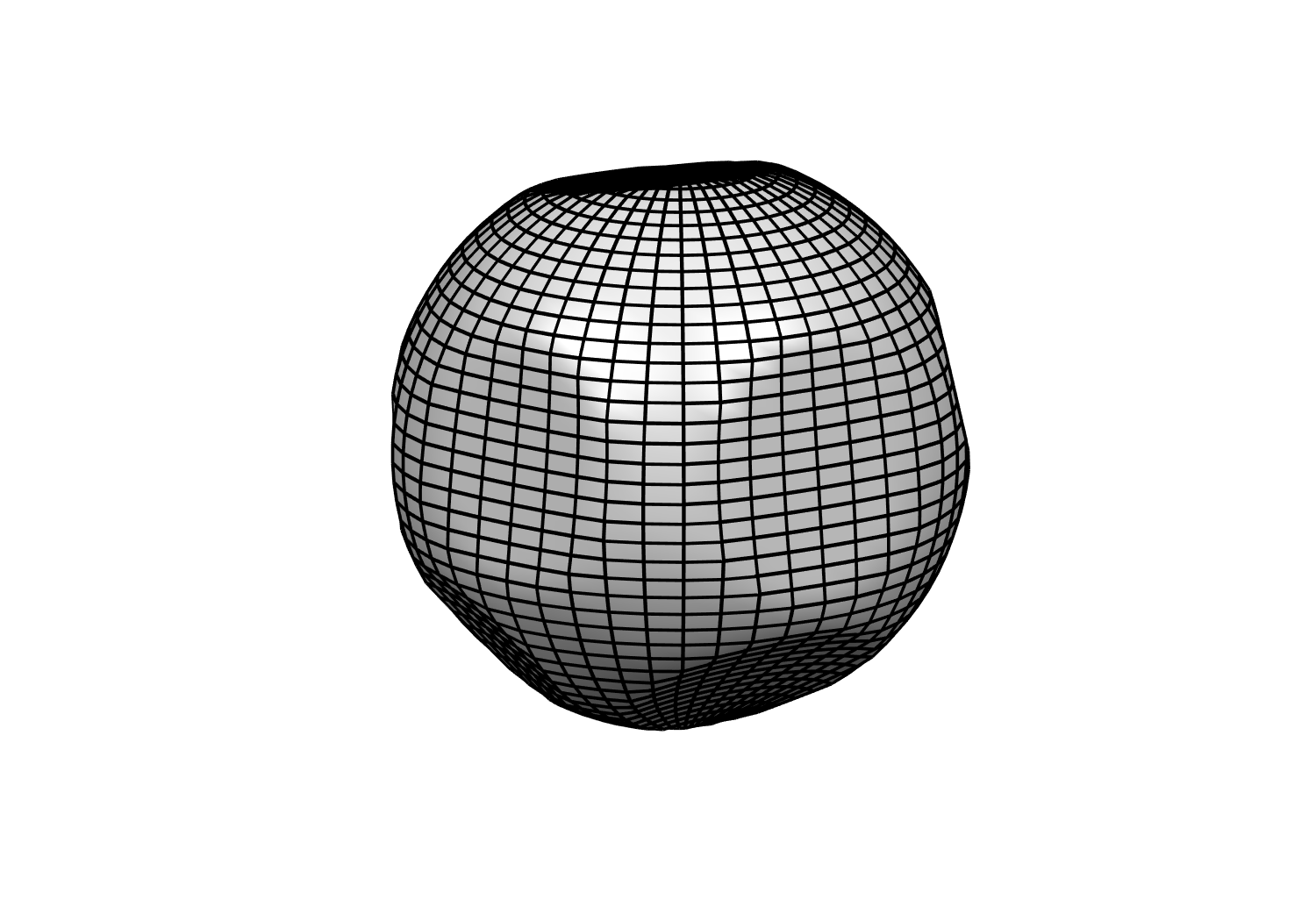}
        &
        \includegraphics[scale=0.57, trim=100 40 85 40, clip]{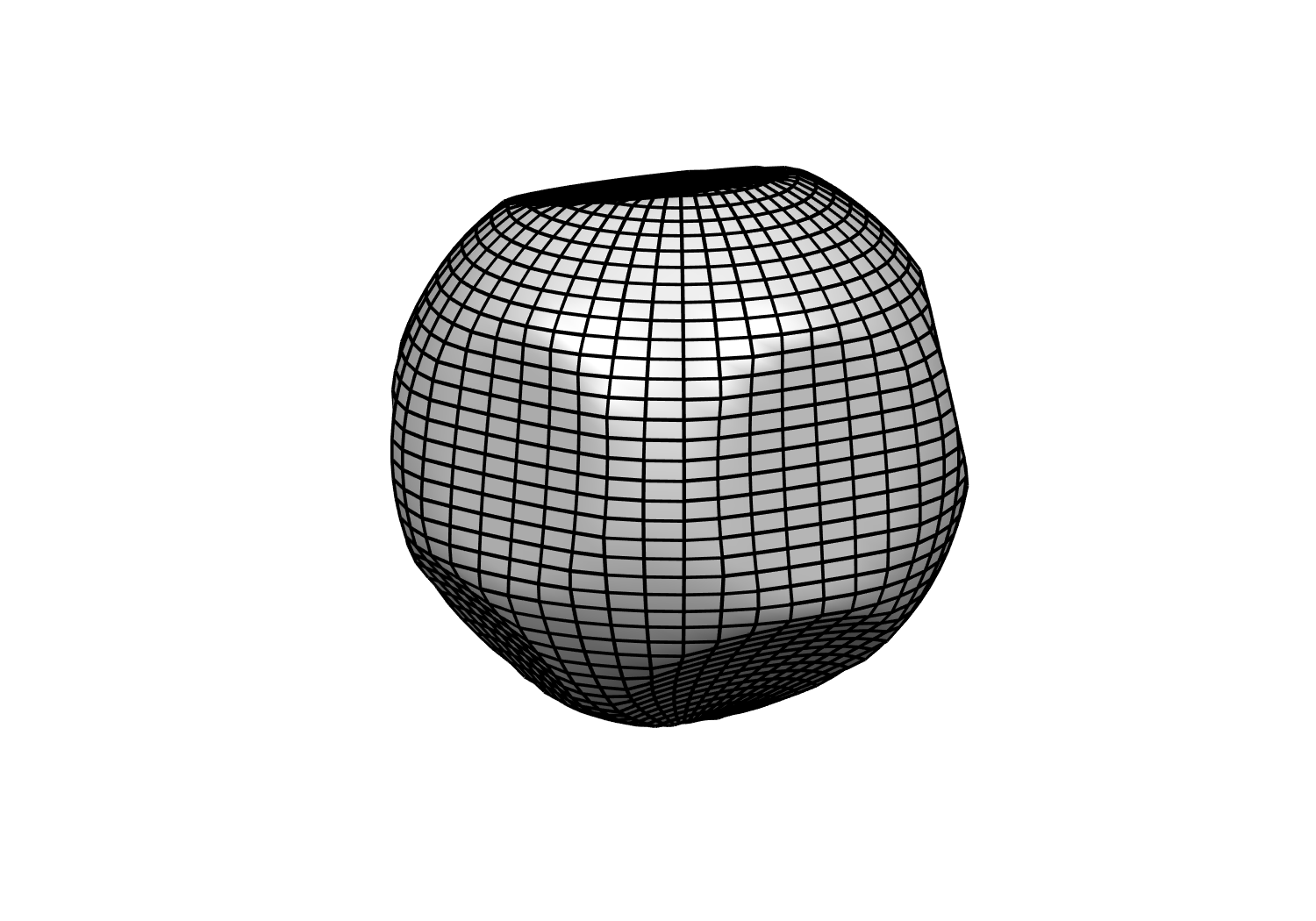}
        &
        \includegraphics[scale=0.57, trim=100 40 85 40, clip]{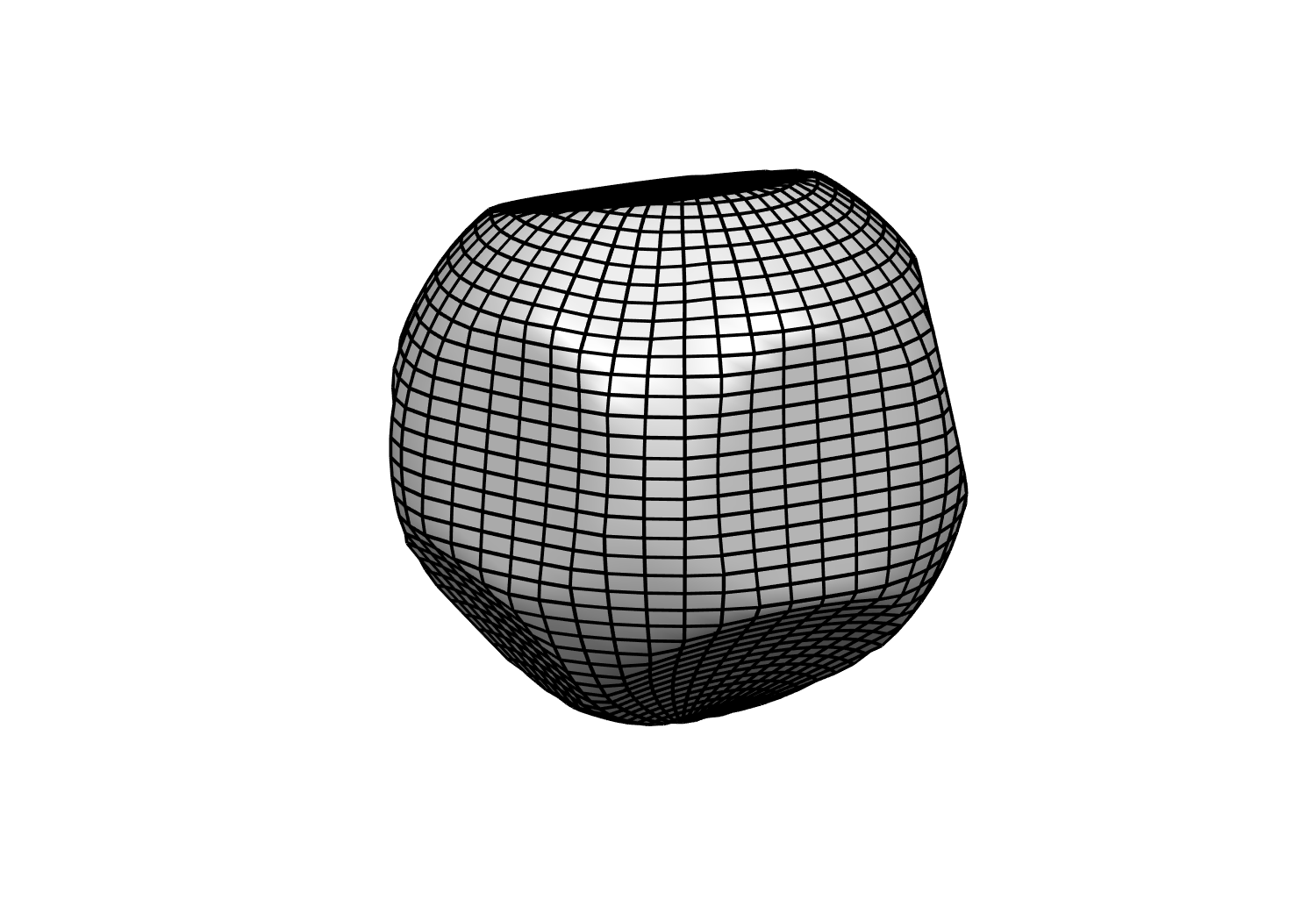}
        \\
        \small{(e) $\rho^{\mbox{\tiny in-die}}_{\mbox{\tiny max}} = 0.8437$}
        &
        \small{(f) $\rho^{\mbox{\tiny in-die}}_{\mbox{\tiny max}} = 0.8770$}
        &
        \small{(g) $\rho^{\mbox{\tiny in-die}}_{\mbox{\tiny max}} = 0.9131$}
        &
        \small{(h) $\rho^{\mbox{\tiny in-die}}_{\mbox{\tiny max}} = 0.9523$}
    \end{tabular}
    \caption{Deformed configuration of a single particle located inside the powder bed of Material 1 at eight different levels of confinement $\rho^{\mbox{\tiny in-die}}_{\mbox{\tiny max}}=\{  0.7323, 0.7573,  0.7841, 0.8128,$ $0.8437, 0.8770,  0.9131, 0.9523  \}$.}
    \label{Fig-SingleParticle3D}
\end{figure}

We specifically study a noncohesive frictionless \cite{Mahmoodi-2010} granular system comprised by 6,512 weightless spherical particles with radius $R = 220 \mu$m, and two sets of material properties, namely (i) {\it Material 1} with Young's modulus $E=5$~GPa, and Poisson's ratio $\nu=0.25$, plastic stiffness $\kappa=150$~MPa, plastic law exponent $m=2.00$, and fracture toughness $K_{Ic}=1.26$~MPa~m$^{1/2}$, and (ii) {\it Material 2} with Young's modulus $E=30$~GPa, and Poisson's ratio $\nu=0.25$, plastic stiffness $\kappa=900$~MPa, plastic law exponent $m=2.00$, and fracture toughness $K_{Ic}=6.19$~MPa~m$^{1/2}$ (see Table~\ref{Table-MaterialProperties}). These materials properties do not correspond to any material in particular but rather represent lower and upper bounds for many pharmaceutical powders, including drugs and excipients (see, e.g., \cite{Mahmoodi-2013,Panelli-2001} and references therein). The granular bed, which is numerically generated by means of a ballistic deposition technique \cite{Jullien-1989}, is constrained by a rigid cylindrical die of diameter $D=10$~mm. Assuming a sufficiently small compaction speed, we consider rate-independent material behavior and we neglect traveling waves, or any other dynamic effect \cite{Gonzalez-2016}. The deformation process is therefore described by a sequence of static equilibrium configurations using the particle mechanics approach presented in Section~\ref{Section-ParticleMechanicsAlgorithm}. In this work we employ 115 quasi-static load steps and we consider 12 unloading points, namely $\rho^{\mbox{\tiny in-die}}_{\mbox{\tiny max}}=\{ 0.6663, 0.6869, 07089, 0.7323, 0.7573,  0.7841, 0.8128, 0.8437,  0.8770,$ $0.9131, 0.9523, 0.9950   \}$. Figure~\ref{Fig-FullSystem} shows the compacted granular bed at $\rho^{\mbox{\tiny in-die}}_{\mbox{\tiny max}} = 0.7323$ and at $\rho^{\mbox{\tiny in-die}}_{\mbox{\tiny max}} = 0.9523$. Figure~\ref{Fig-SingleParticle3D} shows the evolution of deformations of a single particle located inside the powder bed, where the particle deformed configuration is estimated from neighboring particles' displacements and contact radii. The similitude with the experimentally observed shape of die-compacted spherical granules formed from microcrystalline cellulose \cite {Nordstrom-2013} is striking.

We investigate jamming transition, evolution of the mean mechanical coordination number (number of non-zero contact forces between a particle and its neighbors) in Section~\ref{Section-MeanCoord}, punch force and die-wall reaction during compaction and unloading in Section~\ref{Section-PunchForce}, in-die elastic recovery during unloading in Section~\ref{Section-ElasticRecovery}, and residual radial pressure after unloading and ejection pressure in Section~\ref{Section-ResidualEjection}. We also investigate microstructure evolution by studying probability density function of contact forces as well as the anisotropic granular fabric after compaction, unloading and ejection in Section~\ref{Section-NetworkCFAnisotroy}. Finally, we evaluate the evolution of bonding surface area during all stages of die compaction in Section~\ref{Section-BondingArea} and we estimate the Young's modulus and Poisson's ratio of the compacted solid in Section~\ref{Section-YoungPoisson}. We close by depicting a microstructure-mediated process-structure-property-performance interrelationship of the compaction process in Section~\ref{Section-PSPP}.
\subsection{Mean coordination number}
\label{Section-MeanCoord}

\begin{figure}[htbp]
    \centering
    \begin{tabular}{cc}
        \includegraphics[scale=0.64, trim=0 0 30 0, clip]{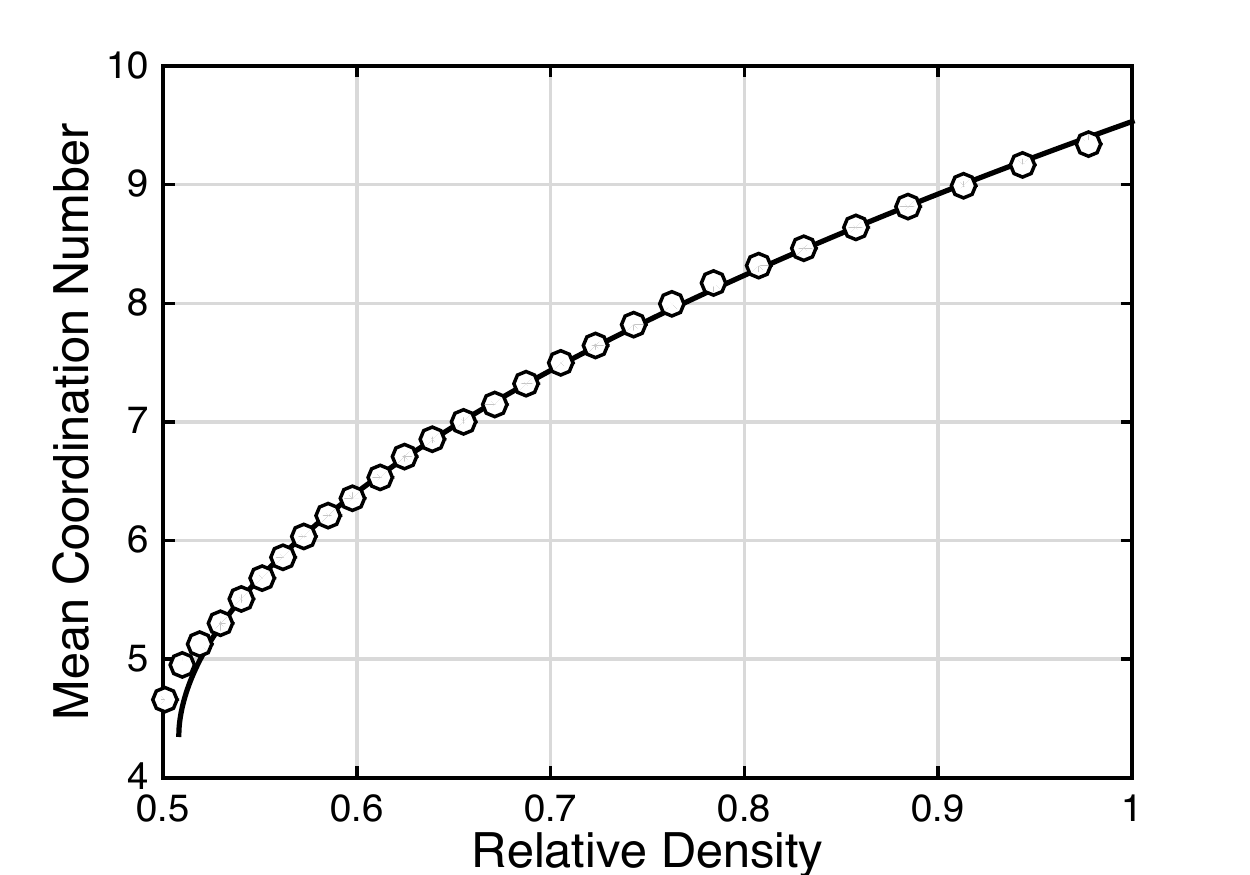}
        &
        \includegraphics[scale=0.64, trim=0 0 30 0, clip]{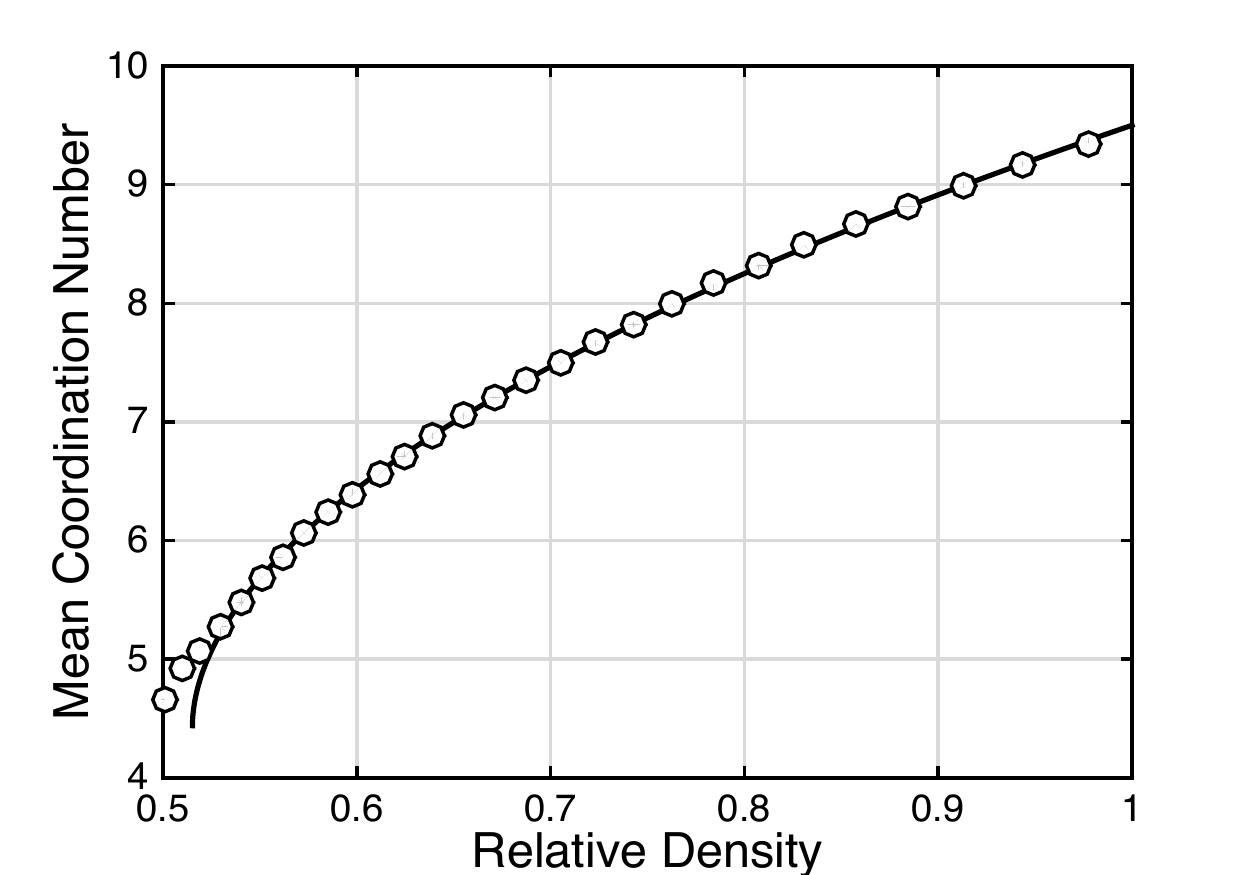}
    \\
    \small{(a) Material 1}
    &
    \small{(b) Material 2}        
    \end{tabular}
    \caption{Mean coordination number as a function of relative density $\rho^{\mbox{\tiny in-die}}_{\mbox{\tiny max}}$. Solid line corresponds to the best fit of equation (\ref{Eqn-CoordinationFit}) to the mean coordination number obtained from the particle contact mechanics simulation of the granular bed (symbols).}
    \label{Fig-CN}
\end{figure}

The mean coordination number $\bar{Z}$ evolves as a power law of the following form
\begin{equation}
    \bar{Z} - \bar{Z}_c
    =
    \bar{Z}_0 (\rho^{\mbox{\tiny in-die}}_{\mbox{\tiny max}}-\rho_{c,\bar{Z}})^{\theta}
    \label{Eqn-CoordinationFit}
\end{equation}
where $\rho_{c,\bar{Z}}$ is the critical relative density, $\bar{Z}_c$ is the minimal average coordination number and $\theta$ is the critical exponent. This well-known critical-like behavior has an exponent consistent with $1/2$ for different pair-interaction contact laws, polydispercity and dimensionality of the problem \cite{OHern-2002,OHern-2003,Durian-1995}. It is known, however, that this power law is a first order approximation to the behavior of a deformable material for which, as demonstrated by Gonzalez et al. \cite{Gonzalez-2016} for elastic materials, the coordination number depends on the level of compressibility, i.e., on Poisson's
ratio, of the particles and thus its scaling behavior is not independent of material properties as previously thought. This more realistic behavior is predicted by nonlocal contact formulations \cite{Gonzalez-2012, Gonzalez-2016} and it will not be the focus of this paper, as it was previously stated. Figure~\ref{Fig-CN} shows the results obtained from the particle contact mechanics simulations and their best fit to equation \eqref{Eqn-CoordinationFit}. Jamming occurs at $\bar{Z}_c=4.366$ and  $\rho_{c,\bar{Z}}=0.5081$ with $\theta=0.5535$ for Material 1, and at $\bar{Z}_c=4.439$ and  $\rho_{c,\bar{Z}}=0.5151$ with $\theta=0.5333$ for Material 2. The fit to numerical results is good not only near jamming but also at large relative densities. It is worth noting that the isostatic condition for frictionless packings implies a critical coordination number equal to 6 and a critical density close to 0.64. In addition, however, there exists a body of work that indicates that $\rho_{c,\bar{Z}}$ depends on the protocol used for obtaining jammed configurations and on the particle-die size ratio, and that monodisperse systems are prone to crystallization \cite{Baranau-2104, Chaudhuri-2010, Schreck-2011, Vagberg-2011}. Here we restrict our discussion to post-jamming behavior and to one preparation protocol. It is also interesting to note that the jamming transition occurs later for $K_{Ic}=0$---cf.  \cite{Gonzalez-2018}, that is $\rho_{c,\bar{Z}}=0.57$ with  $\bar{Z}_c\approx 5$, for the same preparation protocol.

\subsection{Punch force and die-wall reaction}
\label{Section-PunchForce}

\begin{figure}[htbp]
    \centering
    \begin{tabular}{cc}
        \includegraphics[scale=0.64, trim=0 0 30 0, clip]{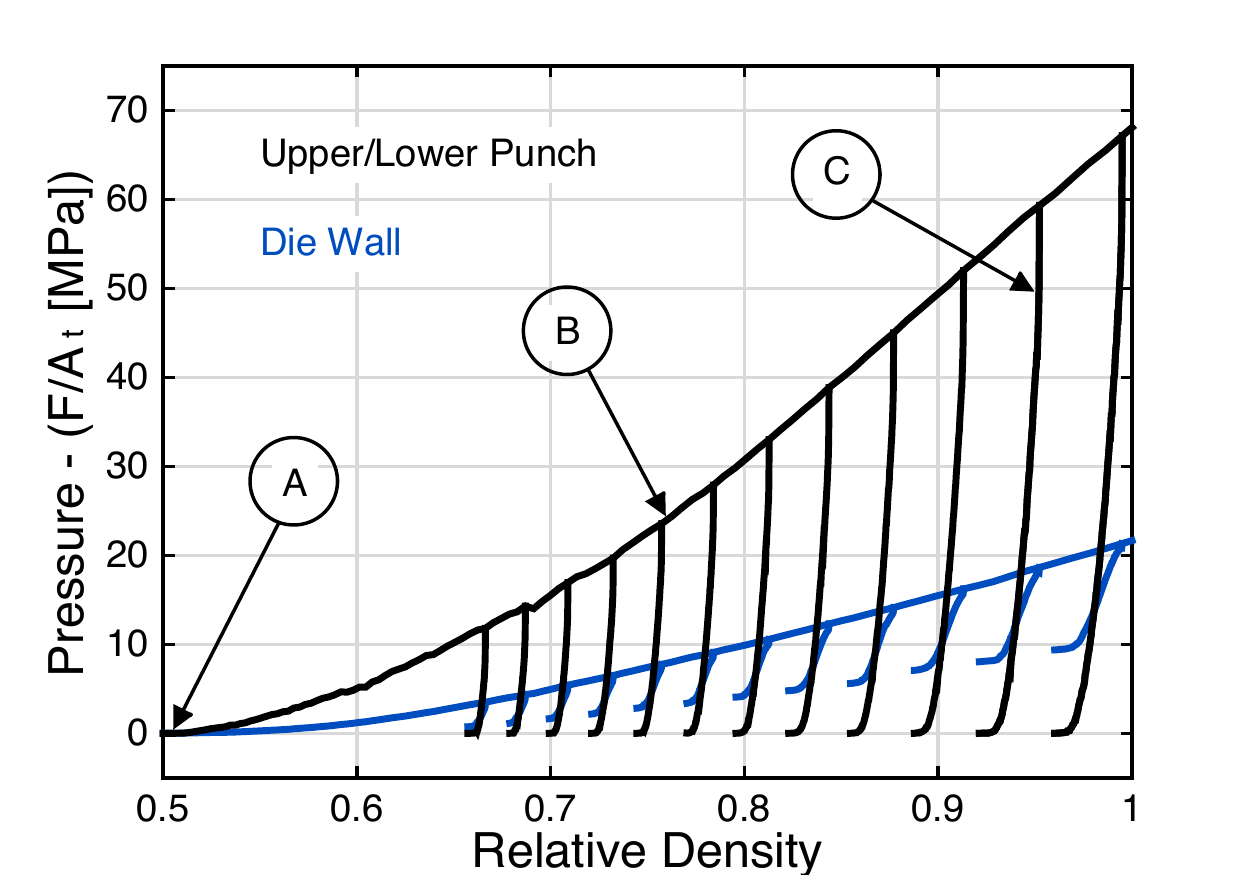}
        &
        \includegraphics[scale=0.64, trim=0 0 30 0, clip]{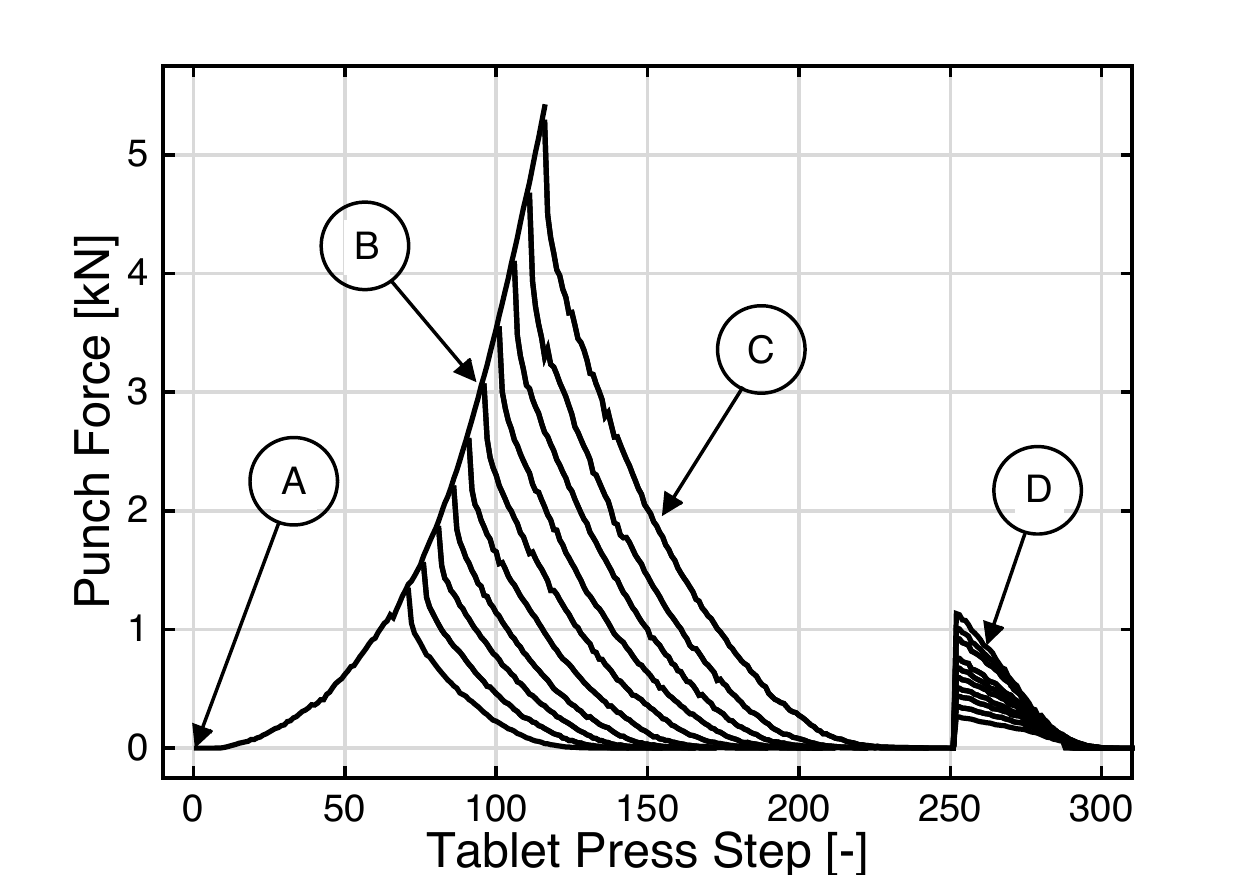}
    \\
    \small{(a) Material 1}
    &
    \small{(b) Material 1}        
    \\
        \includegraphics[scale=0.64, trim=0 0 30 0, clip]{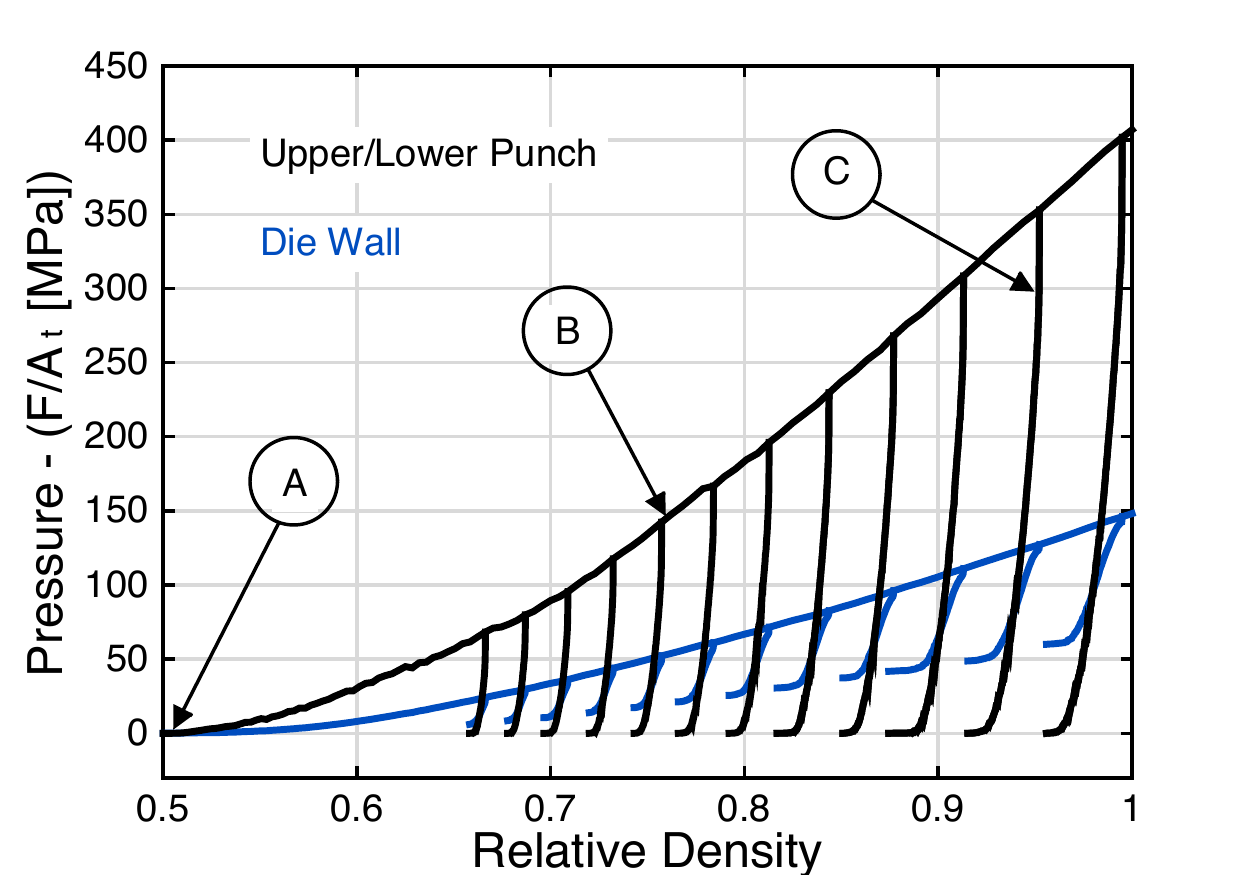}
        &
        \includegraphics[scale=0.64, trim=0 0 30 0, clip]{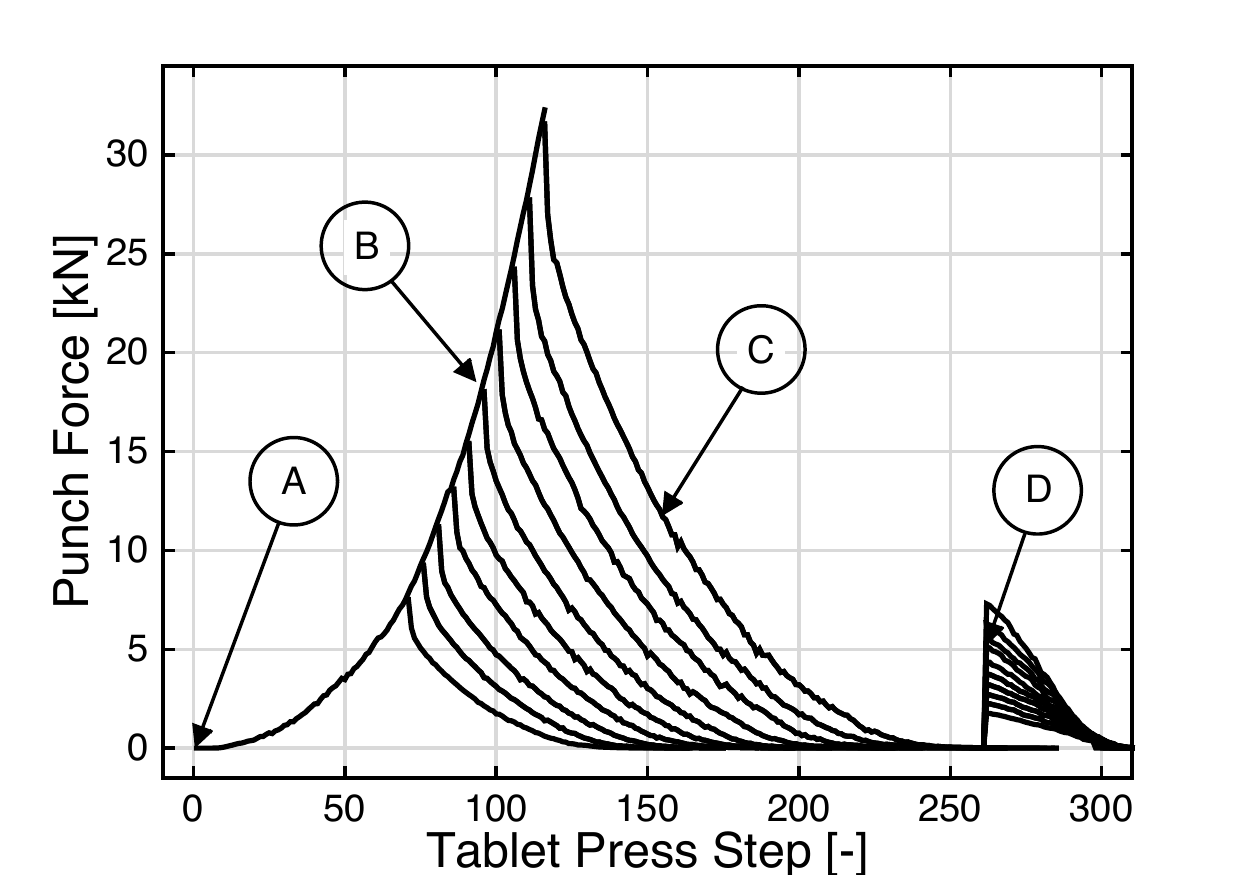}
    \\
    \small{(c) Material 2}
    &
    \small{(d) Material 2}        
    \end{tabular}
    \caption{Punch and die-wall pressures as a function of relative density $\rho^{\mbox{\tiny in-die}}$ (a)\&(c) and  punch force (b)\&(d) as a function of tablet press step during the powder die-compaction process---labels correspond to Fig.~\ref{Fig-CompactionSchematics}.}
    \label{Fig-AppliedPressure}
\end{figure}

The pressures applied by the punches and the reaction at the die wall are macroscopic variables relevant to powder die-compaction that are effectively predicted by the particle contact mechanics simulation. These predictions are presented in Figure~\ref{Fig-AppliedPressure}a for solid compacts compressed at 12 different relative densities. The compaction process is dominated by plastic deformations and formation of solid bridges, while the unloading stage is characterized by elastic recovery and breakage of bonded surfaces. The numerical simulation accounts for these different physical mechanisms and it successfully predicts a residual radial stress after unloading. If there is friction between the solid compact and the die wall during the ejection stage, the residual radial stress will lead to an ejection force. Figure~\ref{Fig-AppliedPressure}b shows the evolution of the punch force during compaction, unloading and ejection (assuming, for simplicity, a friction coefficient of 1). We also note that the compaction pressure follows a power law of the following form
\begin{equation}
     \sigma_{\mbox{\tiny punch}} 
     = 
     K_{\mbox{\tiny P}}  	(\rho^{\mbox{\tiny in-die}}_{\mbox{\tiny max}} - \rho_{c,\bar{Z}})^{\beta_{\mbox{\tiny P}} }
     \label{Eqn-AppliedPressure}	
\end{equation}
where $\rho_{c,\bar{Z}}$ is obtained from the evolution of $\bar{Z}$, and the coefficients $K_{\mbox{\tiny P}} = 210$~MPa and $\beta_{\mbox{\tiny P}} =1.561$ are best-fitted to the numerical results for Material 1---$K_{\mbox{\tiny P}} = 1.265$~GPa and $\beta_{\mbox{\tiny P}} = 1.541$ for Material 2. It is interesting to note that a factor of two in $\kappa$ translates into a factor of two in $K_{\mbox{\tiny P}}$, as noted in \cite{Gonzalez-2018} for $K_{Ic}=0$. As mentioned above, the compaction curves shown in Figure \ref{Fig-AppliedPressure} represent lower and upper bounds for many pharmaceutical powders, including drugs and excipients \cite{Mahmoodi-2013,Panelli-2001}---e.g., ammonium chloride's compaction curve is similar to Material 1, and lactose monohydrate's  compaction curve to Material 2 \cite{Razavi-2018}.

\subsection{In-die elastic recovery}
\label{Section-ElasticRecovery}

\begin{figure}[htbp]
    \centering
    \begin{tabular}{cc}
        \includegraphics[scale=0.64, trim=0 0 30 0, clip]{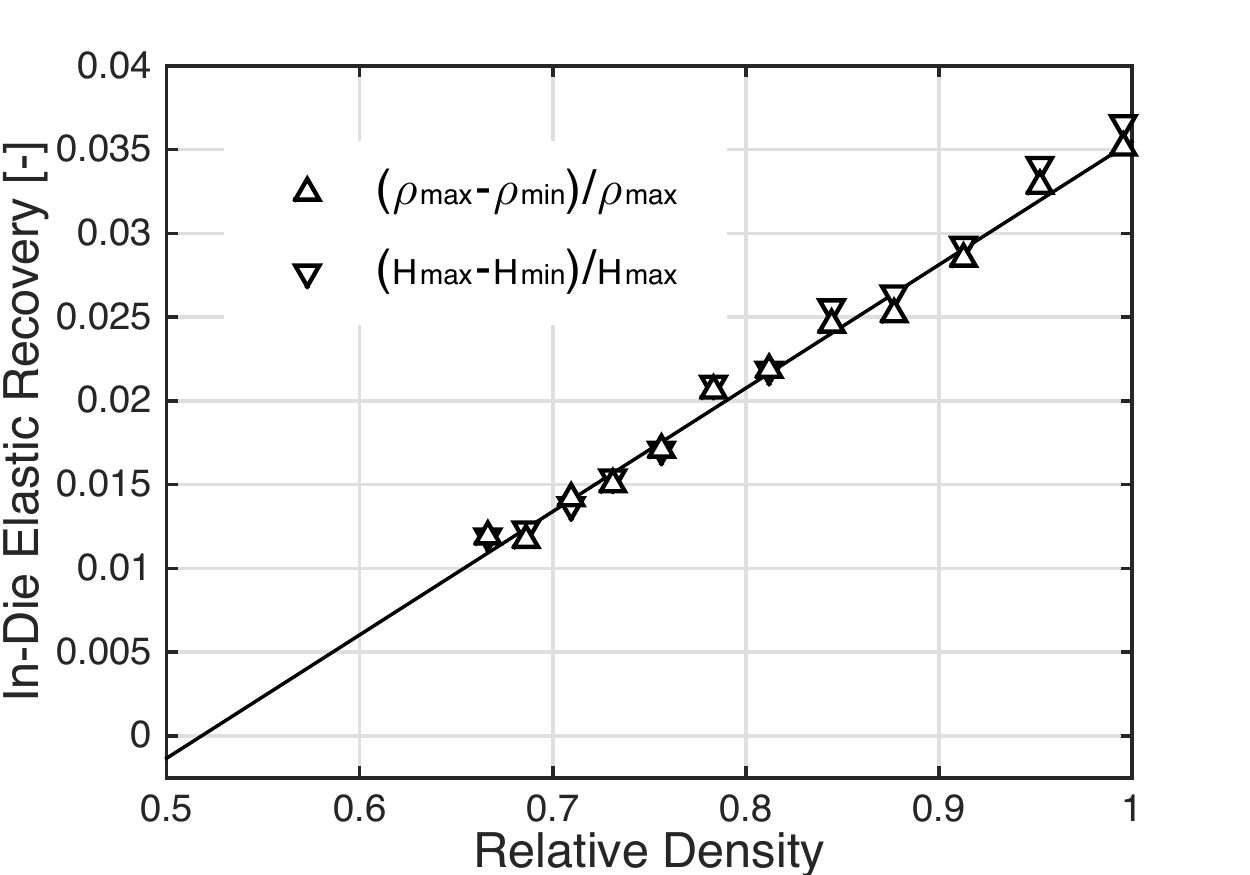}
        &
        \includegraphics[scale=0.64, trim=0 0 30 0, clip]{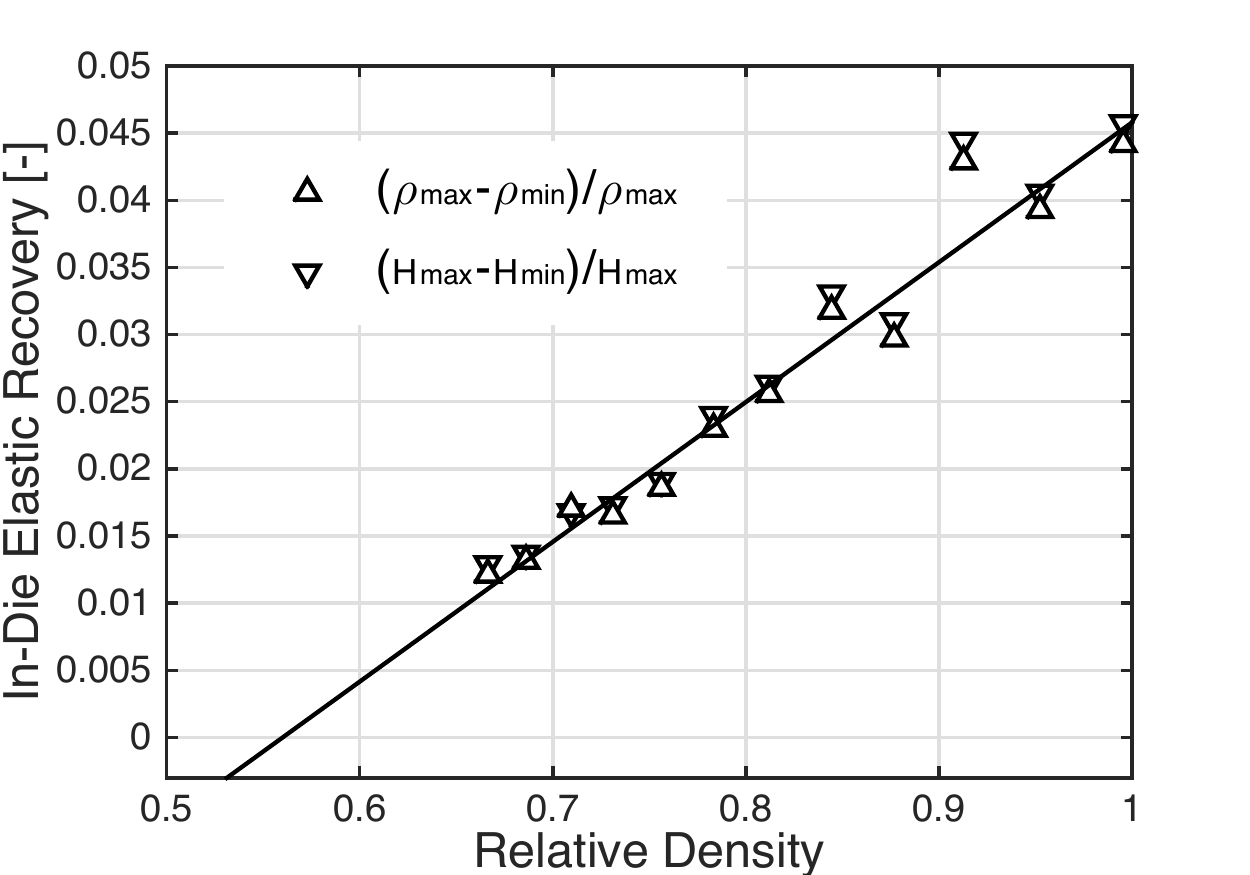}
    \\
    \small{(a) Material 1}
    &
    \small{(b) Material 2}        
    \end{tabular}
    \caption{In-die elastic recovery as a function of relative density $\rho^{\mbox{\tiny in-die}}_{\mbox{\tiny max}}$. Solid line corresponds to the best fit of equation \eqref{Eqn-ElasticRecovery} to the in-die elastic recovery in terms of relative density obtained from the particle contact mechanics simulation of the granular bed (symbols).}
    \label{Fig-ElasticRecovery}
\end{figure}

The in-die elastic recovery is further investigated and two alternative definitions found in the literature are proposed, namely the elastic recovery in terms of relative density $\epsilon_{\rho}$ and in terms of tablet height $\epsilon_{H}$, that is
\begin{equation}
	\epsilon_{\rho} 
	= 
	\frac{\rho_{\mbox{\tiny max}}^{\mbox{\tiny in-die}}-\rho^{\mbox{\tiny in-die}}_{\mbox{\tiny min}}}{\rho^{\mbox{\tiny in-die}}_{\mbox{\tiny max}}} 
	\mbox{~~~~~,~~~~~}
	\epsilon_{H} 
	= 
	\frac{H^{\mbox{\tiny in-die}}_{\mbox{\tiny min}} - H^{\mbox{\tiny in-die}}_{\mbox{\tiny max}}}{H^{\mbox{\tiny in-die}}_{\mbox{\tiny max}}}
\label{Eqn-DefinitionElasticRecovery}
\end{equation}
Naturally, these two definitions are interrelated and, adopting a linear relationship between $\epsilon_{\rho}$ and relative density, one obtains they are the same to fist order in $\epsilon_{\rho}$. Specifically, the following relationships hold
\begin{eqnarray}
	\epsilon_{\rho} 
	&=& 
	\epsilon_{0} \frac{\rho^{\mbox{\tiny in-die}}_{\mbox{\tiny max}} - \rho_{c,\epsilon}}{1-\rho_{c,\epsilon}}
	\label{Eqn-ElasticRecovery}
	\\
	\epsilon_{H} 
	&=& 
	\frac{\epsilon_{0} (\rho^{\mbox{\tiny in-die}}_{\mbox{\tiny max}} - \rho_{c,\epsilon})}{1- \rho_{c,\epsilon}-\epsilon_{0} (\rho^{\mbox{\tiny in-die}}_{\mbox{\tiny max}} - \rho_{c,\epsilon})}
	= 
	\epsilon_{\rho} + \mathcal{O}(\epsilon_{\rho}^2)
\end{eqnarray}
where the in-die elastic recovery at full compaction $\epsilon_{0}=3.550\%$ and the critical relative density $\rho_{c,\epsilon}=0.5180$ are best-fitted to the numerical results for Material 1---$\epsilon_{0}=4.579\%$ and $\rho_{c,\epsilon}=0.5602$ for Material 2. Figure~\ref{Fig-ElasticRecovery} shows the results obtained from the particle mechanics simulations and their fit to the above equations. It is worth noting that these values are in the lower range of many pharmaceutical excipients \cite{Haware-2010,Yohannes-2015}, which highlights the ability of the proposed model to decouple the loading and unloading parts of the compaction curve by properly choose material properties.

\subsection{Residual radial pressure and ejection pressure}
\label{Section-ResidualEjection}

\begin{figure}[htbp]
    \centering
    \begin{tabular}{cc}
        \includegraphics[scale=0.64, trim=5 0 3 0, clip]{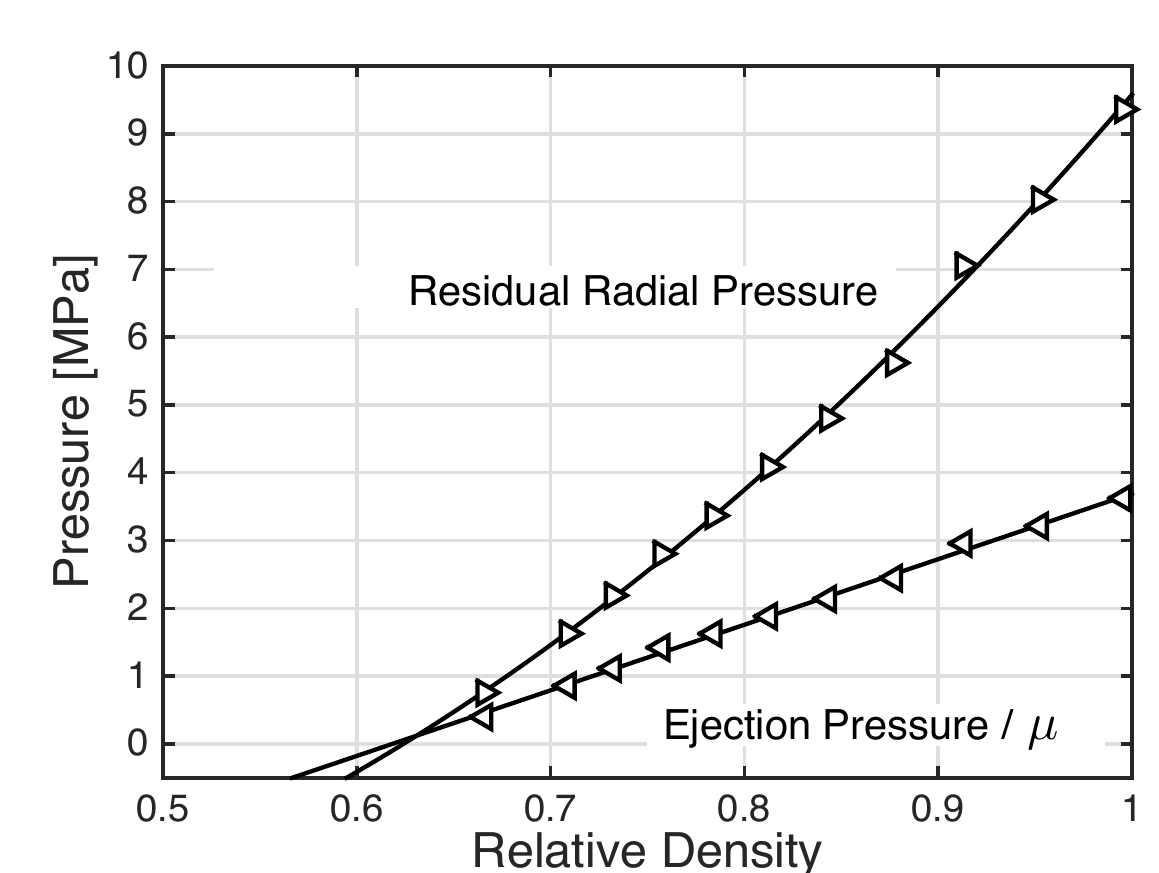}
        &
        \includegraphics[scale=0.64, trim=5 0 3 0, clip]{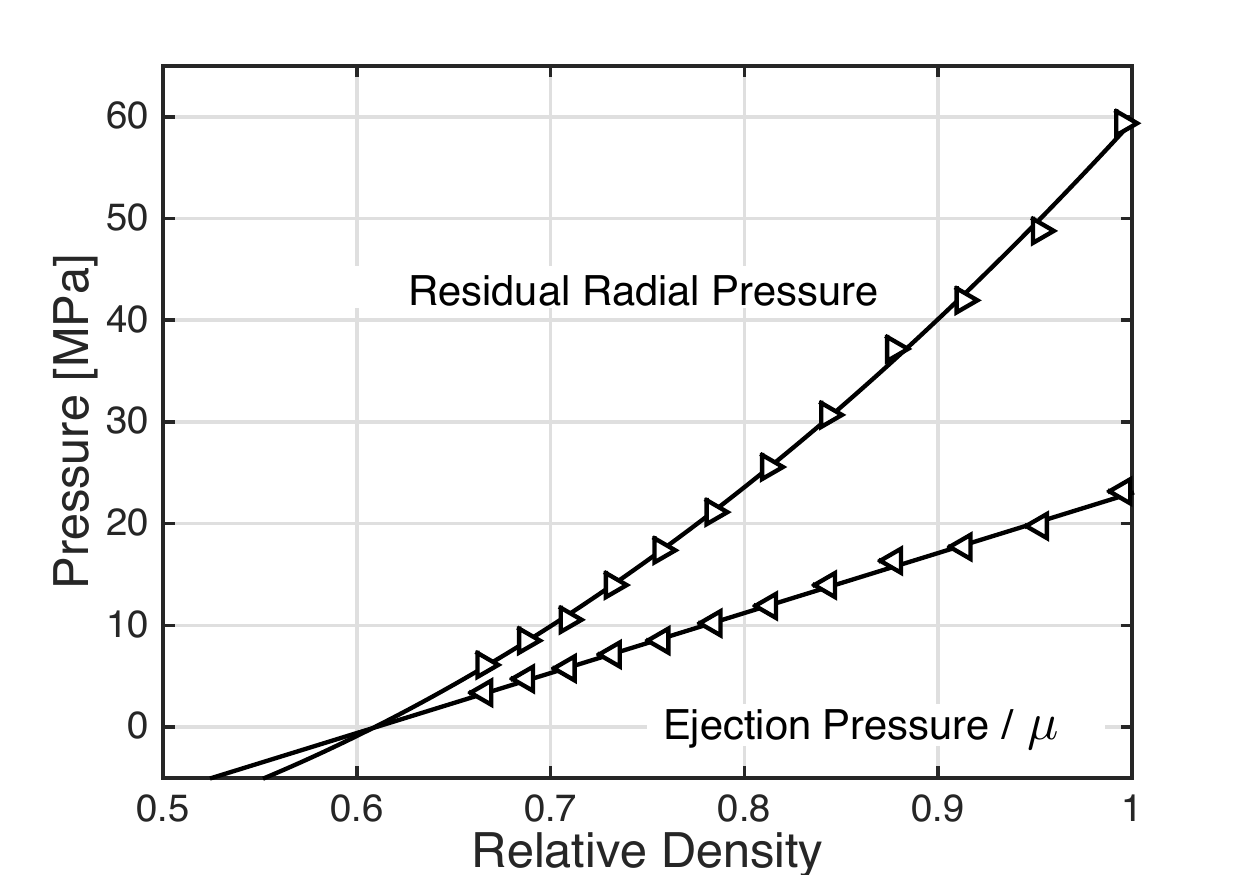}
    \\
    \small{(a) Material 1} 
    &
    \small{(b) Material 2}        
    \end{tabular}
    \caption{Residual radial pressure and ejection pressure as a function of relative density $\rho^{\mbox{\tiny in-die}}_{\mbox{\tiny max}}$. Solid lines correspond to the best fit of equations \eqref{Eqn-ResidualPressure}-\eqref{Eqn-EjectionPressure} to the residual radial pressure and ejection pressure obtained from the particle contact mechanics simulation of the granular bed (symbols).}
    \label{Fig-EjectionPressure}
\end{figure}

The residual radial pressure and ejection pressure are further investigated and the following relations are proposed
\begin{eqnarray}
	\sigma_{\mbox{\tiny residual}}
	\label{Eqn-ResidualPressure}
	&=&  
	\sigma_{\mbox{\tiny res},0} ~ \dfrac{\rho^{\mbox{\tiny in-die}}_{\mbox{\tiny max}}  (\rho^{\mbox{\tiny in-die}}_{\mbox{\tiny max}} - \rho_{c,e})}{1-\rho_{c,e}}
	\\
	\sigma_{\mbox{\tiny ejection}} 
	&=& 
	\mu ~ \dfrac{\sigma_{\mbox{\tiny res},0}~16 W}{\rho_t \pi D^3 } ~ \dfrac{\rho^{\mbox{\tiny in-die}}_{\mbox{\tiny max}} - \rho_{c,e}}{1-\rho_{c,e}}
	\label{Eqn-EjectionPressure}
\end{eqnarray}
where the residual radial radial pressure at full compaction $\sigma_{\mbox{\tiny res},0}=9.719$~MPa and the critical relative density $\rho_{c,e}=0.6196$ are best-fitted to the numerical results for Material 1---$\sigma_{\mbox{\tiny res},0}=59.51$~MPa and $\rho_{c,e}=0.6093$ for Material 2. The two equations presented above are equivalent and obtained by using the relationship between the punch gap and in-die relative density, i.e., $H^{\mbox{\tiny in-die}}_{\mbox{\tiny max}}= 4W/(\pi D^2\rho_t\rho^{\mbox{\tiny in-die}}_{\mbox{\tiny max}})$---where $\rho_t$ is the true density of the material and $W$ is the weight of the powder inside the die with diameter $D$. Figure~\ref{Fig-EjectionPressure} shows the results obtained from the particle mechanics simulations and their fit to the above equations. These values are similar to those observed in many pharmaceutical excipients (see, e.g., \cite{AbdelHamid-2011,Doelker-2004}).

\subsection{Network of contact forces and granular fabric anisotropy}
\label{Section-NetworkCFAnisotroy}

In previous subsections we studied the evolution of macroscopic, effective properties of the compaction process (i.e., punch force and die-wall reaction during compaction and unloading, in-die elastic recovery during unloading, and residual radial pressure after unloading and ejection pressure). Next, we turn attention to the evolution of some of the microstructural features that give rise to such macroscopic behavior. Under increasing confinement, powders support stress by spatial rearrangement and deformation of particles and by the development of inhomogeneous force networks. A force network is typically characterized by the probability distribution of its inter-particle contact forces and their directional orientation.

\begin{figure}[htbp]
    \centering
    \begin{tabular}{ccc}
        \includegraphics[scale=0.56, trim=0 0 12 0, clip]{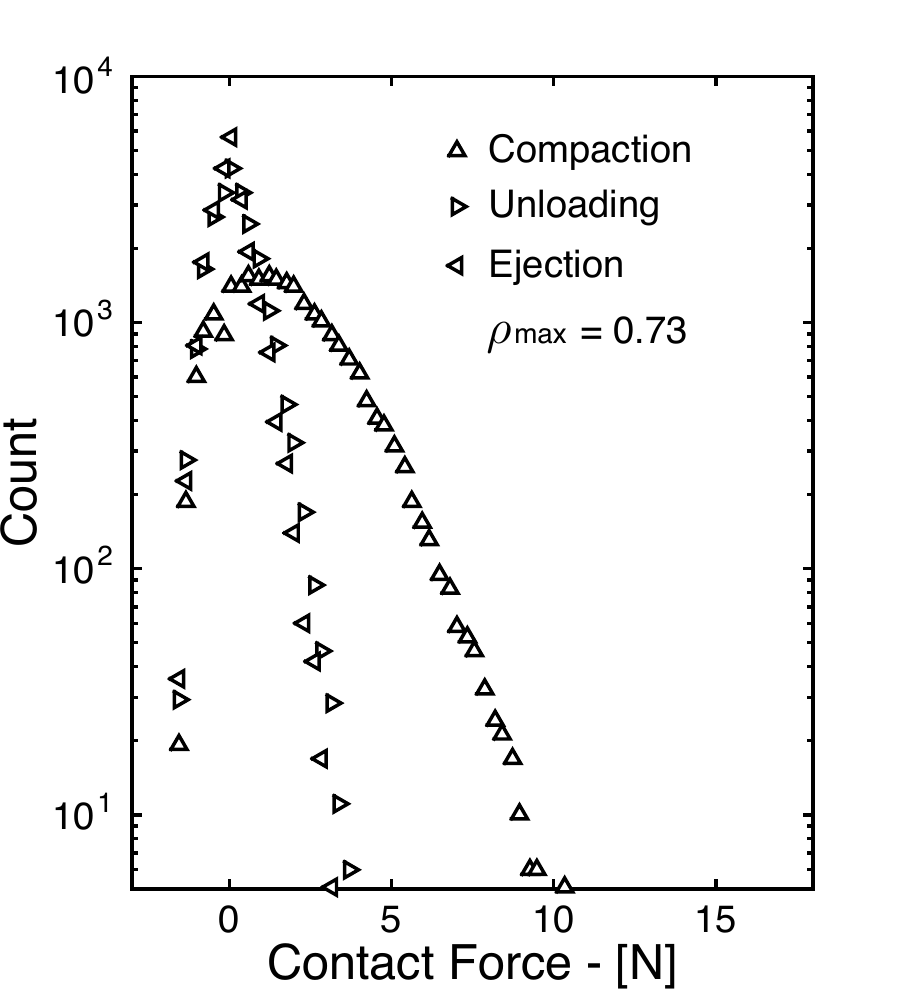}
        &
        \includegraphics[scale=0.56, trim=0 0 12 0, clip]{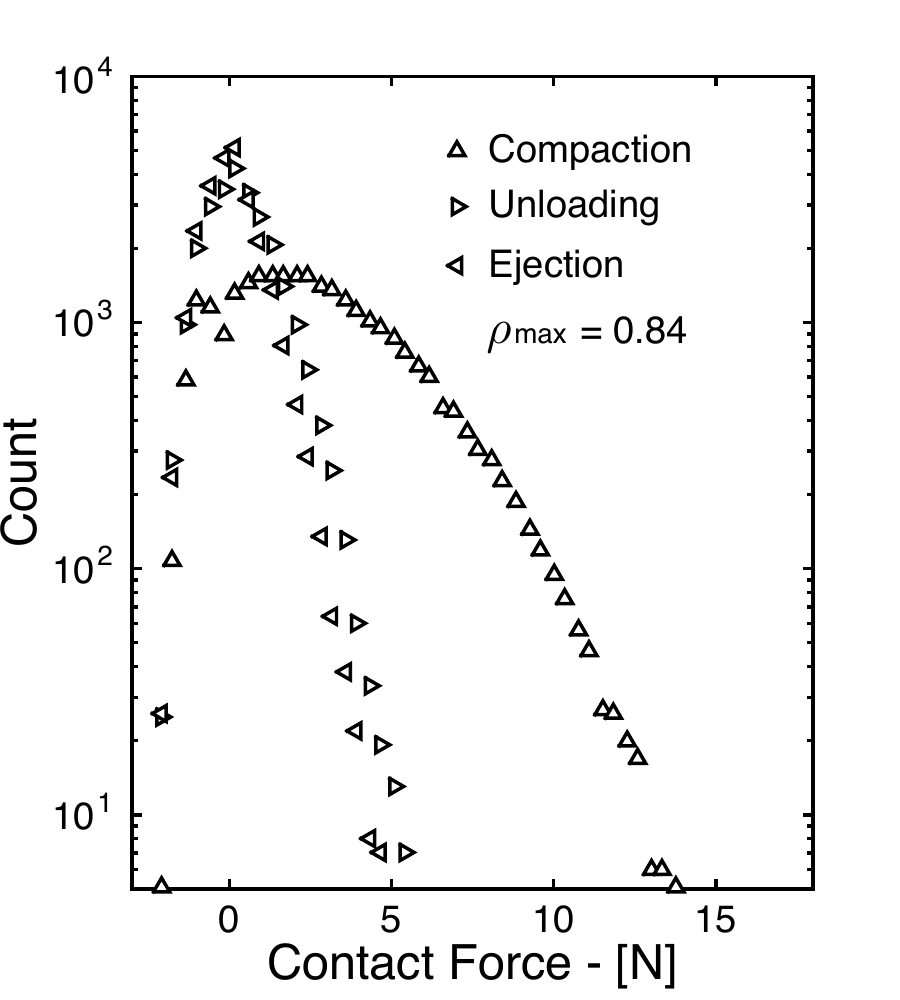}
        &
        \includegraphics[scale=0.56, trim=0 0 12 0, clip]{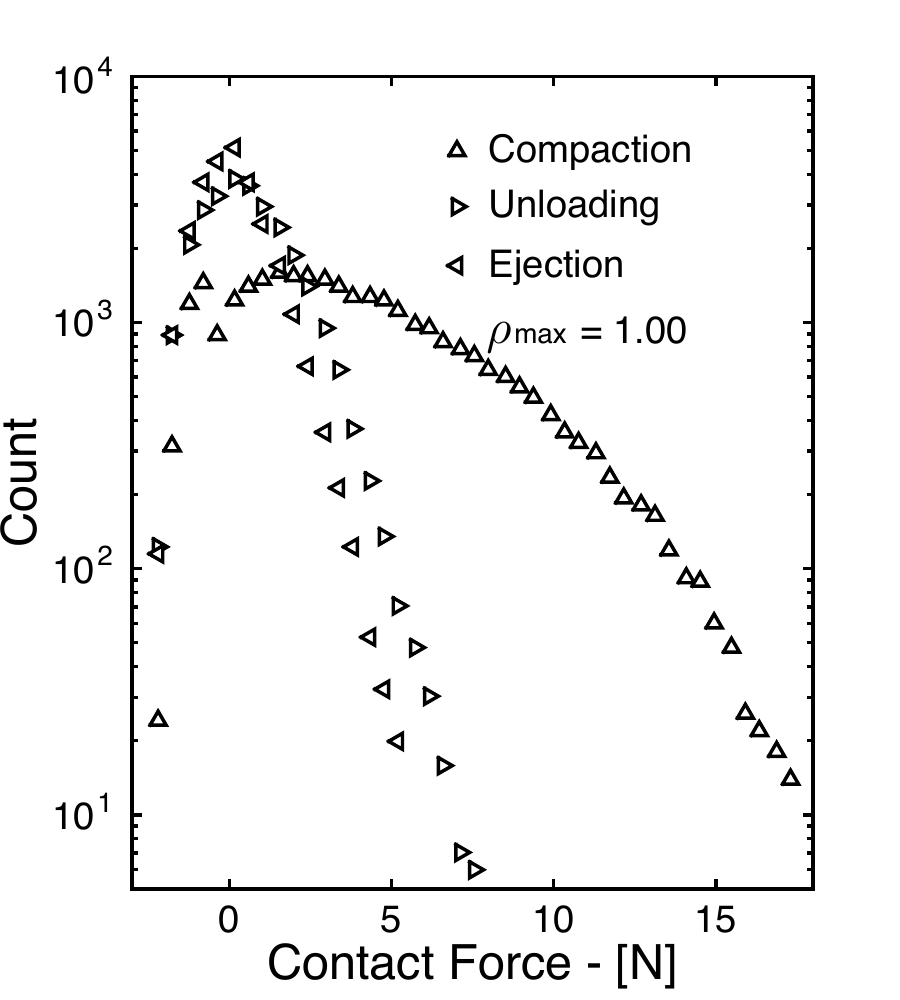}
    \end{tabular}
    \caption{Distribution of contact forces after loading, unloading and ejection at three different levels of compaction  $\rho^{\mbox{\tiny in-die}}_{\mbox{\tiny max}}$ equal to $0.7323$, $0.8437$ and $0.9950$. Broken solid bridges are not included and distributions are obtained from the particle contact mechanics simulation of the granular bed.}
    \label{Fig-ContactForceDistribution}
\end{figure}

For simplicity of exposition, we restrict attention to the behavior of Material 1---results are similar for Material 2. Figure~\ref{Fig-ContactForceDistribution} shows the distribution of contact forces after loading, unloading and ejection at three different levels of compaction. It is worth noting that we show force histograms rather than, as it is typically used in the literature, probability distributions of contact forces non-dimensionalized by their mean value. In turn, it is evident from the figure that: (i) the range of compressive forces increases with relative density more than the range of tensile forces does; (ii) compressive forces reduce significantly in magnitude during unloading, while tensile forces hardly change; and (iii) distributions after unloading and after ejection are very similar and both exhibit some symmetry about zero.

\begin{figure}[htbp]
    \centering
    \begin{tabular}{ccc}
         \includegraphics[scale=0.5, trim=51 13 35 3, clip]{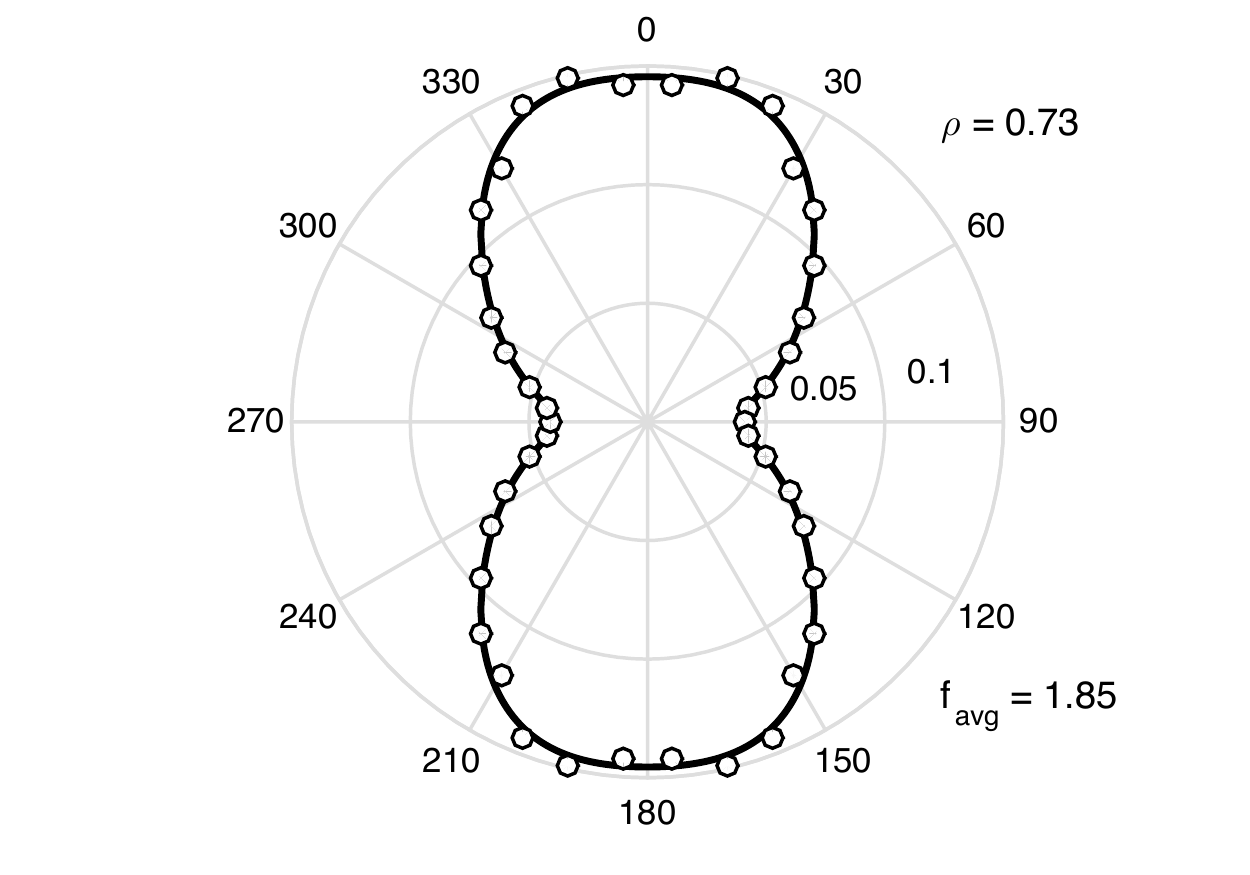}
        &
        \includegraphics[scale=0.5, trim=51 13 35 3, clip]{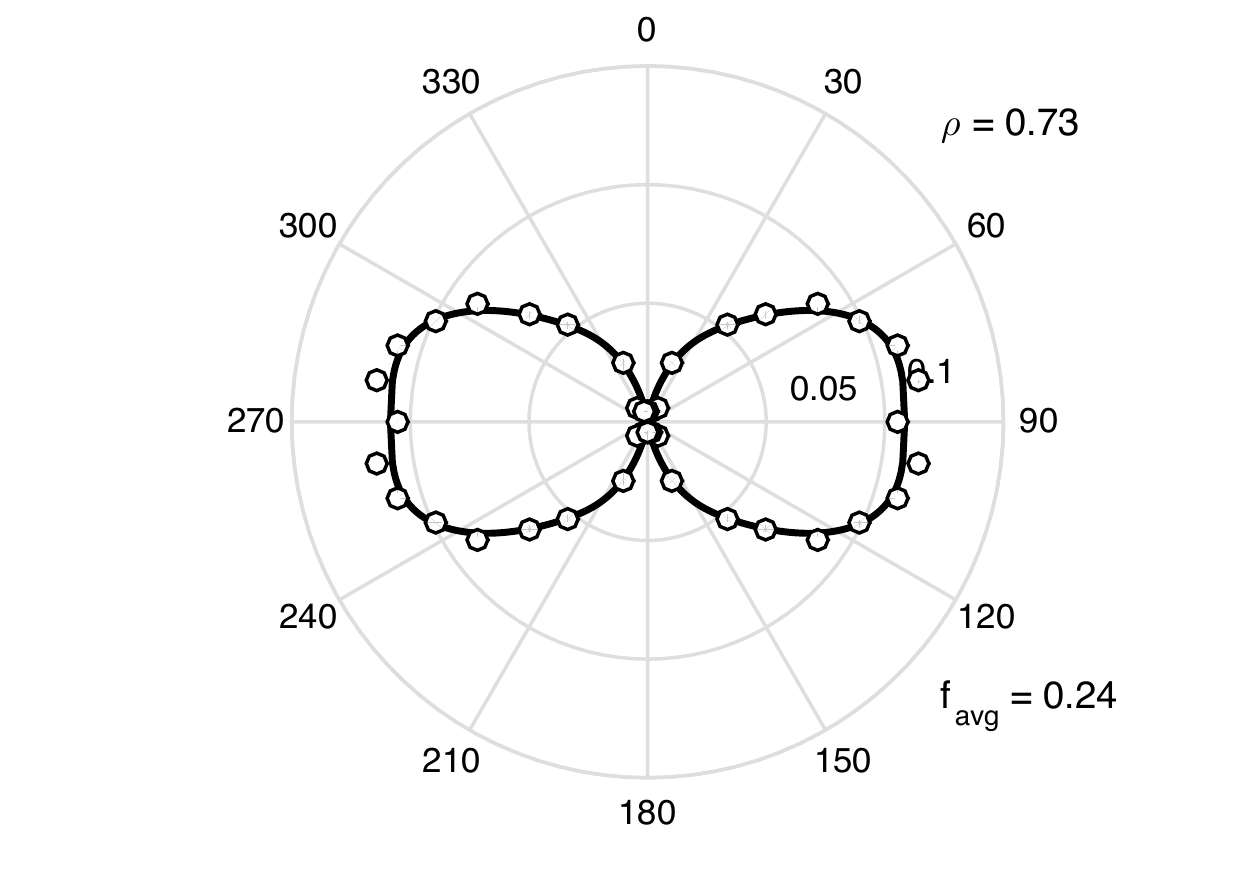}
        &
        \includegraphics[scale=0.5, trim=51 13 35 3, clip]{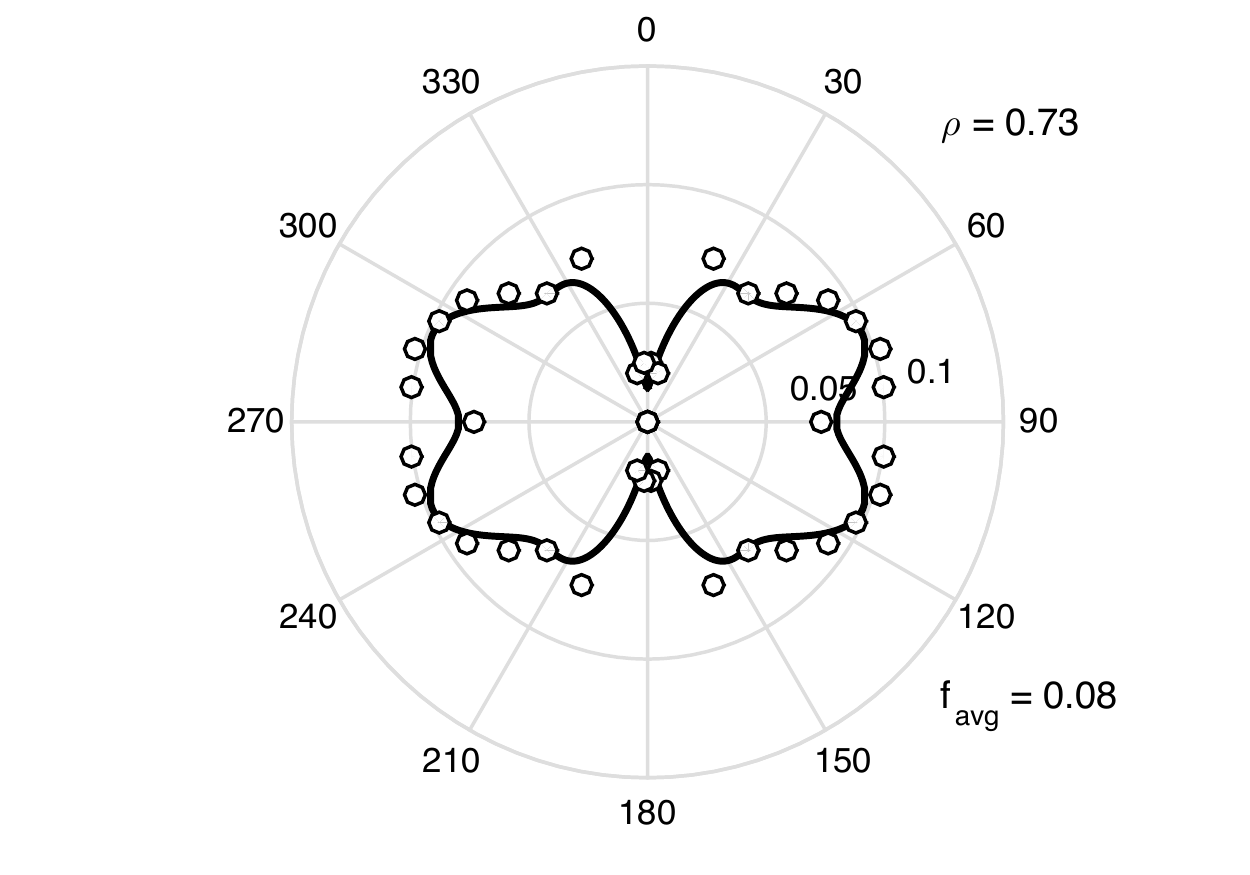}
        \\
	\small{After compaction}
	&
	\small{After unloading}
	&
	\small{After ejection}        
        \\
	\multicolumn{3}{p{1\linewidth}}{\centering \small{(a) Orientation distribution function of the mean contact force ${ f(\theta,\phi) }/{f_{\rm avg}}$.}}
	\\    
        \includegraphics[scale=0.5, trim=51 13 35 3, clip]{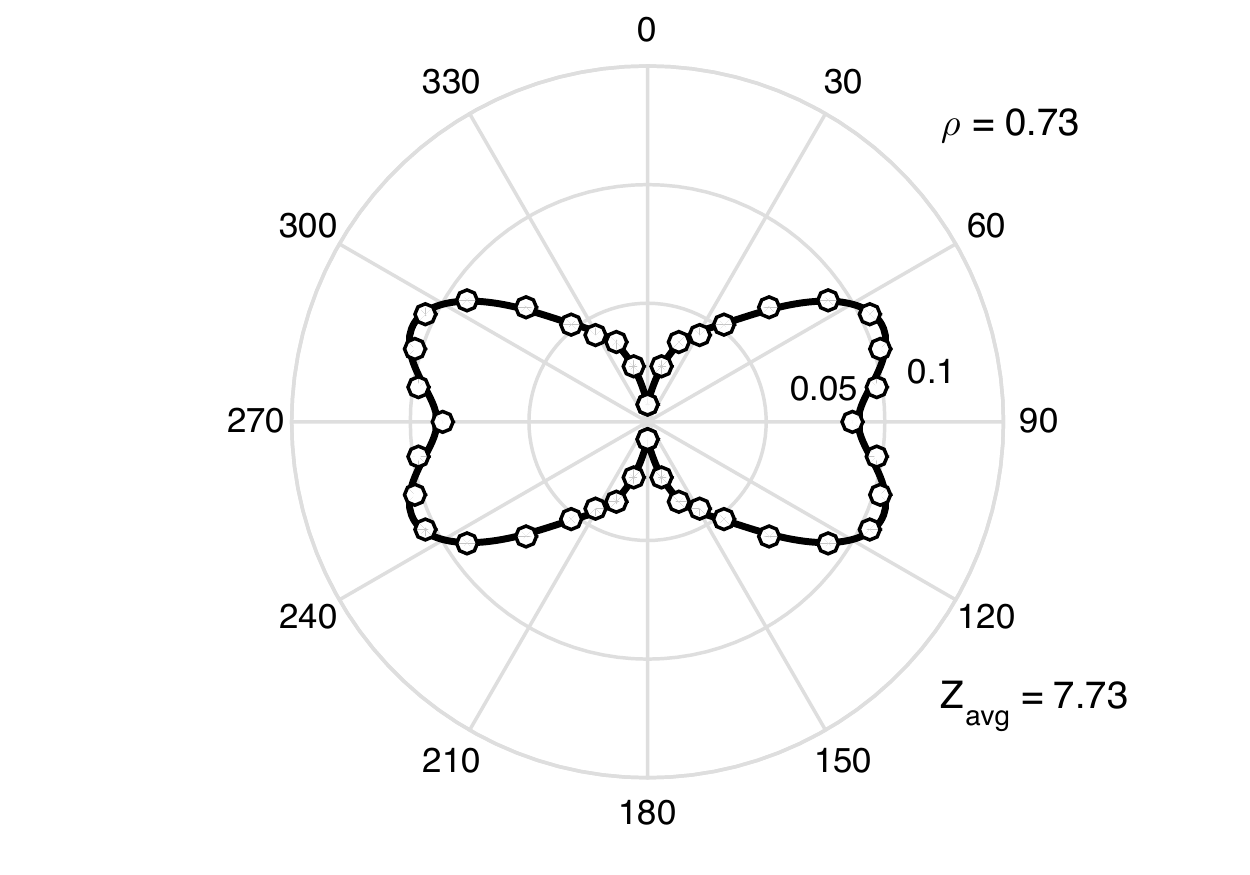}
        &
        \includegraphics[scale=0.5, trim=51 13 35 3, clip]{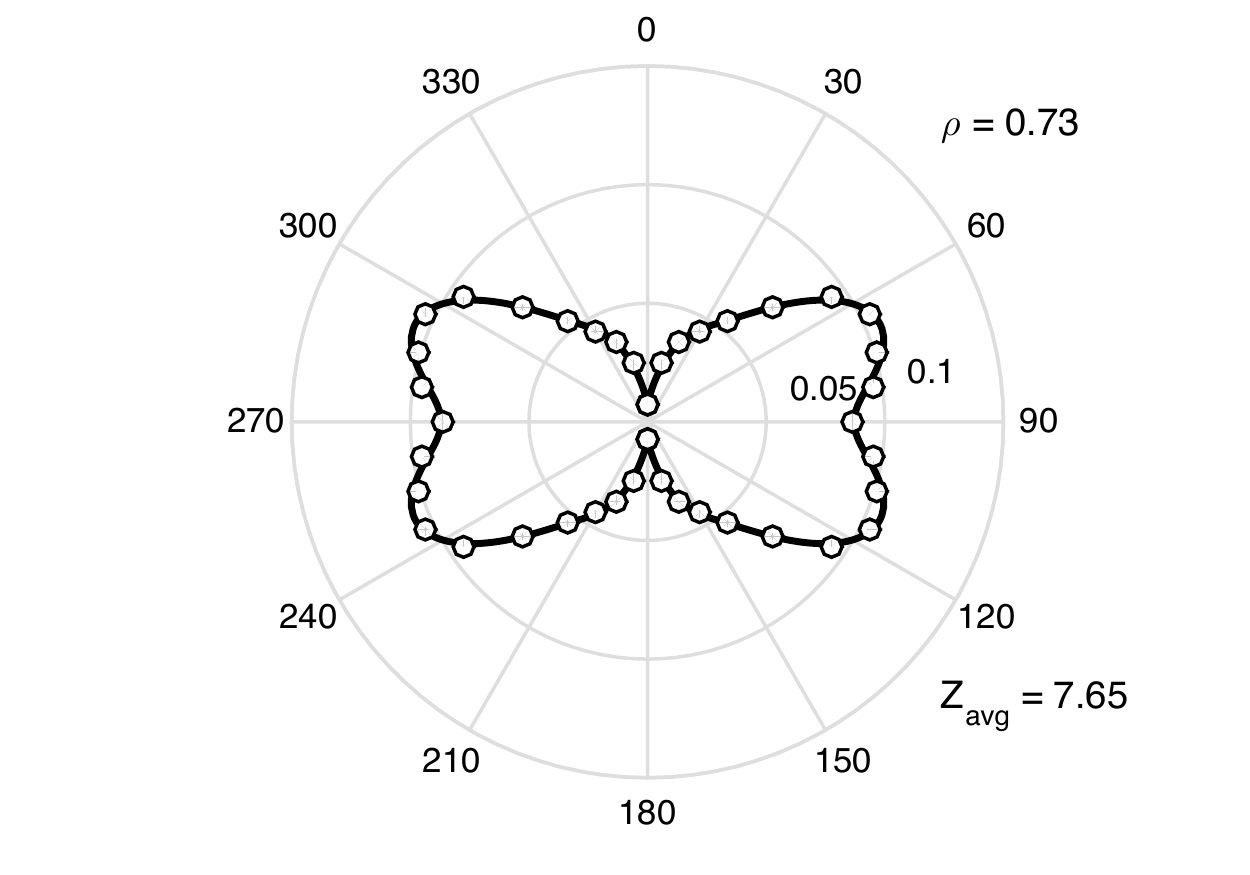}
        &
        \includegraphics[scale=0.5, trim=51 13 35 3, clip]{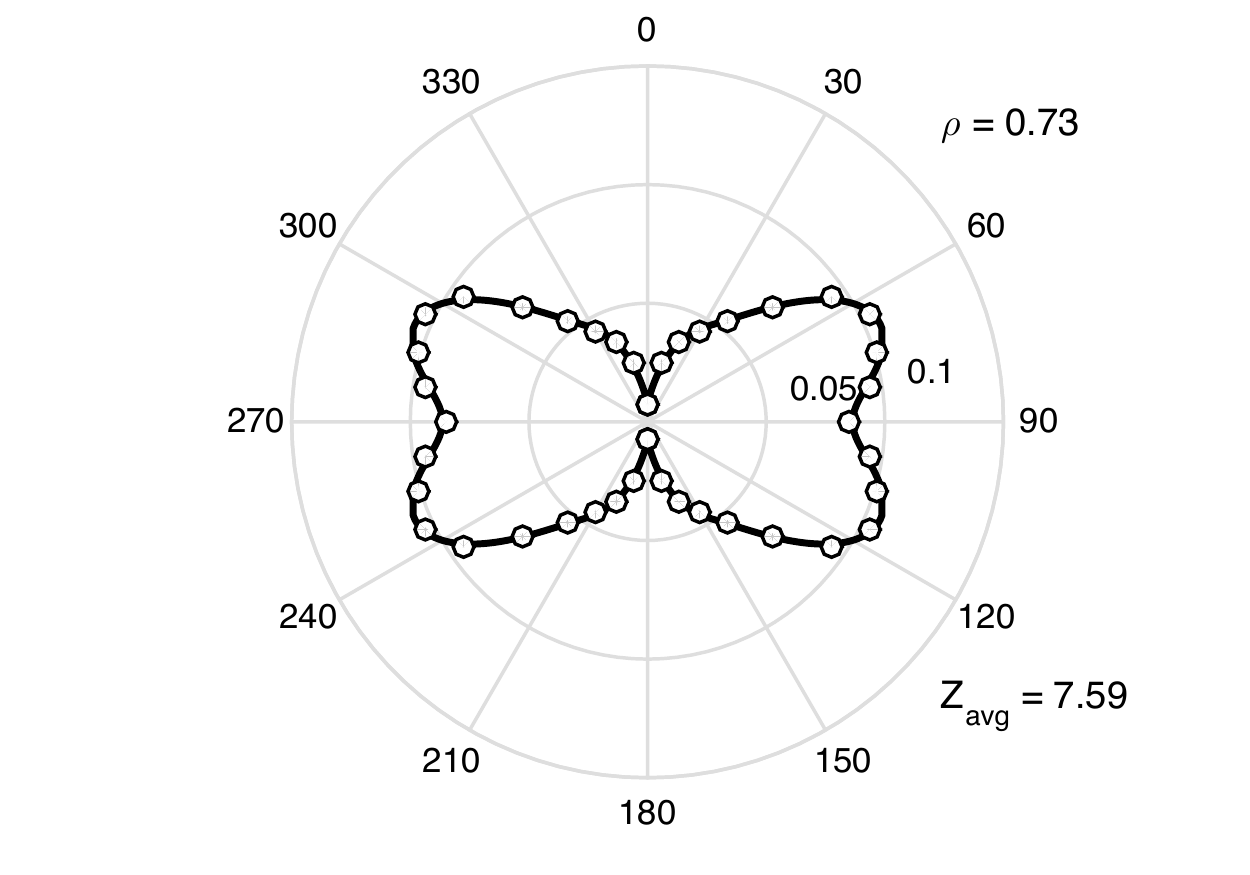}
        \\
	\small{After compaction}
	&
	\small{After unloading}
	&
	\small{After ejection}        
        \\
	\multicolumn{3}{p{1\linewidth}}{\centering \small{(b) Orientation distribution function of contact normals  $\xi(\theta,\phi)$.}}
    \end{tabular}
    \caption{Granular fabric anisotropy, adopting axial symmetry around the direction of compaction, for relative density $\rho^{\mbox{\tiny in-die}}_{\mbox{\tiny max}} = 0.7323$. Solid lines correspond to the best fit of equations \eqref{Eqn-ODF-Normals} and \eqref{Eqn-ODF-Forces} to the distributions obtained from the particle contact mechanics simulation of the granular bed (symbols).}
    \label{Fig-FabricAnisotropy001}
\end{figure}

\begin{figure}[htbp]
    \centering
    \begin{tabular}{ccc}
        \includegraphics[scale=0.5, trim=51 13 35 3, clip]{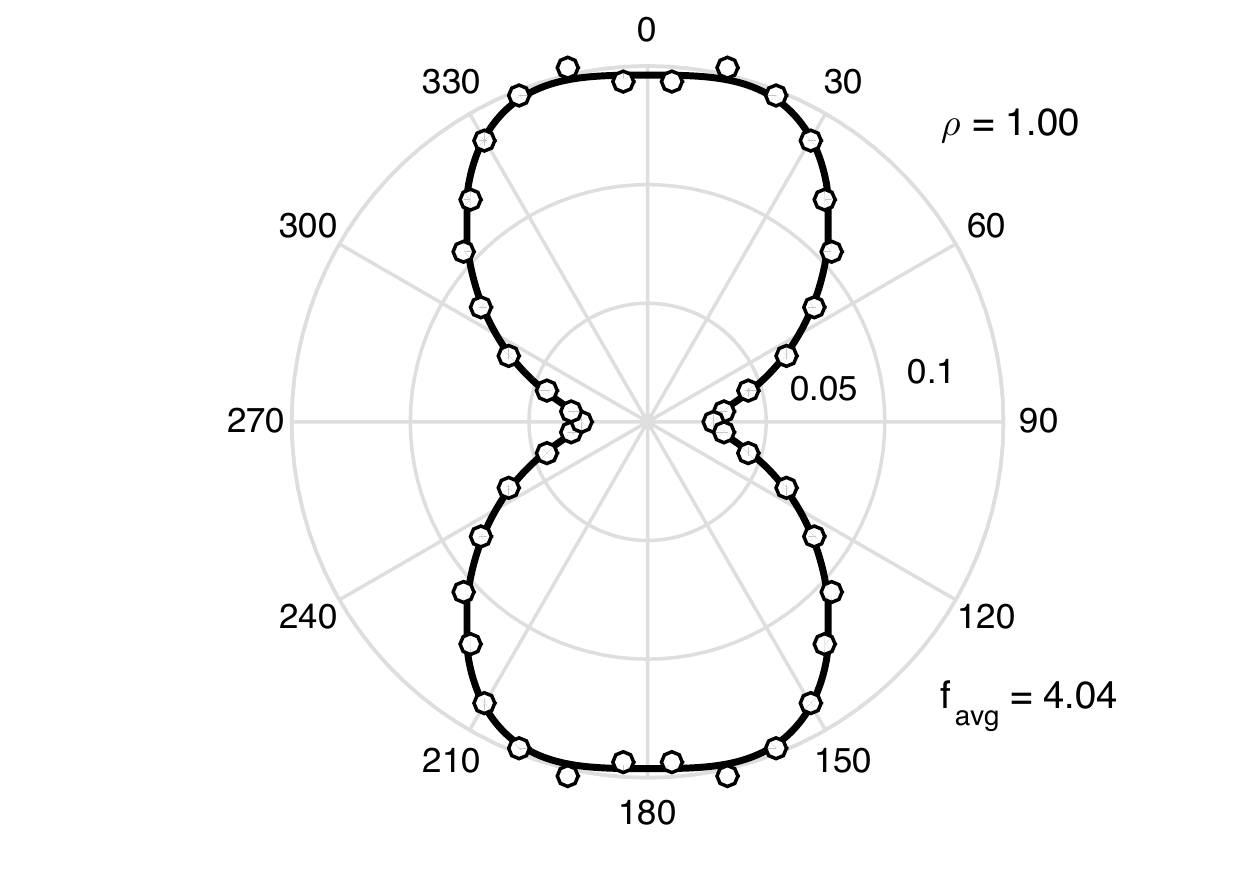}
        &
        \includegraphics[scale=0.5, trim=51 13 35 3, clip]{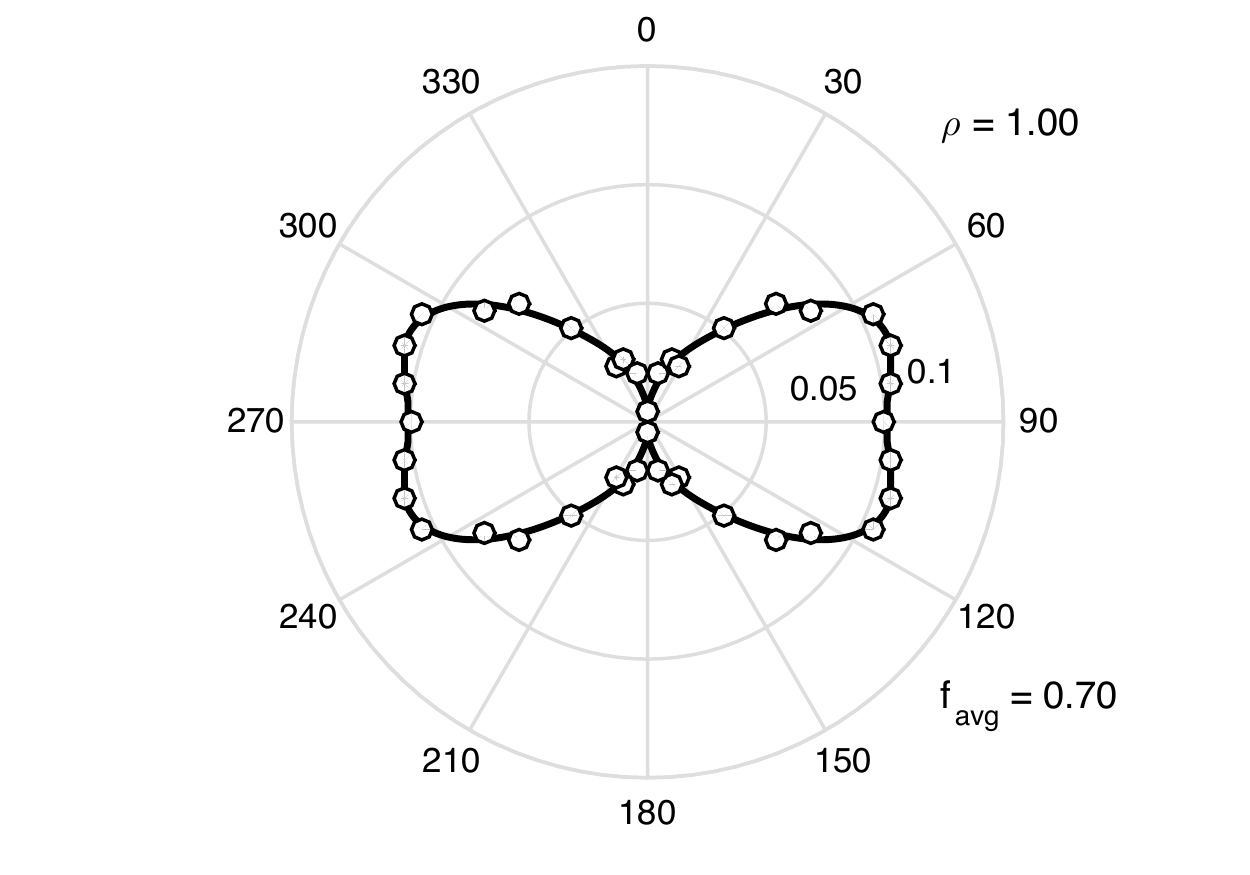}
        &
        \includegraphics[scale=0.5, trim=51 13 35 3, clip]{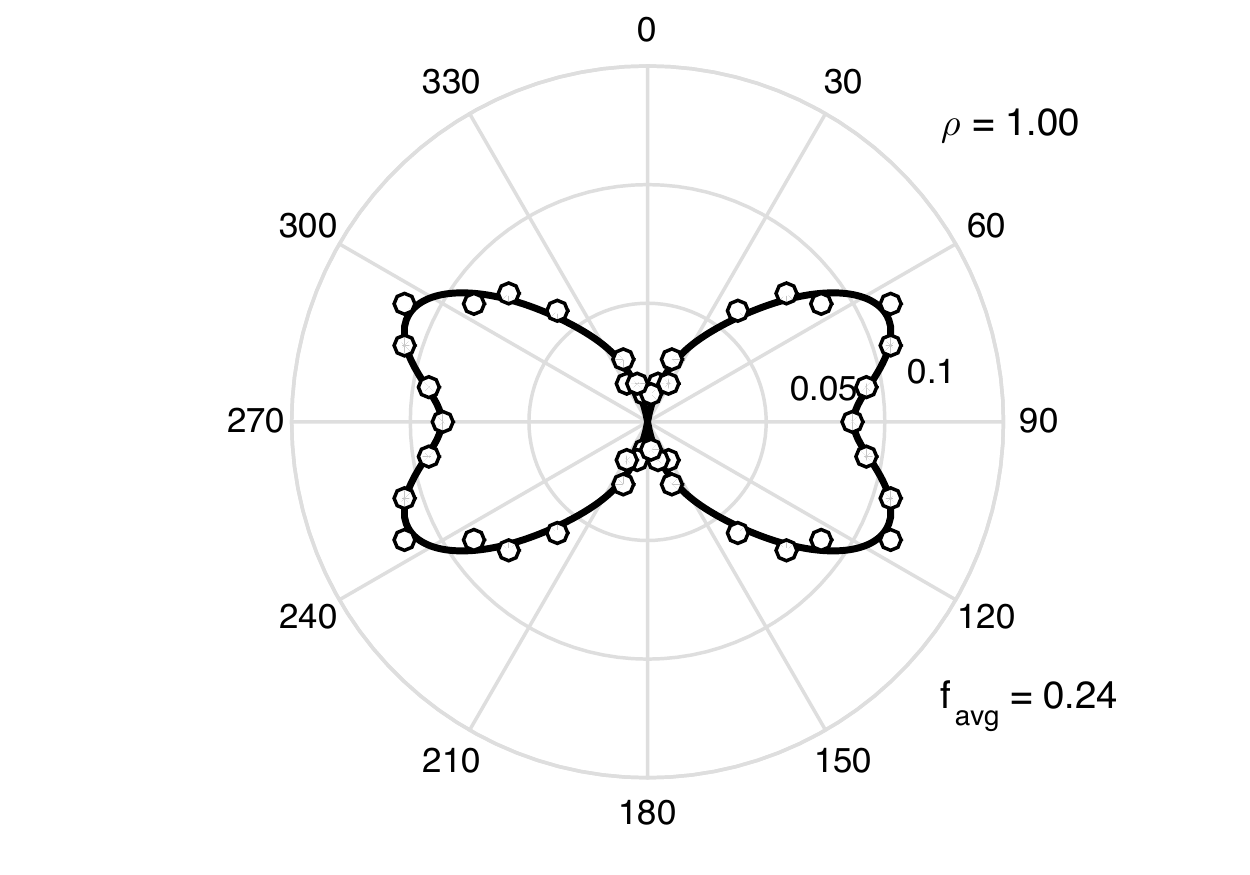}
        \\
	\small{After compaction}
	&
	\small{After unloading}
	&
	\small{After ejection}        
	\\    
	\multicolumn{3}{p{1\linewidth}}{\centering \small{(a) Orientation distribution function of the mean contact force ${ f(\theta,\phi) }/{f_{\rm avg}}$.}}
	\\    
        \includegraphics[scale=0.5, trim=51 13 35 3, clip]{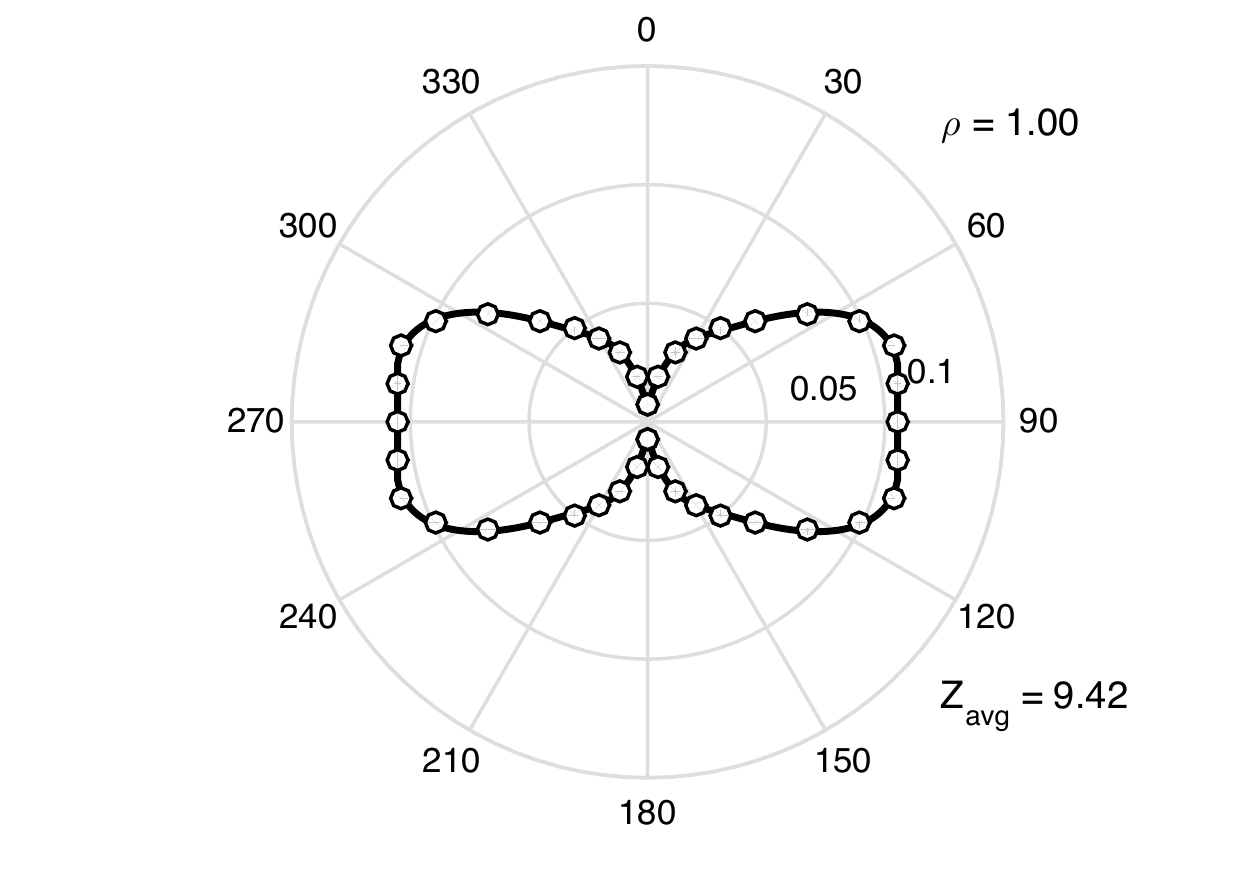}
        &
        \includegraphics[scale=0.5, trim=51 13 35 3, clip]{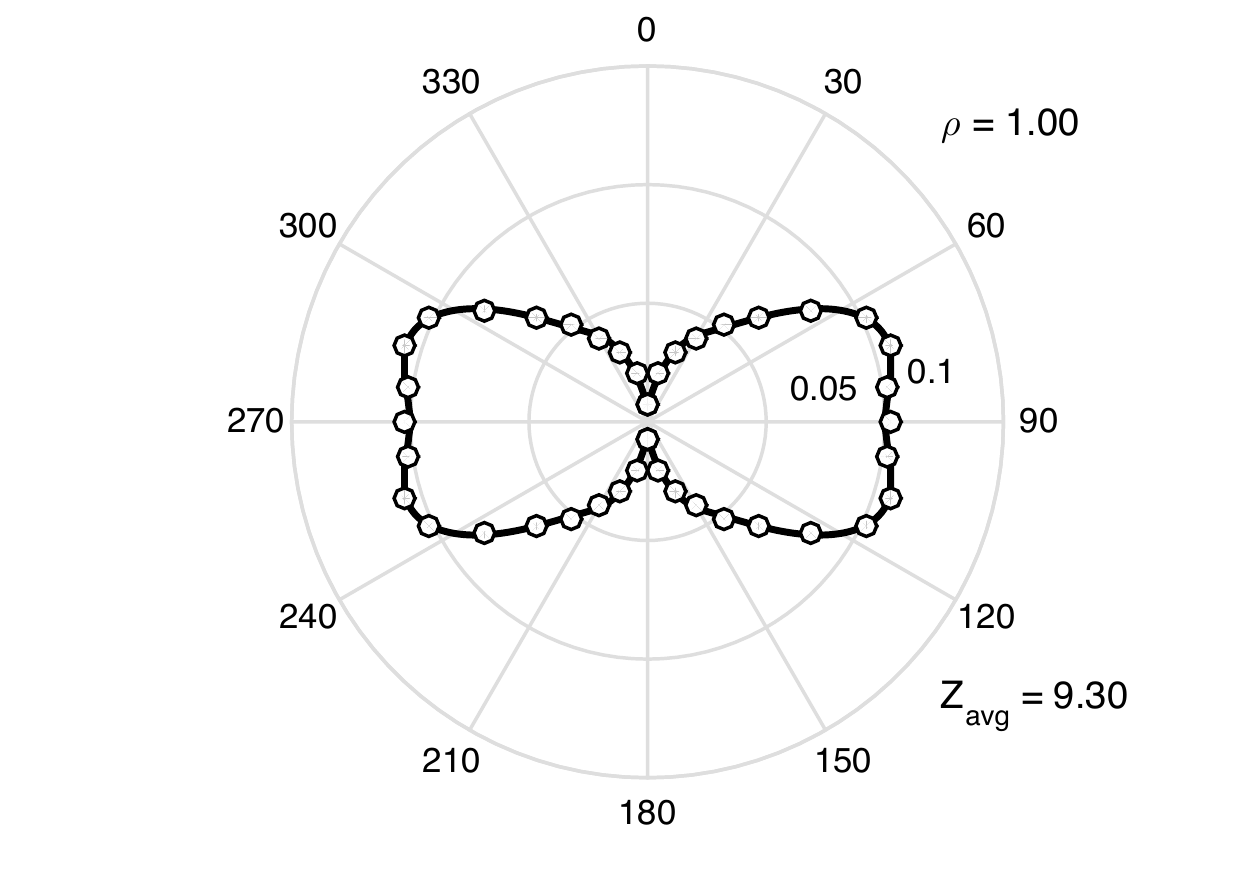}
        &
        \includegraphics[scale=0.5, trim=51 13 35 3, clip]{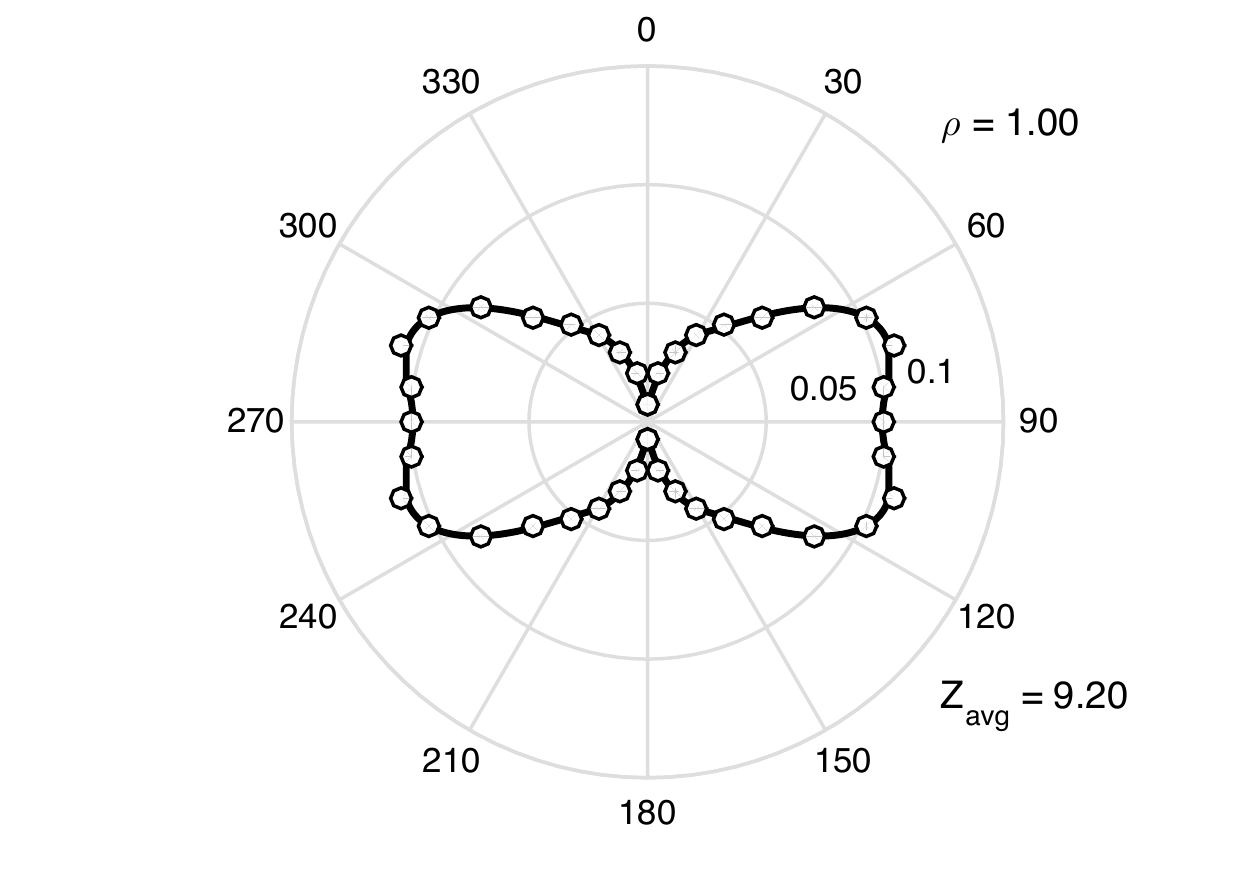}
        \\
	\small{After compaction}
	&
	\small{After unloading}
	&
	\small{After ejection}        
        \\
	\multicolumn{3}{p{1\linewidth}}{\centering \small{(b) Orientation distribution function of contact normals $\xi(\theta,\phi) $.}}
    \end{tabular}
    \caption{Granular fabric anisotropy, adopting axial symmetry around the direction of compaction, for relative density $\rho^{\mbox{\tiny in-die}}_{\mbox{\tiny max}} = 0.9950$. Solid lines correspond to the best fit of equations \eqref{Eqn-ODF-Normals} and \eqref{Eqn-ODF-Forces} to the distributions obtained from the particle contact mechanics simulation of the granular bed (symbols).}
    \label{Fig-FabricAnisotropy002}
\end{figure}

Granular fabric anisotropy is a complementary aspect of stress transmission and it can be determined from particle mechanics descriptions of granular systems under static equilibrium. This anisotropy is influenced by different factors, such as particle shape, die filling protocol, and deformation history. We specifically study the orientation distribution function of contact normals and of the mean contact force. Using spherical coordinates with azimuth and zenith angles $\theta$ and $\phi$, we define the contact orientation vector by
$$
{\bf{n}} = (\sin(\theta) \cos(\phi),\sin(\theta) \sin(\phi), \cos(\theta) )
$$
and, for axial symmetry around the zenith axis or the direction of compaction \cite{Gonzalez-2018b}, the spherical harmonics spectrum of the orientation distribution function of contact normals $
\xi({\bf{n}})=\xi(\theta,\phi)$ \cite{Chang-1990} by
\begin{eqnarray}
&\xi(\theta,\phi) 
=
\dfrac{1}{4 \pi} 
&\left( 1 + \frac{a_{20}}{2} (3 \cos(\theta)^2-1)  + \frac{a_{40}}{8} (35 \cos(\theta)^4-30 \cos(\theta)^2+3)   \right.   \\ \nonumber
&& +  \frac{a_{60}}{16} (231 \cos(\theta)^6 -315 \cos(\theta)^4 + 105 \cos(\theta)^2 - 5)  \\ \nonumber
&& + \left. \frac{a_{80}}{128} (6435 \cos(\theta)^8 - 12012 \cos(\theta)^6 + 6930 \cos(\theta)^4 - 1260 \cos(\theta)^2 + 35)   \right)
\label{Eqn-ODF-Normals}
\end{eqnarray}
with $ \int_0^\pi \int_0^{2\pi} \xi(\theta,\phi) \sin(\theta) \mathrm{d}\theta \mathrm{d}\phi = 1$. Similarly, the orientation distribution function of the mean contact force is assumed to be
\begin{eqnarray}
&\dfrac{ f(\theta,\phi) }{f_{\rm avg}}
=
\dfrac{1}{4 \pi}
& \left(1 + \frac{b_{20}}{2} (3 \cos(\theta)^2-1)  + \frac{b_{40}}{8} (35 \cos(\theta)^4-30 \cos(\theta)^2+3)   \right.   \\ \nonumber
&& + \frac{b_{60}}{16} (231 \cos(\theta)^6 -315 \cos(\theta)^4 + 105 \cos(\theta)^2 - 5)  \\ \nonumber
&& \left.  +	\frac{b_{80}}{128} (6435 \cos(\theta)^8 - 12012 \cos(\theta)^6 + 6930 \cos(\theta)^4 - 1260 \cos(\theta)^2 + 35)   \right)
\label{Eqn-ODF-Forces}
\end{eqnarray}
with $ \int_0^\pi \int_0^{2\pi} f(\theta,\phi) \sin(\theta) \mathrm{d}\theta \mathrm{d}\phi = {f_{\rm avg}}$. Figures~\ref{Fig-FabricAnisotropy001} and \ref{Fig-FabricAnisotropy002} show the orientation distribution function of the mean contact force and of contact normals after compaction, unloading and ejection obtained from the particle contact mechanics simulation of the granular bed at relative densities $\rho^{\mbox{\tiny in-die}}_{\mbox{\tiny max}}$ equal to $0.7323$ and $0.9950$, respectively. It is evident from the figure that: (i) the small number of large forces are oriented in the loading direction after compaction, while the large number of intermediate to small forces are oriented at $\pm60^{\circ}$ from the loading direction; (ii) the orientation distribution of contact normals does not significantly change during unloading and ejection due to the plastic, permanent nature of the deformations; (iii) after unloading, most large, vertically oriented forces are relaxed and, after ejection, most radially oriented forces are relaxed; (iv) compressive residual forces in the ejected solid compact are mostly oriented at $\pm60^{\circ}$ from the loading direction, and there is a small number of tensile residual forces that are oriented in the direction of loading; and (v) the orientation distribution of residual mean contact forces is different for different relative densities $\rho^{\mbox{\tiny in-die}}_{\mbox{\tiny max}}$. 

Table~\ref{Table-FabricAnisotropy} shows the coefficients $a_i$ and $b_i$ determined by fitting equations \eqref{Eqn-ODF-Normals} and \eqref{Eqn-ODF-Forces} to the distributions illustrated in Figures~\ref{Fig-FabricAnisotropy001}-\ref{Fig-FabricAnisotropy002} and obtained from the particle contact mechanics simulation of the granular bed. It is worth noting that a eighth-order approximation is the lowest-order representation of the orientation distributions functions that captures the characteristics of directional distributions at the desired approximation accuracy. The eighth-order expansion of $\xi({\bf{n}})$ takes the form 
$$
	\xi({\bf{n}}) 
	=
	C + C_{ij} n_i n_j + C_{ijkl} n_i n_j n_k n_l + C_{ijklpq} n_i n_j n_k n_l n_p n_q + C_{ijklpqrs} n_i n_j n_k n_l n_p n_q n_r n_s
$$
where four independent coefficients $a_{20}$, $a_{40}$, $a_{60}$ and $a_{80}$ emerge after enforcing symmetry about the zenith axis and $ \int_0^\pi \int_0^{2\pi} \xi(\theta,\phi) \sin(\theta) \mathrm{d}\theta \mathrm{d}\phi = 1$---cf. equation \eqref{Eqn-ODF-Normals}. In a similar manner, four independent coefficients $b_{20}$, $b_{40}$, $b_{60}$ and $b_{80}$ characterize the eighth-order expansion of ${ f({\bf n}) }/{f_{\rm avg}}$---cf. equation \eqref{Eqn-ODF-Forces}. 

\begin{table}[htbp]
    \centering
{\onehalfspacing \footnotesize
    \begin{tabular}{c|cccc|cccc}
        \hline
        \multicolumn{1}{p{0.19\linewidth}|}{\centering }
        &
        \multicolumn{4}{p{0.34\linewidth}|}{\centering Distribution of mean contact force}
        &
        \multicolumn{4}{p{0.34\linewidth}}{\centering Distribution of contact normals}
        \\
	\hline
        $\rho^{\mbox{\tiny in-die}}_{\mbox{\tiny max}} = 0.7323$
        &
        $a_{20}$
        &
        $a_{40}$
        &
        $a_{60}$
        &
        $a_{80}$
        &
        $b_{20}$
        &
        $b_{40}$
        &
        $b_{60}$
        &
        $b_{80}$        
        \\
    	After compaction
	&
	0.9491
	&
	-0.05312
	&
	-0.006645
	&
	-0.06142
	&
	-0.6826
	&
	-0.2216
	&
	0.2513
	&
	-0.2356
	\\
	After unloading
	&
	-0.8806
	&
	-0.09625
	&
	0.02114
	&
	-0.1343
	&
	-0.656
	&
	-0.2475
	&
	0.2506
	&
	-0.2305
	\\
	After ejection
	&
	-0.3673
	&
	-0.1978
	&
	0.07271 
	&
	-0.3248
	&
	-0.6495
	&
	-0.2555
	&
	0.2556
	&
	-0.2324
	\\
        \hline
        $\rho^{\mbox{\tiny in-die}}_{\mbox{\tiny max}} = 0.9950$
        &
        $a_{20}$
        &
        $a_{40}$
        &
        $a_{60}$
        &
        $a_{80}$
        &
        $b_{20}$
        &
        $b_{40}$
        &
        $b_{60}$
        &
        $b_{80}$        
        \\
    	After compaction
	&
	1.137
	&
	-0.2228
	&
	-0.005998
	&
	-0.0717
	&
	-0.8201
	&
	-0.01669
	&
	0.09396
	&
	-0.176
	\\
	After unloading
	&
	-0.8273
	&
	-0.1439
	&
	0.1995
	&
	-0.1348
	&
	-0.776
	&
	-0.086
	&
	0.1392
	&
	-0.197
	\\
	After ejection
	&
	-0.7724
	&
	-0.3899
	&
	0.2906
	&
	-0.2151
	&
	-0.7657
	&
	-0.1047
	&
	0.1529
	&
	-0.2025
	\\
        \hline
    \end{tabular}
    }
    \caption{Material 1. The coefficients correspond to the best fit of equations \eqref{Eqn-ODF-Normals} and \eqref{Eqn-ODF-Forces} to the distributions obtained from the particle contact mechanics simulation of the granular bed.}
    \label{Table-FabricAnisotropy}
\end{table}

\subsection{Bonding surface area}
\label{Section-BondingArea}

\begin{figure}[htbp]
    \centering
    \begin{tabular}{cc}
        \includegraphics[scale=0.64, trim=0 0 30 0, clip]{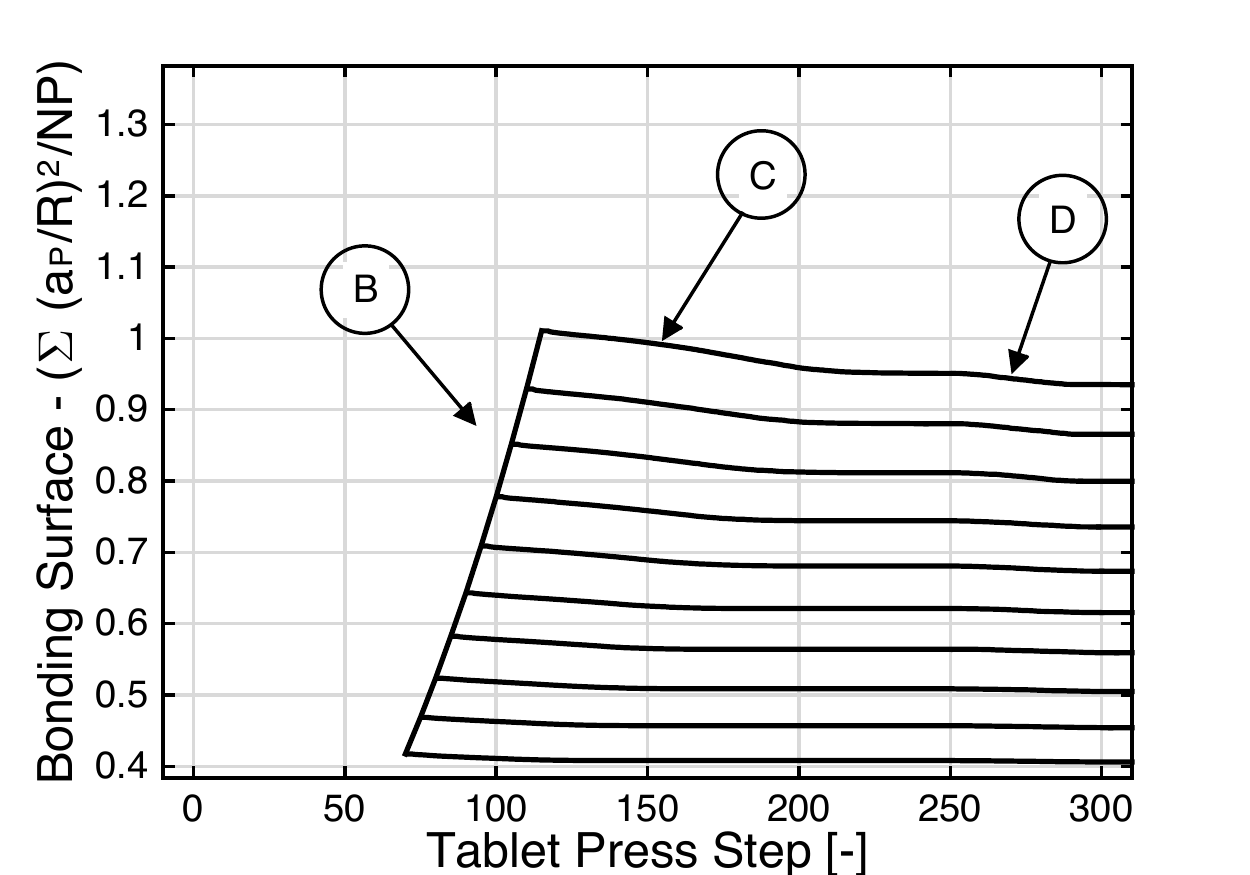}
        &
        \includegraphics[scale=0.64, trim=0 0 30 0, clip]{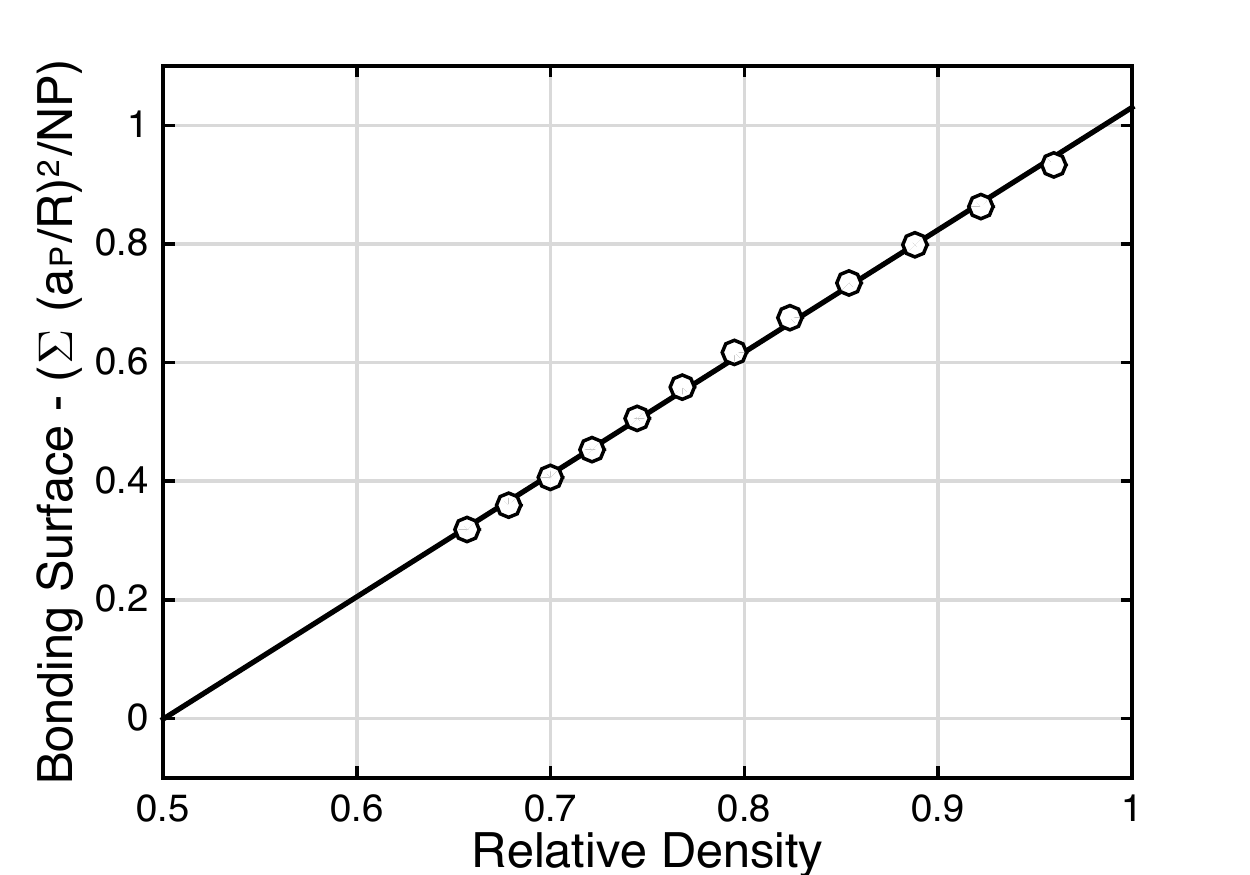}
    \\
    \small{(a) Material 1}
    &
    \small{(b) Material 1}        
    \\
        \includegraphics[scale=0.64, trim=0 0 30 0, clip]{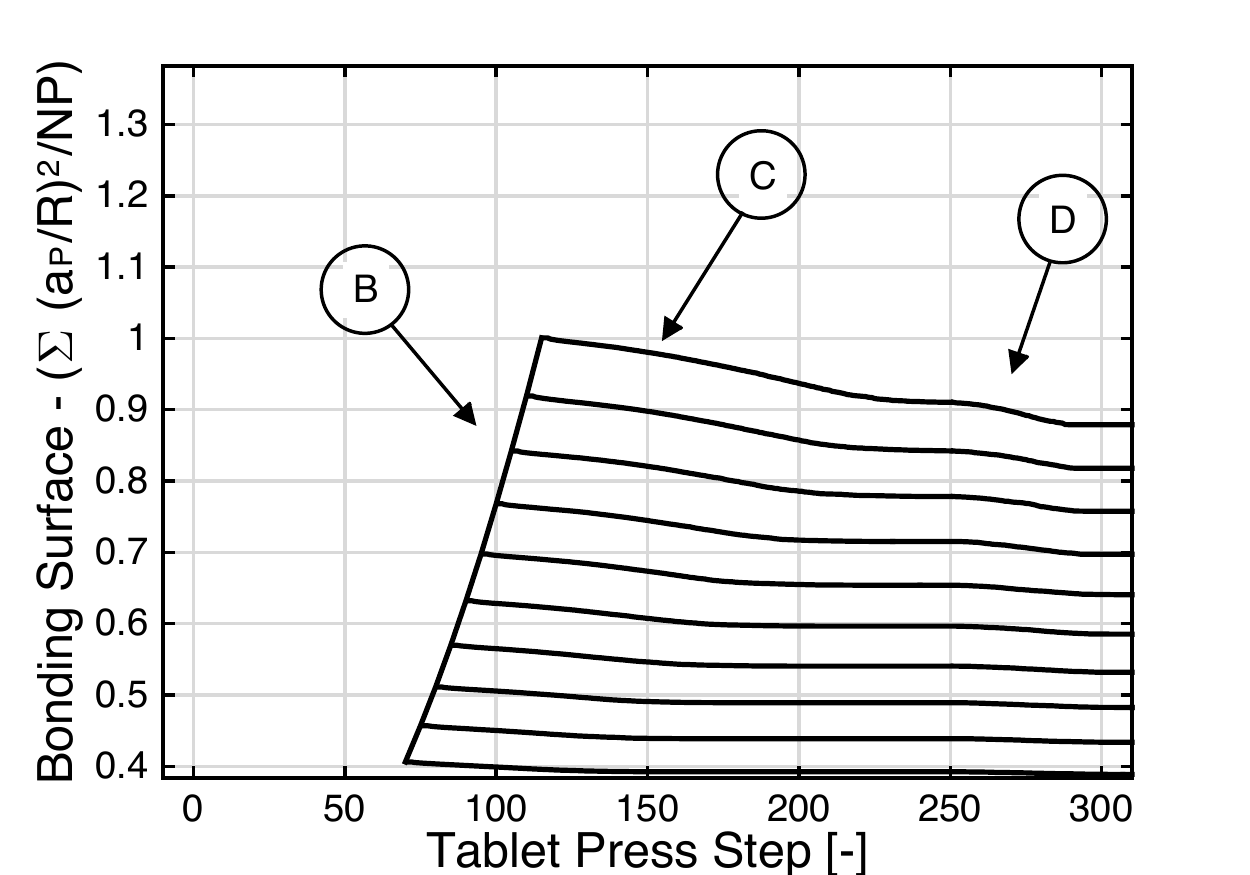}
        &
        \includegraphics[scale=0.64, trim=0 0 30 0, clip]{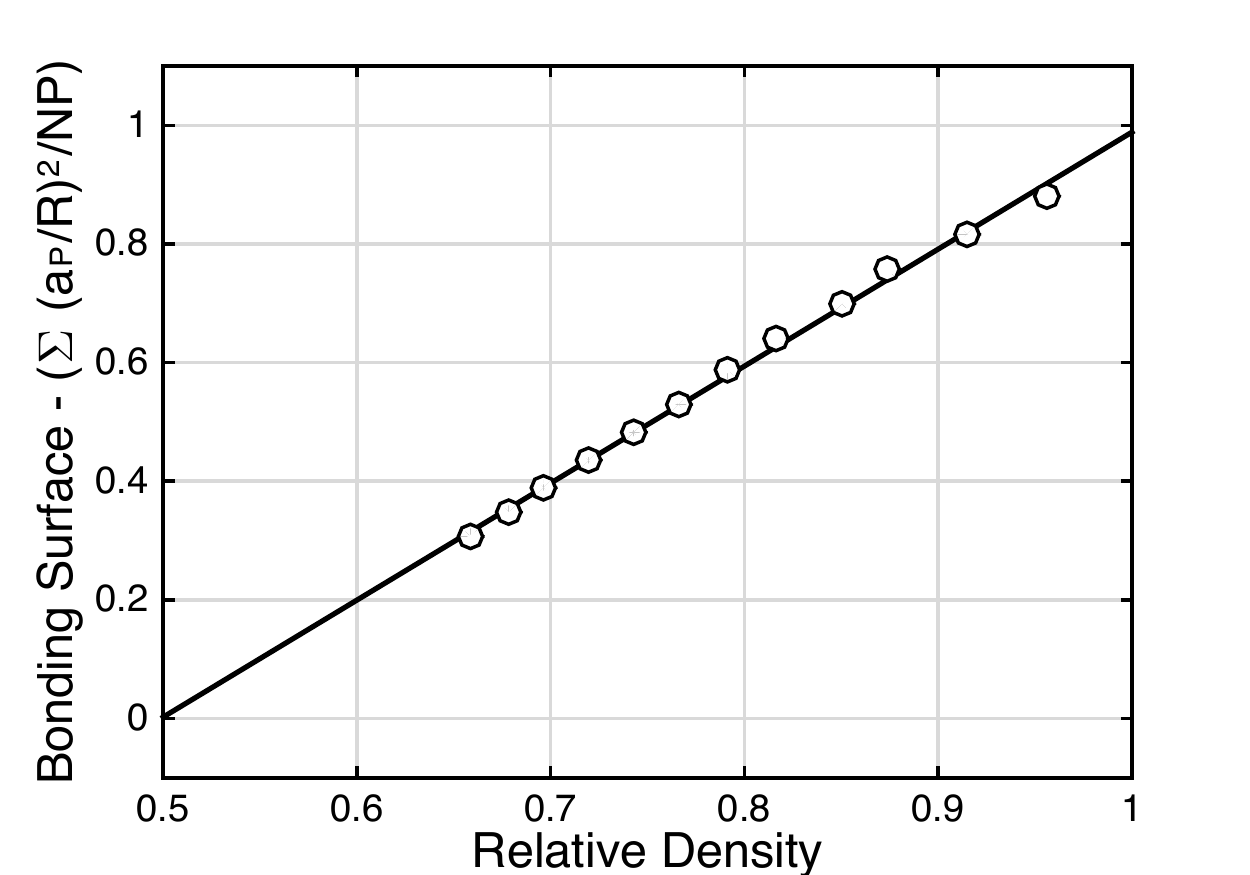}
    \\
    \small{(c) Material 2}
    &
    \small{(d) Material 2}            
    \end{tabular}
    \caption{Bonding surface area. (a)\&(c) bonding surface parameter $\bar{A}_{\mbox{\tiny b}}$ as a function of tablet press step for loading, unloading and ejection; (b)\&(d) bonding surface parameter as a function of relative density $\rho^{\mbox{\tiny in-die}}_{\mbox{\tiny min}}$. Solid line in (b)\&(d) corresponds to the best fit of equation \eqref{Eqn-BondingSurface} to the bonding surface parameter obtained from the particle contact mechanics simulation of the granular bed (symbols).}
    \label{Fig-BondingSurface}
\end{figure}

As discussed above, the quantitative elucidation of strength formation requires not only the identification of the deformation and bonding mechanisms of interest but also of the bonding surface involved in the process. Here we assume that an upper bound for the bonding surface involved in the formation of solid bridges is the particle-to-particle contact area created during compaction. Therefore, we study the bonding surface area by defining a parameter $\bar{A}_{\mbox{\tiny b}}$ that is proportional to the ratio between the total bonding surface and that total available surface in the powder bed, i.e., $\bar{A}_{\mbox{\tiny b}}=(\sum a_{\mbox{\tiny P}}^2)/(R^2 N_P)$ with $N_P$ being the number of particles in the monodisperse bed. Specifically, we investigate the evolution of the bonding surface parameter $\bar{A}_{\mbox{\tiny b}}$ during all stages of die compaction (see Figure~\ref{Fig-BondingSurface}) and we identify that the following relationship holds
\begin{equation}
	\bar{A}_{\mbox{\tiny b}} 
	=
	\bar{A}_{\mbox{\tiny b},0} \frac{\rho^{\mbox{\tiny in-die}}_{\mbox{\tiny min}} - \rho_{c,b}}{1-\rho_{c,b}}
	\label{Eqn-BondingSurface}
\end{equation}
where the bonding surface parameter of a fully dense tablet $\bar{A}_{\mbox{\tiny b},0} = 1.030$ and the critical relative density $\rho_{c,b}=0.5004$ are best-fitted to the numerical results for Material 1---$\bar{A}_{\mbox{\tiny b},0} = 0.9883$ and $\rho_{c,b}=0.4988$ for Material 2.

It bears emphasis that the formation of bonding surface area is controlled by both plastic deformations and elastic deformations. It is known that extensive particle elasticity could cause a drastic decrease in tablet strength, due to the breakage of solid bridges and thus reduction of bonding surface area. Figure~\ref{Fig-BondingSurface} shows that during unloading and ejection the lost in bonding surface area is larger for higher relative densities, i.e., for higher elastic recovery (cf. Figure~\ref{Fig-ElasticRecovery}). These trends are consistent with the behavior of many pharmaceutical excipients  obtained from permeametry measurements \cite{Adolfsson-1999,Nystrom-1986}. Finally, it is interesting to note that $\rho_{c,\bar{Z}} \approx \rho_{c,b}$.

\subsection{Young's modulus and Poisson's ratio of the compacted solid}
\label{Section-YoungPoisson}

\begin{figure}[htbp]
    \centering
    \begin{tabular}{cc}
        \includegraphics[scale=0.64, trim=0 0 30 0, clip]{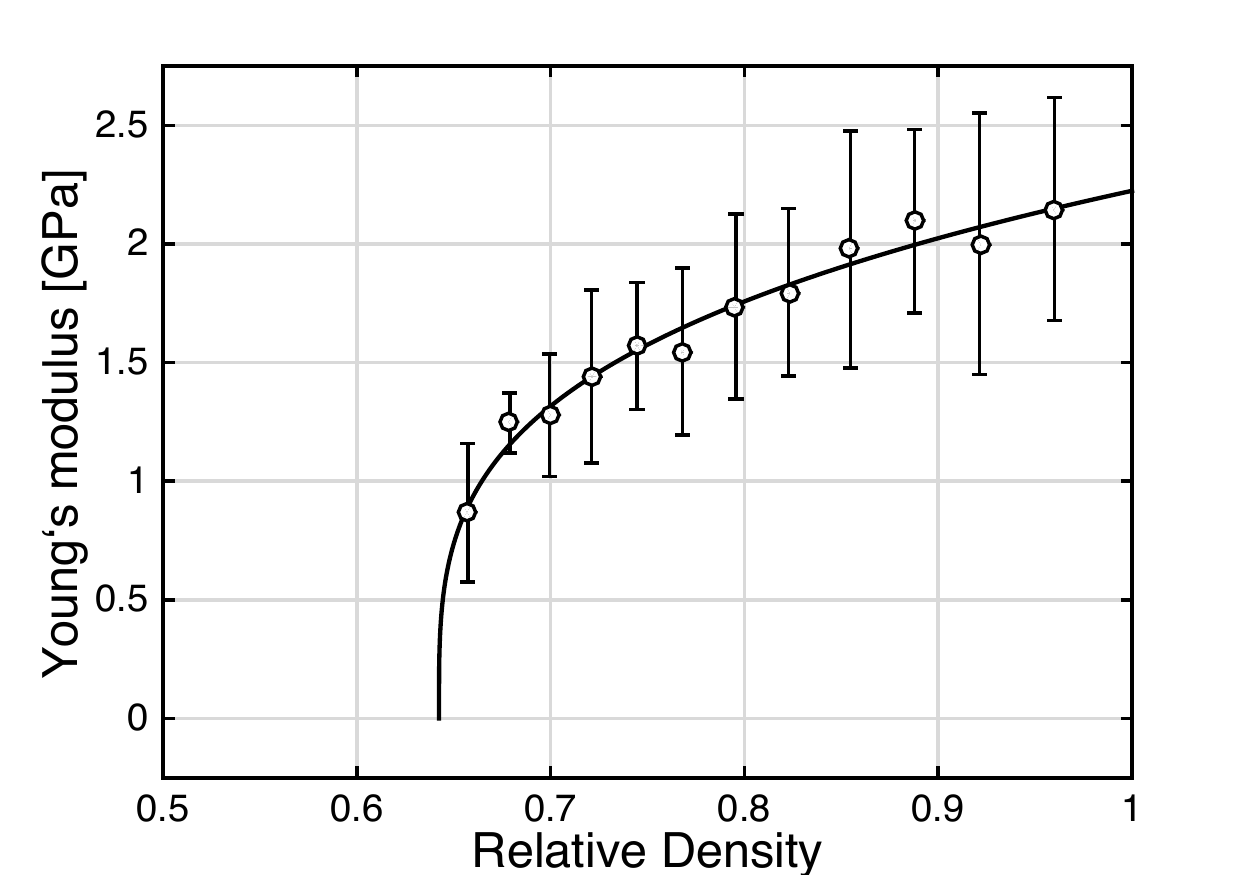}
        &
        \includegraphics[scale=0.64, trim=0 0 30 0, clip]{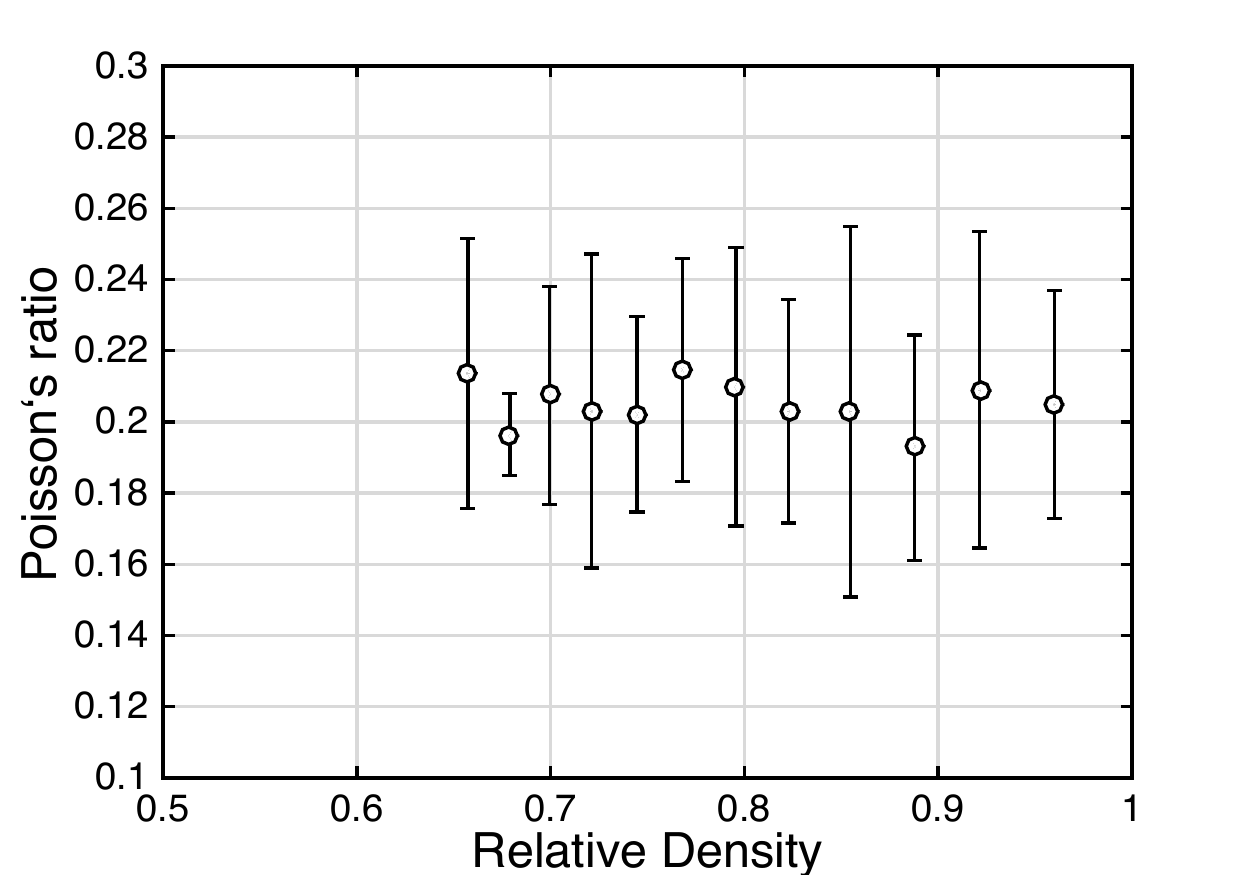}
    \\
    \small{(a) Material 1}
    &
    \small{(b) Material 1}        
    \\
        \includegraphics[scale=0.64, trim=0 0 30 0, clip]{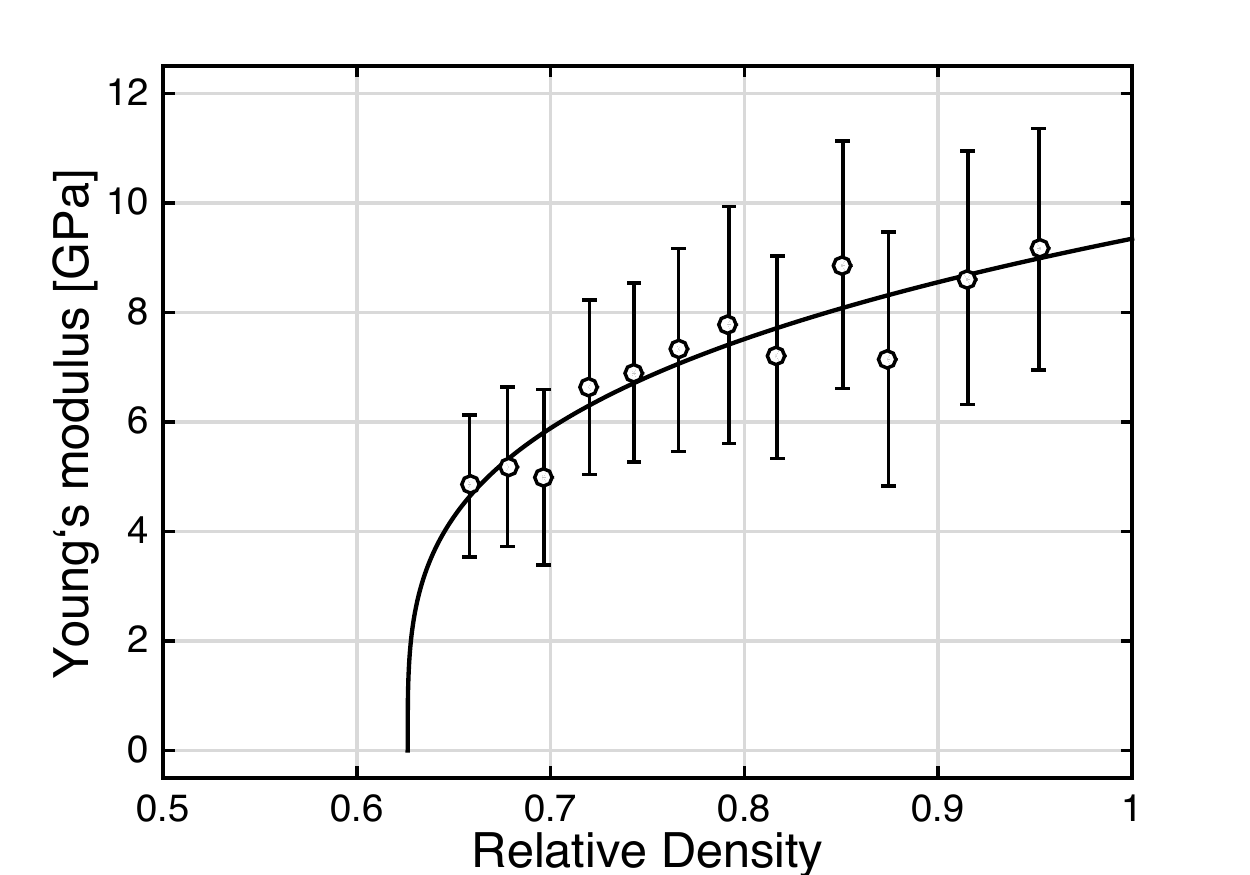}
        &
        \includegraphics[scale=0.64, trim=0 0 30 0, clip]{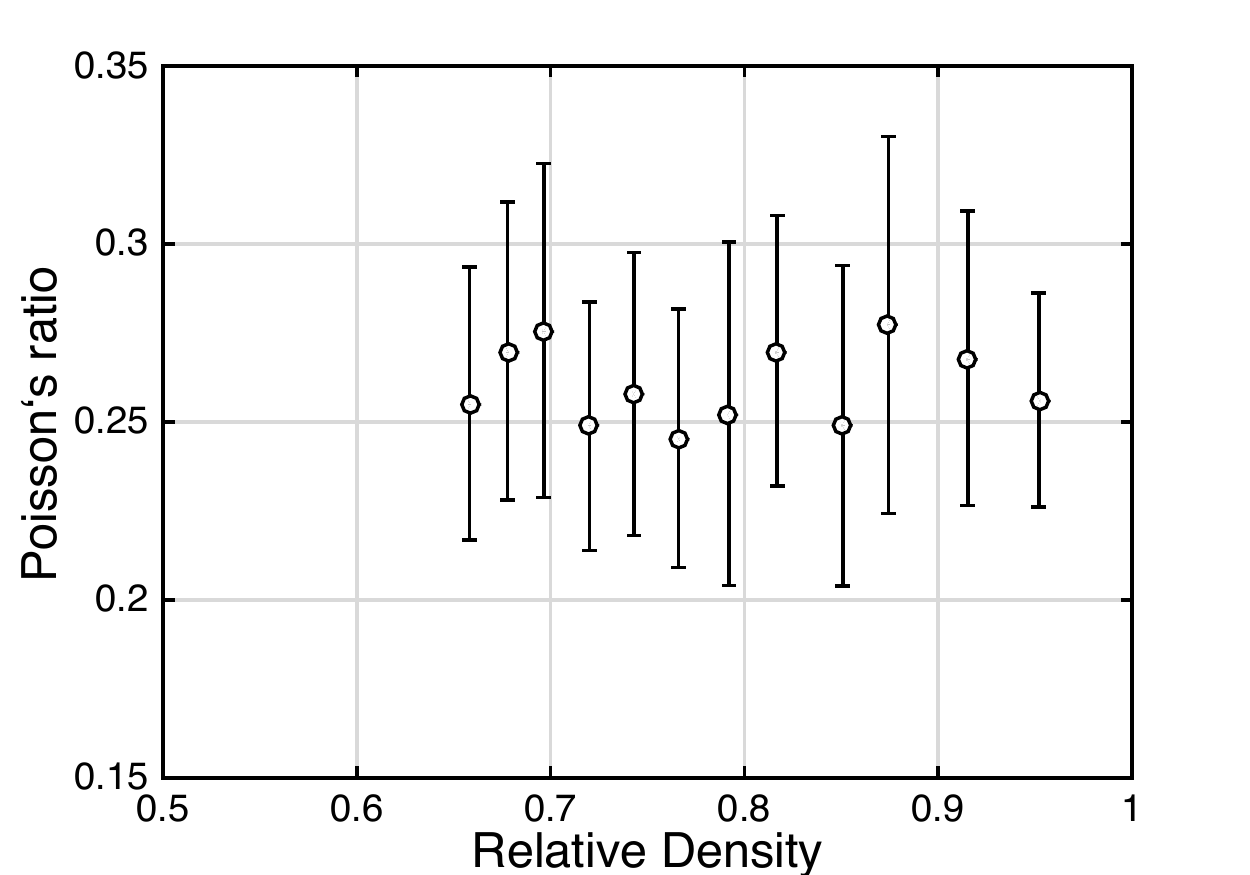}
    \\
    \small{(c) Material 2}
    &
    \small{(d) Material 2}        
    \end{tabular}
    \caption{Young's modulus and Poisson's ratio. (a)\&(c) Young's modulus of the compacted solid as a function of relative density $\rho^{\mbox{\tiny in-die}}_{\mbox{\tiny min}}$; (b)\&(d) Poisson's ratio of the compacted solid as a function of relative density $\rho^{\mbox{\tiny in-die}}_{\mbox{\tiny min}}$. Solid line in (a)\&(c) corresponds to the best fit of equation \eqref{Eqn-Young} to the bonding surface parameter obtained from the particle contact mechanics simulation of the granular bed (symbols).}
    \label{Fig-YoungPoisson}
\end{figure}

The Young's modulus of the compacted solid, obtained from central one-third of the unloading curve using Hooke's law  \cite{Han-2008,Swaminathan-2016}, is given by
\begin{equation}
	E_{\mbox{\tiny tablet}} 
	=
	E_{0} \left[ \frac{\rho^{\mbox{\tiny in-die}}_{\mbox{\tiny min}} - \rho_{c,E}}{1-\rho_{c,E}} \right]^n
\label{Eqn-Young}
\end{equation}
where the Young's modulus of a fully dense tablet $E_{0}=2.224$~GPa, the exponent $n = 0.2873$ and the critical relative density $\rho_{c,E} = 0.6423$ are best-fitted to the numerical results for Material 1---$E_{0}= 9.345$~GPa, $n = 0.285$ and $\rho_{c,E} = 0.6262$ for Material 2. These values shown in Figure~\ref{Fig-YoungPoisson} are in agreement with those obtained for two grades on lactose, at different lubrication levels, using an ultrasound transmission technique \cite{Razavi-2016,Razavi-2018}. Equation \eqref{Eqn-Young} is a semi-empirical relationship derived by Phani and Niyogi for porous solids \cite{Phani-1987}. The exponent $n$ is regarded as a material constant  dependent on particle morphology and pore geometry of the material, which is also suggested by these results for which the packings are identical and the values of $n$ are equal. Finally, it is also interesting to note that $\rho_{c,e} \approx \rho_{c,E}$.

As point out above, the granular bed develops anisotropic mechanical properties during loading, unloading and ejection. In this section, for simplicity, we assume isotropic behavior and thus determine two elastic properties, i.e., Young's modulus and Poisson's ratio, from the unloading curve. This assumption can be relaxed and, if a transversely isotropic material is assumed, five anisotropic continuum properties can be determined from the loading curve \cite{Gonzalez-2018b}. The extension of this analysis to unloading and ejection stages, though beyond the scope of this work, is currently being pursued by the author.

\subsection{Microstructure-mediated process-structure-property-performance interrelationship}
\label{Section-PSPP}

\begin{figure}[htbp]
    \centering
        \includegraphics[scale=0.60]{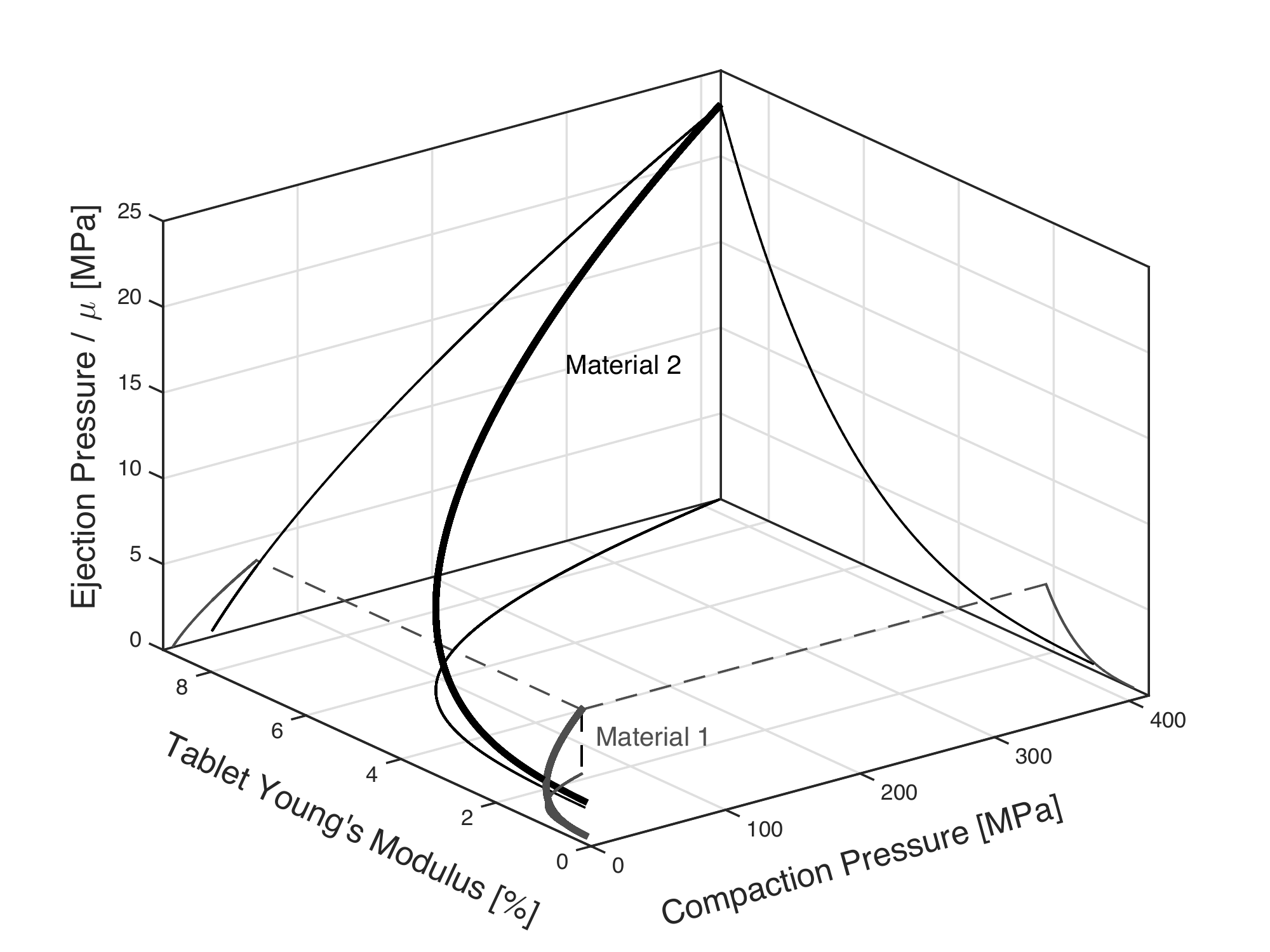}
    \caption{Interrelationship between process variables (namely compaction and ejection pressures), material properties (i.e., $E$, $\nu$, $\kappa$, $m$ and $K_{Ic}$) and critical quality attributes of the compact (namely the tablet Young's modulus). Materials 1 and 2 are depicted in gray and in black, respectively.}
    \label{Fig-TabletCQA}
\end{figure}

In the spirit of Olson's design framework that integrates process, structure, property, and performance \cite{Olson-1997}, as well as of the Quality by Design (QbD) principles recently adopted by the U.S. Food and Drug Administration (FDA) \cite{Lawrence-2008,Lee-2015}, we study next the interrelationship between two process variables (namely compaction and ejection pressures), particle-scale material properties (i.e., $E$, $\nu$, $\kappa$, $m$ and $K_{Ic}$) and a critical quality attribute of the compacted solid product (namely the tablet Young's modulus). This relationship is derived from microstructure formation and evolution predicted with the proposed particle mechanics approach and the generalized loading-unloading contact laws. Figure~\ref{Fig-TabletCQA} uses equations \eqref{Eqn-AppliedPressure}, \eqref{Eqn-EjectionPressure}, \eqref{Eqn-DefinitionElasticRecovery} and \eqref{Eqn-Young} to represent this interrelationship for Materials 1 and 2. It is worth noting that this interrelationships can not only be used to design a material with targeted quality attributes (see, e.g., \cite{Sun-2001,Tye-2005,Razavi-2015,Razavi-2016,Razavi-2018}) but it can also be used to control the manufacturing process to assure such quality attributes are achieved (see, e.g., \cite{Su-2018}). 

\section{Concluding remarks}
\label{Section-ConcludingRemarks}

We have reported three-dimensional particle mechanics static calculations that enabled us to predict microstructure evolution during the three most important steps of die-compaction of solid tablets, namely during compaction, unloading, and ejection \cite{nanoHUB-2017}. Specifically, we have simulated the compaction, inside a rigid cylindrical die, of monodisperse elasto-plastic spherical particles capable of forming solid bridges. To this end, we have developed and employed generalized loading-unloading contact laws for elasto-plastic spheres with bonding strength. The proposed loading-unloading contact laws are continuous at the onset of unloading by means of a regularization term, in the spirit of a cohesive zone model, that introduces a small, controllable error in the solid bridge breakage force and the critical contact surface. This continuity property is in sharp contrast with the behavior of standard mechanistic loading and unloading contact theories, which exhibit a discontinuity at the onset of unloading when particles form solid bridges during plastic deformations. In addition, these generalized contact laws are explicit in terms of the relative position between the particles, and are updated incrementally to account for strain path dependency. Furthermore, the three-dimensional particle mechanics static calculations show that the formulation is numerically robust, efficient, and mechanistically sound.

We have exemplified the effectiveness and versatility of the particle mechanics approach by studying two sets of material properties, which do not correspond to any material in particular but rather represent lower and upper bounds for many pharmaceutical powders, including drugs and excipients. These simulations reveal the evolution, up to relative densities close to one, of (i) mean mechanical coordination number, (ii) punch force and die-wall reaction, (iii) in-die elastic recovery, (iv) ejection pressure, (v) network of contact forces and granular fabric anisotropy, (vi) bonding surface area, (vii) Young's modulus and Poisson's ratio of the compacted solid. Our results are quantitatively similar to those experimentally observed in many pharmaceutical formulations \cite{AbdelHamid-2011, Adolfsson-1999, Doelker-2004, Haware-2010, Mahmoodi-2013, Nystrom-1986, Panelli-2001, Phani-1987, Razavi-2016, Razavi-2018, Yohannes-2015}. Moreover, the evolution during compaction of these process variables (such as punch, die-wall and ejection pressures, and in-die elastic recovery) and compact attributes (such as Young's modulus and Poisson's ratio) is in remarkable agreement with the formulae reported in the literature---i.e., (semi-)empirical relationships developed over the last decades \cite{Durian-1995, Gonzalez-2016, Gonzalez-2018, OHern-2002, OHern-2003, Phani-1987, Razavi-2016, Su-2018}. Furthermore, these relationships have enabled the development of microstructure-mediated process-structure-property-performance interrelationships for QbD product development and process control. Table~\ref{Table-SummaryEvolution} summarizes these findings. It is evident from the table that a small number of parameters, with well-defined physical meaning, is required to describe such evolution with relative density. The systematic investigation of the relationship between these parameters with particle-level material properties (i.e., $E$, $\nu$, $\kappa$, $m$ and $K_{Ic}$), particle morphology (such as particle size distribution) and process variables (such as tablet weight, dimensions and composition) is a worthwhile direction of future research---see \cite{Gonzalez-2018} for a systematic study of particles with hardening plastic behavior, but no elastic unloading and formation of bonding strength. It is worth noting that the development of these relationships and, by extension, of predictive contact mechanics formulations for highly confined systems, is key to better design, optimize and control many manufacturing processes widely used in pharmaceutical, energy, food, ceramic and metallurgical industries.

\begin{table}[htbp]
    \centering \footnotesize
{\def\arraystretch{2.5}
    \begin{tabular}{l|l|c|c|cc}
    	\hline
    	Property
    	& Evolution during compaction & Eqn.
    	& Parameter & Material 1 & Material 2
    	\\
        \hline
        Coordination number
        &
        $    \bar{Z} = \bar{Z}_c
            +
            \bar{Z}_0 (\rho^{\mbox{\tiny in-die}}_{\mbox{\tiny max}}-\rho_{c,\bar{Z}})^{\theta}
        $
        &
        \eqref{Eqn-CoordinationFit}
        &
        $\bar{Z}_c$
        &
        $4.366$
        &
        $4.439$
        \\
        & & &
		$\rho_{c,\bar{Z}}$
        &
		$0.5081$
		&
		$0.5151$
		\\
		& & &
		$\theta$
		&
		$0.5535$
		&
		$0.5333$
		\\
        \hline
		Punch pressure
		&
        $     \sigma_{\mbox{\tiny punch}} 
             = 
             K_{\mbox{\tiny P}}  	(\rho^{\mbox{\tiny in-die}}_{\mbox{\tiny max}} - \rho_{c,\bar{Z}})^{\beta_{\mbox{\tiny P}} }
        $
        &
        \eqref{Eqn-AppliedPressure}
        &
        $K_{\mbox{\tiny P}}$
        &
        $210$~MPa
        &
        $1.265$~GPa
        \\
        & & &
        $\beta_{\mbox{\tiny P}}$
        &
        $1.561$
        &
        $1.541$
        \\
        \hline
        In-die elastic recovery
        &
        $ 	\epsilon_{\rho} 
        	= 
        	\epsilon_{0} \frac{\rho^{\mbox{\tiny in-die}}_{\mbox{\tiny max}} - \rho_{c,\epsilon}}{1-\rho_{c,\epsilon}}
        	\approx
        	\epsilon_{H} 
       	$
       	&
       	\eqref{Eqn-ElasticRecovery}
       	&
       	$\epsilon_{0}$
       	& 
       	$3.55\%$
       	&
       	$4.579\%$
       	\\
       	& & &
       	$\rho_{c,\epsilon}$
       	&
       	$\approx \rho_{c,\bar{Z}}$
       	&
       	$\approx \rho_{c,\bar{Z}}$
       	\\
       	\hline
       	Residual radial pressure
       	&
       	$\sigma_{\mbox{\tiny residual}}=  
       	\sigma_{\mbox{\tiny res},0} ~ \frac{\rho^{\mbox{\tiny in-die}}_{\mbox{\tiny max}}  (\rho^{\mbox{\tiny in-die}}_{\mbox{\tiny max}} - \rho_{c,e})}{1-\rho_{c,e}}$
       	&
       	\eqref{Eqn-ResidualPressure}
       	&
       	$\sigma_{\mbox{\tiny res},0}$
       	&
       	$9.719$~MPa
       	&
       	$59.51$~MPa
       	\\
       	Ejection pressure
       	&
       	$\sigma_{\mbox{\tiny ejection}}= 
       	 \mu ~ \frac{\sigma_{\mbox{\tiny res},0}~16 W}{\rho_t \pi D^3 } ~ \frac{\rho^{\mbox{\tiny in-die}}_{\mbox{\tiny max}} - \rho_{c,e}}{1-\rho_{c,e}}$
       	&
       	\eqref{Eqn-EjectionPressure}
       	&
       	$\rho_{c,e}$
       	&
       	$0.6196$
       	&
       	$0.6093$
        \\
        \hline  
		In-die minimum relative density
		&
		$\rho^{\mbox{\tiny in-die}}_{\mbox{\tiny min}}
		=
		\rho^{\mbox{\tiny in-die}}_{\mbox{\tiny max}}
		(1-\epsilon_{\rho})
		$
		&
		\eqref{Eqn-DefinitionElasticRecovery}
		&
		&
		&
        \\
        \hline  
        Bonding surface area
        &
        $	\bar{A}_{\mbox{\tiny b}} 
        	=
        	\bar{A}_{\mbox{\tiny b},0} \frac{\rho^{\mbox{\tiny in-die}}_{\mbox{\tiny min}} - \rho_{c,b}}{1-\rho_{c,b}}
        $
        &
        \eqref{Eqn-BondingSurface}
        &
        $\bar{A}_{\mbox{\tiny b},0}$
        &
        $1.030$
        &
        $0.9883$
        \\
        & & &
        $\rho_{c,b}$
        &
        $\approx \rho_{c,\bar{Z}}$
        &
        $\approx \rho_{c,\bar{Z}}$        
        \\
        \hline
        Young's modulus of compact
        &
		$E_{\mbox{\tiny tablet}} 
		=
		E_{0} \left[ \frac{\rho^{\mbox{\tiny in-die}}_{\mbox{\tiny min}} - \rho_{c,E}}{1-\rho_{c,E}} \right]^n$
		&
		\eqref{Eqn-Young}
		&
		$E_{0}$
		&
		$2.224$~GPa
		&
		$9.345$~GPa
		\\
		& & &
		$n$
		&
		$0.2873$
		&
		$0.285$
		\\
		& & &
		$\rho_{c,E}$
		&
		$\approx \rho_{c,e}$       
		&
		$\approx \rho_{c,e}$       
        \\
        \hline    
        Poisson's ratio of compact
        & $\nu_{\mbox{\tiny tablet}}$ & & &
        $\approx 0.21$
        &
        $\approx 0.26$
        \\
        \hline 	
    \end{tabular}
    }
    \caption{Summary of microstructure formation and evolution during compaction, unloading and ejection. Material 1: $E=5$~GPa, $\nu=0.25$, $\kappa=150$~MPa, $m=2$, $K_{Ic}=1.26$~MPa m$^{1/2}$. Material 2: $E=30$~GPa, $\nu=0.25$, $\kappa=900$~MPa, $m=2$, $K_{Ic}=6.19$~MPa m$^{1/2}$.}
    \label{Table-SummaryEvolution}
\end{table}

\section*{Acknowledgements}
The author gratefully acknowledges the support received from the National Science Foundation grant number CMMI-1538861, from Purdue University's startup funds, and  from the Network for Computational Nanotechnology (NCN) and nanoHUB.org. The author also thanks Prof. Alberto Cuiti\~{n}o for interesting discussions.

\bibliographystyle{plainnat}

\end{document}